\documentclass[12pt]{article}

\usepackage{amsfonts}
\usepackage{amsmath}
\usepackage{epsfig}
\usepackage{latexsym}
\usepackage{graphicx}
\usepackage{color}
\usepackage{amssymb}
\numberwithin{equation}{section}
\allowdisplaybreaks
\usepackage{epstopdf}

\def\spa#1{\phantom{\fbox{\rule[-#1cm]{0cm}{0cm}}}}

\def\be{\begin{equation}}
\def\ee{\end{equation}}
\def\bea{\begin{eqnarray}}
\def\eea{\end{eqnarray}}
\def\bequ{\begin{equation}}
\def\eequ{\end{equation}}

\def\del{\partial}

\renewcommand{\thefootnote}{\fnsymbol{footnote}}

\newcommand{\eq} {equation}
\newcommand{\eqa} {eqnarray}
\newcommand{\NN} {\mbox {$\nonumber$}}

\begin{document}

\hfuzz=100pt
\title{S-duality invariant perturbation theory\\
improved by holography 
}
\author{
Abhishek Chowdhury$^{a}$\footnote{abhishekAThri.res.in},$\ $  
Masazumi Honda$^{b}$\footnote{masazumi.hondaATweizmann.ac.il} $\ $
and$\ $ Somyadip Thakur$^{c}$\footnote{somyadipATtheory.tifr.res.in}
  \spa{0.5} \\
\\
$^{a}${\small{\it Harish-Chandra Research Institute,}}
\\ {\small{\it Chhatnag Road, Jhusi, Allahabad 211019, India}} \\
$^{b}${\small{\it Department of Particle Physics and Astrophysics,}}
\\ {\small{\it Weizmann Institute of Science, Rehovot 7610001, Israel}} \\
$^{c}${\small{\it Tata Institute of Fundamental Research, Mumbai 400005, India}} \\
}
\date{\small{July 2016}}

\maketitle
\thispagestyle{empty}
\centerline{}
\vspace{-3em}
%
\begin{abstract}
We study anomalous dimensions of unprotected low twist operators
in the four-dimensional $SU(N)$ $\mathcal{N}=4$ supersymmetric Yang-Mills theory.
We construct a class of interpolating functions 
to approximate the dimensions of the leading twist operators
for arbitrary gauge coupling $\tau$.
The interpolating functions
are consistent with previous results 
on the perturbation theory, holographic computation and full S-duality.
We use our interpolating functions 
to test a recent conjecture by the $\mathcal{N}=4$ superconformal bootstrap
that upper bounds on the dimensions are saturated 
at one of the duality-invariant points $\tau =i$ and $\tau =e^{i\pi /3}$.
It turns out that
our interpolating functions have maximum at $\tau =e^{i\pi /3}$,
which are close to the conjectural values 
by the conformal bootstrap.
In terms of the interpolating functions,
we draw the image of conformal manifold  
in the space of the dimensions.
We find that
the image is almost a line
despite the conformal manifold is two-dimensional.
We also construct interpolating functions 
for the subleading twist operator
and study level crossing phenomenon
between the leading and subleading twist operators.
Finally
we study the dimension of the Konishi operator in the planar limit. 
We find that our interpolating functions match 
with numerical result obtained by Thermodynamic Bethe Ansatz very well.
It turns out that
analytic properties of the interpolating functions
reflect an expectation on a radius of convergence
of the perturbation theory.
\end{abstract}
\vfill
\noindent {\small 
HRI/ST/1601, TIFR/TH/16-18, WIS/06/16-JUN-DPPA  }
\renewcommand{\thefootnote}{\arabic{footnote}}
\setcounter{footnote}{0}
\newpage
\setcounter{page}{1}
\tableofcontents

\section{Introduction}
In the last two decades,
there has been huge progress in understanding 
the four-dimensional $\mathcal{N}=4$ supersymmetric Yang-Mills theory (SYM).
The $\mathcal{N}=4$ SYM is supposed to have many special properties 
such as superconformal symmetry \cite{Avdeev:1980bh,Grisaru:1980nk,Caswell:1980ru,Sohnius:1981sn,Howe:1983sr},
UV finiteness \cite{Howe:1983sr,Mandelstam:1982cb,Brink:1982wv}, 
S-duality \cite{Goddard:1976qe,Montonen:1977sn,Osborn:1979tq,Sen:1994yi,Vafa:1994tf},
integrability \cite{Beisert:2003yb,Beisert:2010jr},
dual conformal symmetry \cite{Drummond:2006rz} and so on. 
The $\mathcal{N}=4$ SYM also provides 
the canonical example of the AdS/CFT correspondence \cite{Maldacena:1997re,Gubser:1998bc,Witten:1998qj},
where the $\mathcal{N}=4$ SYM is dual to 
type IIB superstring on $AdS_5 \times S^5$.
While the AdS/CFT correspondence 
has stimulated our understanding on the $\mathcal{N}=4$ SYM
and vice versa,
most of the progress is based on weak coupling perturbation theory, planar limit or protected observables\footnote{
To our knowledge,
only exceptions so far
are the conformal bootstrap \cite{Rattazzi:2008pe,Rychkov:2009ij,ElShowk:2012ht,ElShowk:2012hu,Beem:2013qxa}
and Monte Carlo simulation 
with appropriate regularizations \cite{Honda:2010nx,Honda:2011qk,Catterall:2012yq,Honda:2013nfa}.
}.

Recently it has turned out that
the conformal bootstrap approach \cite{Rattazzi:2008pe,Rychkov:2009ij,ElShowk:2012ht,ElShowk:2012hu} 
is very powerful tool also for the $\mathcal{N}=4$ SYM \cite{Beem:2013qxa} 
(see also \cite{Alday:2013opa,Alday:2014qfa,Alday:2014tsa,Alday:2016htq}).
Indeed 
the $\mathcal{N}=4$ superconformal bootstrap \cite{Beem:2013qxa} 
gives strong constraints
on dimensions of unprotected leading twist\footnote{
Twist is dimension minus spin.
} operators with various spins, 
which are $SU(4)_R$ singlets and non-BPS primary operators
belonging to long representation of the $\mathcal{N}=4$ superconformal group.
The leading twist operators in the $\mathcal{N}=4$ SYM at classical level
are the so-called twist-two operators defined by
\begin{\eq}
\mathcal{O}_M =  {\rm Tr}\phi^I D^M \phi^I , \qquad M=0,2,4,\cdots ,
\label{eq:twist2}
\end{\eq}
where $\phi^I$ is the adjoint scalar.
The $\mathcal{N}=4$ superconformal bootstrap 
finds rigorous bounds on the dimensions within numerical errors,
which are fully non-perturbative even for finite $N$ and 
independent of the complex gauge coupling 
$\tau = \frac{\theta}{2\pi}+\frac{4\pi i}{g_{\rm YM}^2}$.
The authors in \cite{Beem:2013qxa}
have also conjectured that 
the upper bounds on the dimensions are 
saturated at either $\tau =i$ or $\tau =e^{i\pi /3}$,
which are the duality invariant points 
under $\mathbf{S}$-transformation and $(\mathbf{T}\cdot\mathbf{S})$-transformation,
respectively.

Main purpose of this paper is
to study the dimensions of the leading twist operators
in the $SU(N)$ $\mathcal{N}=4$ SYM 
by a completely different approach. We find approximate formula of the dimensions
by resumming the perturbative data 
with help of 
the S-duality \cite{Goddard:1976qe,Montonen:1977sn,Osborn:1979tq,Sen:1994yi,Vafa:1994tf}
and AdS/CFT correspondence \cite{Maldacena:1997re,Gubser:1998bc,Witten:1998qj}.
We approximate the dimensions through interpolating functions
which interpolates between two perturbative expansions 
around two different points in parameter space. 
The standard approach is to apply (two-point) Pad\'e approximation, 
which is a rational function encoding the two expansions up to some orders.  
Recently Sen constructed another type of interpolating function, 
which has the form of a Fractional Power of Polynomial (FPP) \cite{Sen:2013oza}.  
A more general form of the interpolating function with the form of Fractional Powers of Rational function (FPR) has been constructed out 
by one of the current authors \cite{Honda:2014bza}.
It has turned out that these interpolating functions usually provide
better approximations than each perturbative expansion in intermediate
regime of the parameter.
See \cite{Sen:2013oza,Honda:2014bza,Asnin:2007rw,Banks:2013nga,Pius:2013tla,Alday:2013bha,Honda:2015ewa}
for various applications\footnote{
  There are other types of interpolating functions
  \cite{Kleinert:2001ax} and \cite{Yukalov:2015wca}, 
which are not special cases of the FPR.  
  }.

In this paper, 
we construct a class of interpolating functions to approximate the anomalous dimensions,
which are consistent with 
known results on the perturbation theory, holographic computation and full S-duality $SL(2,\mathbb{Z})$.
In other words,
our modular invariant interpolating functions reproduce the correct weak coupling expansion and large 't Hooft coupling limit in the planar limit.
Such attempt was initiated in \cite{Beem:2013hha},
which has constructed interpolating functions invariant 
under one particular element of $SL(2,\mathbf{Z})$ 
either $\mathbf{S}$ or $(\mathbf{T}\cdot \mathbf{S})$-transformations.
After a while,
Alday and Bissi constructed  
a class of interpolating functions \cite{Alday:2013bha},
which is similar to FPP \cite{Sen:2013oza}
but invariant under all elements of $SL(2,\mathbf{Z})$.
Here we construct a new class of interpolating functions
by combining the ideas of FPR \cite{Honda:2014bza} and Alday-Bissi \cite{Alday:2013bha},
and further imposing a consistency with the previous holographic results.

Our interpolating functions give predictions 
for arbitrary values of $N$ and the complex gauge coupling $\tau$.
We compare our result with the recent results 
by the $\mathcal{N}=4$ superconformal bootstrap
and test the conjecture that 
the upper bounds on the dimensions are saturated 
at one of the duality-invariant points $\tau =i$ and $\tau =e^{i\pi /3}$.
As a conclusion
we find that
when we expect reasonable approximation by the interpolating functions,
the interpolating functions have 
their maximal values at $\tau =e^{i\pi /3}$, close to the conjectural values 
of the $\mathcal{N}=4$ superconformal bootstrap. 

In terms of the interpolating functions,
we also study an image of conformal manifold 
in the space of the dimensions of the leading twist operators.
We find that despite varying the coupling $\tau$ in the (real) two-dimensional region,
the image is a very narrow line,
which is almost one-dimensional. 
The narrow line is almost straight for $N=2$
as in the result by different interpolating functions \cite{Beem:2013hha}
while it is somewhat curved for $N\geq 3$ contrary to \cite{Beem:2013hha}.

We also construct interpolating functions for the subleading twist operator
and study level crossing phenomenon
between the leading and subleading twist operators.
We use the terminology ``level crossing" in the following two senses.
As we increase the coupling,
the dimensions of the leading and subleading operators approach each other with two possibilities. Firstly, there is no operator mixing due to additional protected symmetries and their dimensions cross over.
Secondly, operator mixing occurs and the dimensions of new eigenstates repel.
We refer to the both as level crossing.
We check that
the interpolating functions for the leading and subleading twist operators with spin-0
do not cross each other for finite $N$.
Namely the level crossing in the first sense does not occur for finite $N$.
For large but finite $N$,
we observe that
the dimension of the leading twist operator becomes very close to the subleading one around $\tau =i$ and $\tau =e^{\pi i /3}$.
This implies that 
the level crossing in the second sense occurs for large but finite $N$.

We also study the dimension of the Konishi operator\footnote{
Note that the Konishi operator is no longer the leading twist operator
for large 't Hooft coupling in the planar limit.
} in the planar limit. 
We construct interpolating functions consistent 
with the weak coupling expansion and holographic computation.
We compare our result with numerical data obtained 
by Thermodynamic Bethe Ansatz (TBA) and
find that our interpolating functions match very well with the TBA result.
We also discuss that
analytic property of the interpolating function
reflects expectations on radius of convergence
from the weak coupling perturbation theory.

This paper is organized as follows.
In section \ref{sec:anomdim} 
we briefly explain the previous results
obtained by the weak coupling perturbation theory, holographic computation,
and superconformal bootstrap.
In section \ref{sec:interpolating} 
we begin with introducing interpolating functions
constructed previously.
Then we construct our interpolating functions for the anomalous dimensions,
which are consistent 
with the known results on the weak coupling perturbation theory,
holographic results and full S-duality.
Finally we discuss 
which of our interpolating functions
would give the best approximation.
Section \ref{sec:results_finite} is the main section of this paper.
We compare our results with the recent results 
by the $\mathcal{N}=4$ superconformal bootstrap.
We also draw the image of the conformal manifold 
in the space of the dimensions of the leading twist operators 
and study the level crossing phenomenon
between the leading and subleading twist operators for finite $N$.
In section \ref{sec:results_planar}  
we study the dimension of the Konishi operator in the planar limit.
Section \ref{sec:conclusion} is devoted to conclusion and discussions.

\section{Previous results on leading twist operators}
\label{sec:anomdim}
In this paper 
we mainly study the dimensions of the leading twist operators
in the 4d $SU(N)$ $\mathcal{N}=4$ SYM
by using the technique of interpolating functions.
Before introducing the interpolating functions, 
we review some  relevant previous results know in the literature.

The leading twist operators
under consideration
are $SU(4)_R$ singlet and non-BPS primary operators
belonging to long representation of the $\mathcal{N}=4$ superconformal group.
At classical level,
these operators are 
so-called twist-two operators:
\[
\mathcal{O}_M =  {\rm Tr}\phi^I D^M \phi^I ,
\]
where $\phi^I$ are the  adjoint scalars in the $\mathcal{N}=4$ SYM and
$I$ is an index in the $\bf{6}$ of $SU(4)_R$.
The leading twist operator 
has the dimension $(2+M)$ classically 
but receives quantum corrections:
\begin{\eq}
\Delta_M  (\tau ,N ) =2+M+\gamma_M (\tau ,N) ,
\end{\eq}
depending on the complex gauge coupling
\begin{\eq}
\tau = \frac{\theta}{2\pi} +\frac{i}{g} ,\quad 
{\rm with}\ \  g=\frac{g_{\rm YM}^2}{4\pi} .
\end{\eq}
It is known that 
its weak coupling perturbative expansion is independent of $\theta$
while non-perturbative corrections depend on $\theta$ generically\footnote{
There is a statement that
the two-point function of Konishi operator
does not receive instanton corrections \cite{Bianchi:2001cm}.
Recently it is stated in \cite{Alday:2016tll} that
the instanton correction starts at $\mathcal{O}(g^4 )$ .
}.
It is expected that 
the $\mathcal{N}=4$ SYM possesses 
the S-duality \cite{Goddard:1976qe,Montonen:1977sn,Osborn:1979tq,Sen:1994yi,Vafa:1994tf}
described by the $SL(2,\mathbb{Z})$ transformation
\begin{\eq}
\mathbf{h}\cdot \tau = \frac{a\tau +b}{c\tau +d} ,\quad {\rm where }\ ad-bc=1,\ a,b,c,d\in\mathbb{Z} ,
\label{eq:duality}
\end{\eq}
which is a combination of $\mathbf{S}$- and $\mathbf{T}$-transformations:
\begin{\eq}
\mathbf{S}\cdot \tau = -\frac{1}{\tau} ,\quad
\mathbf{T}\cdot \tau = \tau +1 .
\end{\eq}
Note that
there are two special values of $\tau$:
\begin{\eq}
\tau =\tau_S = i ,\quad
\tau =\tau_{TS} =e^{i\pi /3},
\end{\eq}
which are invariant under $\mathbf{S}$-transformation and $(\mathbf{T}\cdot\mathbf{S})$-transformation, respectively.
In this paper 
we assume 
$SL(2,\mathbb{Z})$ invariance of 
the dimensions of the leading-twist operators:
\begin{\eq}
\Delta_M (\mathbf{h}\cdot \tau ) =\Delta_M (\tau ) ,\quad 
\gamma_M (\mathbf{h}\cdot \tau ) =\gamma_M (\tau ),
\end{\eq}
and construct the interpolating functions based on this assumption. 

\subsection{Weak coupling expansion}
In perturbative regime,
the leading twist operator is 
the twist-two operator $\mathcal{O}_M$ \eqref{eq:twist2},
which is the Konishi operator especially for $M=0$ (see 
e.g. \cite{Bianchi:2001cm,Eden:2012fe}).
The anomalous dimension of the twist-two operator
has been computed up to four-loop for $M=0,2$ and three-loop for $M=4$
in the weak coupling perturbation theory\footnote{
If it was limited to the planar limit,
there are higher order computations \cite{Bajnok:2012bz,Gromov:2014bva,Marboe:2014gma}.
} \cite{Kotikov:2004er,Kotikov:2007cy,Fiamberti:2007rj,Fiamberti:2008sh,Bajnok:2008qj,Velizhanin:2008jd,Velizhanin:2009gv}:
\begin{\eqa}
\gamma_0 (\tau ,N)
&=& \frac{3N}{\pi} g -\frac{3N^2}{\pi^2} g^2 +\frac{21N^3}{4\pi^3} g^3
 +\Biggl[ -39 +9\zeta (3) -45\zeta (5) \left( \frac{1}{2} +\frac{6}{N^2} \right)  \Biggr] \frac{N^4 g^4}{4\pi^4} 
+\mathcal{O}(g^5 ) ,\NN\\
\gamma_2 (\tau ,N)
&=& \frac{25N}{6\pi} g -\frac{925N^2}{216\pi^2} g^2 +\frac{241325N^3}{31104\pi^3} g^3 \NN\\
&& +\Biggl[ -\frac{8045275}{2187} +\frac{114500\zeta (3)}{81} -\frac{25000\zeta (5)}{9}
+\frac{8400+28000\zeta (3) -100000\zeta (5)}{3N^2}   \Biggr] \frac{N^4 g^4}{(4\pi )^4}  \NN\\
&&+\mathcal{O}(g^5 ) , \NN\\
\gamma_4 (\tau ,N)
&=& \frac{49N}{10\pi} g -\frac{45619N^2}{9000\pi^2} g^2 +\frac{300642097N^3}{32400000\pi^3} g^3 
+\mathcal{O}(g^4 ) .
\label{eq:weak}
\end{\eqa}
These data
will be used to construct our interpolating functions
in subsequent sections.

\subsection{Supergravity limit}
The $\mathcal{N}=4$ SYM is expected to be dual to
type IIB supergravity on $AdS_5 \times S^5$
at large 't Hooft coupling $\lambda =gN$ \cite{Maldacena:1997re,Gubser:1998bc,Witten:1998qj} in the planar limit. 
In the planar limit,
the anomalous dimension of the twist-two operator $\mathcal{O}_M$ \eqref{eq:twist2}
typically grows as $\sim\lambda^{1/4}$ \cite{Gubser:1998bc}
and the twist-two operator is no longer leading twist operator for large 't Hooft coupling.
The leading twist operator in the supergravity limit is a double trace operator
which has a schematic form
\begin{\eq}
{\rm tr}(\phi^{(i} \phi^{j)} ) D^M {\rm tr}(\phi^{(i} \phi^{j)} ) ,
\label{eq:20prime}
\end{\eq}
where ${\rm tr}(\phi^{(i} \phi^{j)})$ is 
symmetric traceless part of ${\rm tr}(\phi^{i} \phi^{j})$
and chiral primary operator belonging to $\mathbf{20'}$ representation\footnote{
This has the Dynkin label $[0,2,0]$.
} of $SU(4)_R$.
The double trace operator \eqref{eq:20prime} is not protected in general
but the large-$N$ factorization implies that
the dimension becomes the sum of the protected single trace operators and therefore protected in the planar limit.
One can compute the anomalous dimension of the double trace operator \eqref{eq:20prime} by the supergravity \cite{Dolan:2001tt,D'Hoker:1999jp,Arutyunov:2000ku}
and then the one of the leading twist operator in the supergravity limit is
\begin{\eq}
\gamma_0^{\rm SUGRA} (N) =2 - \frac{16}{N^2} ,\quad
\gamma_2^{\rm SUGRA} (N) =2 - \frac{4}{N^2} ,\quad
\gamma_4^{\rm SUGRA} (N) =2 - \frac{48}{25N^2} .
\label{eq:result_gravity}
\end{\eq}
Note that
the first terms are easily understood by the large-$N$ factorization.

\subsection{The $\mathcal{N}=4$ superconformal bootstrap}
In the past few years 
the $\mathcal{N}=4$ superconformal bootstrap approach \cite{Beem:2013qxa}
has obtained a relatively satisfying 
 upper bounds on the dimensions of the unprotected leading twist operators
by studying the four-point function 
(see also \cite{Alday:2013opa,Alday:2014qfa,Alday:2014tsa})
\begin{\eq}
\langle \mathcal{O}_{\bf 20'}^{I_1}(x_1 ) \mathcal{O}_{\bf 20'}^{I_2}(x_2 )
\mathcal{O}_{\bf 20'}^{I_3}(x_3 ) \mathcal{O}_{\bf 20'}^{I_4}(x_4 )\rangle ,
\end{\eq}
where $\mathcal{O}_{\bf 20'}^{I}$ is
a superconformal primary scalar operator of dimension two
in energy-momentum tensor multiplets
transforming as $\mathbf{20'}$ representation in $SU(4)_R$.
The $\mathcal{N}=4$ superconformal symmetry
allows us to describe the four-point function
in terms of the $\mathcal{N}=4$ superconformal block \cite{Eden:2000bk,Dolan:2001tt,Arutyunov:2001mh,Eden:2001ec}.

\begin{table}[t]
\begin{center}
  \begin{tabular}{|c||c | c| c|  }
  \hline                                           & $SU(2)$ & $SU(3)$ & $SU(4)$  \\
\hline \hline Strict upper bound on $\gamma_0$   &   1.05    &   1.38     &    1.59      \\
\hline Corner value on $\gamma_0$   &   0.93    &   1.24    &    1.47      \\ 
\hline \hline Strict upper bound on $\gamma_2$   &    1.32    &   1.66     &   1.80      \\
\hline Corner value on $\gamma_2$   &   1.28    &   1.60     &   1.75     \\ 
\hline \hline Strict upper bound on $\gamma_4$   &    1.55    &   1.80     &   1.89         \\
\hline Corner value on $\gamma_4$   &   1.53    &   1.79     &   1.88        \\ \hline
  \end{tabular}
\end{center}
\caption{Bounds and corner values from Superconformal bootstrap \cite{Beem:2013qxa}.}
\label{tab:bound}
\end{table}
In 
 \cite{Beem:2013qxa} 
the upper bounds on the dimensions of the leading twist operators
with spin-0, 2 and 4, were obtained 
which are rigorous within numerical errors.
Exclusion plots on the anomalous dimensions $(\gamma_0 , \gamma_2 , \gamma_4 )$ 
are presented in fig.1 of \cite{Beem:2013qxa}.
While the shape of not-excluded region roughly looks like a cube,
its precise shape is complicated function of $(\gamma_0 ,\gamma_2 ,\gamma_4 )$.
The ``bound" values are listed in table \ref{tab:bound} for  each maximal value of $(\gamma_0 ,\gamma_2 ,\gamma_4 )$ in the not-excluded region.

The ``bound" values in table \ref{tab:bound} are somewhat conservative.
This is because 
if actual values of $(\gamma_0 ,\gamma_2 )$ were not equal to the bound values in table \ref{tab:bound}
(namely smaller than the bound values) for example,
then possible value of $\gamma_4$ would generically be more strongly constrained. 
Hence we might have better estimates from fig.1 of \cite{Beem:2013qxa} than the bound values in table \ref{tab:bound}.
The authors in \cite{Beem:2013qxa} have conjectured that
this better estimate is given by the value at the corner of the cube-like region and that 
this is saturated by values of $\gamma_M$ at one of duality invariants points $\tau =\tau_S$ or $\tau =\tau_{TS}$.
This conjecture essentially claims the following two things:
\begin{enumerate}
\item The corner value obtained by the conformal bootstrap 
is saturated by the maximal value of $\gamma_M (\tau ,N)$.

\item The maximal value of $\gamma_M (\tau ,N)$ in the physical region of $\tau$
is given by $\tau =\tau_S$ or $\tau =\tau_{TS}$.
\end{enumerate}
The first point of this conjecture is closely related to 
whether the constraints from the conformal bootstrap is sufficiently strong or not.
Namely, the upper bound of the bootstrap is 
greater than all the possible values of the anomalous dimension in general
and may have a gap from the maximal value in principle.
However, if the upper bound is maximally strong, 
then there is no such gap and 
the upper bound is the same as the maximal value
though it is currently unclear if this is true.  
Regarding the second point,
we do not know a priori which value of the coupling realizes the maximal value
but it is natural to expect that
such special thing happens in some special values in the $\tau$-space,
which are only the duality invariant points $\tau =\tau_S$ and $\tau =\tau_{TS}$
to our knowledge.
Main purpose of this paper
is to test the conjecture
by using the interpolating functions.

\section{Interpolating functions}
\label{sec:interpolating}
In this section we introduce some classes of interpolating functions constructed 
in the literature  \cite{Sen:2013oza,Beem:2013hha,Alday:2013bha,Honda:2014bza} and then in the remaining part of this section we 
construct new class of interpolating functions 
for the anomalous dimensions 
of the leading twist operators.
We impose the following conditions to the interpolating functions: 
\begin{enumerate}
\item Real for ${\rm Im}\tau \geq 0$.

\item 
       Small-$g$ expansion agrees with the weak coupling expansion of $\gamma_M (g)$ up to certain order.

\item Invariant under the full $SL(2,\mathbb{Z})$ duality \eqref{eq:duality}:
$ \gamma_M (\mathbf{h}\cdot \tau ) = \gamma_M (\tau ) $.

\item Reproduce the holographic result 
in the planar limit at large 't Hooft coupling\footnote{
Note that
$S$-duality does not automatically imply the holographic matching 
and this condition is not redundant
since the S-duality acts on $g$ rather than $\lambda =gN$. 
To see this explicitly,
let us consider a $S$-duality invariant quantity $f(g,N)$, 
with the 't Hooft expansion $f(g,N ) = \sum_{k=0}^\infty f_{2k}(\lambda )/N^{2k}$.
Then the $S$-duality implies
$\sum_{k=0}^\infty \frac{f_{2k}(\lambda )}{N^{2k}}$ 
$=$ 
$\sum_{k=0}^\infty \frac{f_{2k}(N^2 /\lambda )}{N^{2k}}$ . 
In the leading planar limit,
the LHS has a contribution only from genus-0
while the RHS may receive all genus corrections.
Thus the matching of the small-$\lambda$ expansion
does not imply the holographic matching in general.
}.
\end{enumerate}

\subsection{Interpolating functions without $S$-duality (FPR)}
\label{sec:standardFPR}
Before considering the S-duality invariant interpolating functions,
we introduce usual interpolating functions, 
which can be applied to problems without S-duality.
Suppose 
that we would like to approximate a function $F(g)$,  
which has the small-$g$ expansion around $g=0$
and large-$g$ expansion around $g=\infty$ taking the forms 
\begin{\eq}
F(g) 
= g^a ( s_0 +s_1 g +s_2 g^2 +\cdots )
= g^b ( l_0 +l_1 g^{-1} +l_2 g^{-2} +\cdots ) .
\label{eq:asymptotics}
\end{\eq}
The author in \cite{Honda:2014bza} constructed
the following type of interpolating function for the function $F(g)$:
\begin{\eq}
F_{m,n}^{(\alpha )} (g)
= s_0 g^a \Biggl[ \frac{ 1 +\sum_{k=1}^p c_k g^k}{1 +\sum_{k=1}^q d_k g^k }  \Biggr]^\alpha ,
\label{eq:FPR}
\end{\eq}
where 
\begin{\eq}
p = \frac{1}{2} \left( m+n+1 -\frac{a-b}{\alpha} \right) ,\quad
q = \frac{1}{2} \left( m+n+1 +\frac{a-b}{\alpha} \right) .
\end{\eq}
Here the coefficients $c_k$ and $d_k$ are determined
such that power series expansions around $g=0$ and $g=\infty$
agree with the ones of $F(g)$ up to $\mathcal{O}(g^{a+m+1})$ and $\mathcal{O}(g^{b-n-1})$, respectively.
By construction,
the interpolating function reproduces 
both the small-$g$ and large-$g$ expansions of $F(g)$.
Since this interpolating function is described by
Fractional Power of Rational function, we call this FPR.
Note that we need 
\begin{\eq}
p,q \in \mathbb{Z}_{\geq 0} ,
\end{\eq}
which leads us to
\begin{\eq}
\alpha = \left\{ \begin{matrix}
\frac{a-b}{2\ell +1}  & {\rm for} & m+n:{\rm even} \cr
\frac{a-b}{2\ell}  & {\rm for} & m+n:{\rm odd} \end{matrix} \right. ,\quad
{\rm with}\ \ell \in\mathbb{Z} .
\end{\eq}
If we take $2\ell+1=a-b$ for $a-b \in \mathbb{Z}$ and $m+n$ to be even, 
then this becomes the Pad\'e approximant:
\begin{\eq}
F_{m,n}^{(1 )} (g)
= s_0 g^a \frac{ 1 +\sum_{k=1}^p c_k g^k}{1 +\sum_{k=1}^q d_k g^k } ,
\end{\eq}
while taking $2\ell+1=m+n+1$ ($2\ell=m+n+1 $) for even (odd) $m+n$ gives 
the Fractional Power of Polynomial (FPP):
\begin{\eq}
F_{m,n}^{(1/(m+n+1) )} (g)
= s_0 g^a \Biggl( 1 +\sum_{k=1}^{m+n+1} c_k g^k  \Biggr)^{\frac{b-a}{m+n+1}} ,
\end{\eq}
recently constructed in \cite{Sen:2013oza}.
In next subsection
we will introduce interpolating functions invariant under the full S-duality
inspired by the FPR.
In sec.~\ref{sec:results_planar}
we will use the FPR to study the dimension of the Konishi operator in the planar limit.

\subsection{Modular invariant interpolating functions}
Here we introduce interpolating functions,
which are consistent with the weak coupling expansion \eqref{eq:weak} and full S-duality.
Such an attempt was initiated in \cite{Beem:2013hha},
where the author constructed interpolating functions invariant 
under one specific element of $SL(2,\mathbf{Z})$ such as $\mathbf{S}$- and $(\mathbf{T}\cdot \mathbf{S})$-transformations.
Then Alday-Bissi constructed a class of interpolating functions,
which are similar to FPP 
but invariant under all the elements of $SL(2,\mathbf{Z})$, 
namely modular invariant interpolating functions \cite{Alday:2013bha}.
Here we would like to have a new class of 
modular invariant interpolating functions,
whose form is similar to FPR.

\subsubsection{Alday-Bissi's interpolating function}
Alday and Bissi constructed 
the following type of interpolating function \cite{Alday:2013bha} 
\begin{\eq}
\bar{F}_m^{(s)} (\tau ) = \left( \sum_{k=1}^m c_k E_{s+k} (\tau ) \right)^{-\frac{1}{s+m}} ,
\label{eq:AB}
\end{\eq}
where the coefficient $c_k$ is determined such that
expansion of $\bar{F}_m^{(s)}$ around $g=0$ agrees with
the one of $\gamma_M (\tau )$ up to $\mathcal{O}(g^{m+1} )$.
The building block $E_s (\tau )$ is the non-holomorphic Eisenstein series defined by\footnote{
Note that $s$ can be non-integer and $E_s (\tau )$ has a pole at $s=1$.
Hence we take $s>1$.
}
\begin{\eq}
E_s (\tau ) =\frac{1}{2} \sum_{m,n\in\mathbb{Z}-\{0, 0\}}
\frac{1}{|m+n\tau |^{2s}} ({\rm Im}\tau )^s .
\end{\eq}
Because the Eisenstein series is invariant under the duality transformation \eqref{eq:duality},
the whole interpolating function $\bar{F}_m^{(s)}$ 
is invariant under the full $S$-duality.
The Eisenstein series $E_s (\tau )$ has the weak coupling expansion
\begin{\eq}
E_s (\tau )
= \zeta (2s) g^{-s} +\frac{\sqrt{\pi}\Gamma (s-1/2) }{\Gamma (s)} \zeta (2s-1)g^{s-1}
 +f_s^{\rm np}(q) ,
\end{\eq}
where $f_s^{\rm np}(q)$ is the non-perturbative contribution
containing powers of $q=e^{2\pi i\tau}$ 
(see app.~\ref{app:eisen} for details).
Hence, we easily find that
the expression inside of the bracket of $F_m^{(s)}$ has the small-$g$ expansion
\begin{\eq}
\sum_{k=1}^m c_k E_{s+k} (\tau )
=\sum_{k=1}^m c_k \zeta (2s+2k) g^{-s-k} +\mathcal{O}(g^s )  
=g^{-(s+m)}\sum_{k=1}^m c_k \zeta (2s+2k) g^{m-k} +\mathcal{O}(g^s )  .
\end{\eq}
Thus an appropriate choice of $c_k$ correctly gives the weak coupling expansion of $\gamma_M (\tau )$.
Since the interpolating function is similar to FPP,
it is natural to consider
FPR-like duality invariant interpolating functions as in next subsection.

\subsubsection{FPR-like duality invariant interpolating function}
We propose FPR-like generalization\footnote{
We can also construct FPR-like generalization of the interpolating functions of \cite{Beem:2013hha},
which is invariant under the particular elements of $SL(2,\mathbb{Z})$,
but we do not use it here.
} of the Alday-Bissi's interpolating function:
\begin{\eq}
\tilde{F}_m^{(s,\alpha )} (\tau ) 
= \Biggl[ \frac{\sum_{k=1}^p c_k E_{s+k} (\tau )}{\sum_{k=1}^q d_k E_{s+k} (\tau )} \Biggr]^\alpha ,
\label{eq:FPR_wo_gravity}
\end{\eq}
where we determine the coefficients $c_k$ and $d_k$ such that
expansion of $\tilde{F}_m^{(s,\alpha )}$ around $g=0$ agrees\footnote{
Note that $m$ should be $m\geq 2$
since we need two coefficients at least 
for this interpolating function.
} with
the one of $\gamma_M (\tau )$ up to $\mathcal{O}(g^{m+1} )$.
Matching at $\mathcal{O}(g)$ leads us to
\begin{\eq}
\alpha (-p +q ) =1,\quad
\left( \frac{c_{p}\zeta (2s+2p)}{d_{q}\zeta (2s+2q)} \right)^\alpha = s_1 .
\end{\eq}
Since the interpolating function is invariant 
under $c_k ,d_k \rightarrow \lambda c_k ,\lambda d_k $,
we can take
\begin{\eq}
d_{q} =1 ,
\end{\eq}
without loss of generality.
Imposing matching at other orders leads
\begin{\eq}
p+q-1 = m ,
\end{\eq}
and hence we find
\begin{\eq}
p = \frac{1}{2}\left( m+1-\frac{1}{\alpha }\right) ,\quad
q = \frac{1}{2}\left( m+1+\frac{1}{\alpha }\right) .
\end{\eq}
We also require $p,q \in \mathbb{Z}_{\geq 1}$,
which implies
\begin{\eq}
\alpha = \left\{ \begin{matrix}
\frac{1}{2\ell }  & {\rm for} & m:{\rm odd} \cr
\frac{1}{2\ell +1}  & {\rm for} & m:{\rm even} \end{matrix} \right. ,\quad
{\rm with}\ \ell \in\mathbb{Z} .
\end{\eq}
Note that 
although the interpolating function \eqref{eq:FPR_wo_gravity} is inspired by FPR,
this does not include the Alday-Bissi's interpolating function \eqref{eq:AB} 
as some special case.
In Appendix \ref{app:anotherFPR}
we also construct another type of FPR-like interpolating function invariant under the S-duality,
which includes the Alday-Bissi's interpolating function as a special case.
In next subsection
we will further improve the interpolating functions of the type \eqref{eq:FPR_wo_gravity} by holography.

\subsection{Further improvement by holographic computation}
In previous subsection
we have introduced the FPR-like interpolating functions
consistent with the weak coupling expansion and full S-duality
but not necessarily with the holographic result \eqref{eq:result_gravity}. Here we impose further consistency 
with the holographic computation.
Let us consider
\begin{\eq}
F_m^{(s,\alpha )} (\tau ) 
= \Biggl[ \frac{\sum_{k=1}^p c_k E_{s+k} (\tau )}{\sum_{k=1}^q d_k E_{s+k} (\tau )} \Biggr]^\alpha ,
\label{eq:main_interpolation}
\end{\eq}
which is formally the same as \eqref{eq:FPR_wo_gravity}.
However,
we determine the coefficients $c_k$ and $d_k$ {\it except} $d_1$ such that
expansion of $F_m^{(s,\alpha )}$ around $g=0$ agrees with
the one of $\gamma_M (\tau )$ up to $\mathcal{O}(g^{m+1} )$.
Matching at $\mathcal{O}(g)$ gives
\begin{\eq}
\alpha (-p +q ) =1,\quad
\left( \frac{c_{p}\zeta (2s+2p)}{d_{q}\zeta (2s+2q)} \right)^\alpha = s_1 .
\end{\eq}
Without loss of generality, we can again take $d_{q} =1$.
The remaining coefficient $d_1$ is determined as follows.
Let us consider 't Hooft expansion of the interpolating function\footnote{
Since we do not know $f_4 (\lambda )$,
we take $f_4 (\lambda )=0$ for simplicity.
}:
\begin{\eq}
 F_m^{(s,\alpha )} \left( \frac{iN}{\lambda} \right) 
= f_0 (\lambda ) +\frac{f_2 (\lambda )}{N^2} +\frac{f_4 (\lambda )}{N^4} +\cdots .
\end{\eq}
Then we determine $d_1$  to satisfy
\begin{\eq}
\lim_{\lambda\rightarrow\infty} \left( f_0 (\lambda ) +\frac{f_2 (\lambda )}{N^2} \right)
= \gamma_M^{\rm SUGRA} (N) ,
\end{\eq}
where $\gamma_M^{\rm SUGRA}$ is the result in the supergravity limit
given by \eqref{eq:result_gravity}.
Imposing matching of other orders leads us to
\begin{\eq}
p+q-2 = m ,
\end{\eq}
and therefore we get
\begin{\eq}
p = \frac{1}{2}\left( m+2-\frac{1}{\alpha }\right) ,\quad
q = \frac{1}{2}\left( m+2+\frac{1}{\alpha }\right) .
\end{\eq}
We also require $p,q \in \mathbb{Z}_{\geq 1}$,
which constrains $\alpha$ as
\begin{\eq}
\alpha = \left\{ \begin{matrix}
\frac{1}{2\ell +1 }  & {\rm for} & m:{\rm odd} \cr
\frac{1}{2\ell }  & {\rm for} & m:{\rm even} \end{matrix} \right. ,\quad
{\rm with}\ \ell \in\mathbb{Z} .
\end{\eq}
In this paper
we apply the interpolating function \eqref{eq:main_interpolation} 
to approximate the dimensions of the leading twist operators.
By construction,
the interpolating functions should give good approximations
around $g=0$ for any $(\theta ,N)$, its $SL(2,\mathbb{Z})$ transformations
and the supergravity limit.
It is a priori unclear
how nice the approximations are beyond these regimes.
In general this depends on details of the interpolating functions,
which are specified by the parameters $(m,s,\alpha )$.
Since we know information on the weak coupling expansions 
up to three or four loops,
the numbers of possible $(m,\alpha )$ are finite
but
we have still infinite choices of $s$,
which provide infinite choices of interpolating functions as well.
We would like to know
which $(m,s,\alpha )$ gives the best approximation or
reduce the number of candidates.
In next subsection
we will discuss which interpolating function should give the best approximation
by imposing some physical consistencies.
 
As we argued,
we impose the constant behavior \eqref{eq:result_gravity} to the interpolating functions
in the large $(\lambda ,N)$ limit.
One might wonder 
whether one can construct another modular invariant interpolating functions,
which have the same weak coupling expansions
but the different behaviours $\sim\lambda^{1/4}$ in that regime as in the Konishi operator \eqref{eq:twist2}.
This may not make sense physically
since the dimension of \eqref{eq:twist2} would not be modular invariant\footnote{
There is a proposal that
the Konishi operator belongs to a $SL(2,\mathbb{Z})$ multiplet \cite{Tseng:2002pe}.
But this proposal seems to assume the statement of \cite{Bianchi:2001cm} that
the dimension of the Konishi operator does not receive instanton corrections,
which does not agrees with the recent calculation in \cite{Alday:2016tll}.
}
but this may be useful in future for constructing interpolating functions for other quantities,
which are modular invariants and have different behaviours 
in the classical string regime.
In Appendix~\ref{app:lambda14},
we try to construct a class of modular interpolating functions,
with the same weak coupling expansion and 
$\lambda^{1/4}$ behaviour in the classical string limit.

\subsection{Further constraints on interpolating function}
\label{sec:constraints}
In the previous subsection
we have seen that
we can construct enormous number of interpolating functions \eqref{eq:main_interpolation},
which are consistent with the weak coupling expansions, S-duality and holographic results. 
This situation leads to ``landscape problem of interpolating functions" as 
pointed out in \cite{Honda:2014bza}.
Namely, it is a priori unclear which interpolating function gives the best approximation.
In this subsection we discuss
which value of $(m,s,\alpha )$ would give the best approximation.
As a result,
we will effectively find the best value of $(m,s,\alpha )$ for every spin. 
Because this subsection is not necessary to understand the main results of this paper,
you can skip this subsection if you are interested only in the results.

\subsubsection{Choice of $m$}
By definition,
our interpolating function $F_m^{(s,\alpha )}(\tau )$ reproduces
the correct weak coupling expansion up to $m$-loop correction. 
In general
the best value of $m$ depends 
on details of problems and other parameters of interpolating functions.
Probably most important point on this is
convergence property of the weak coupling expansion.
Namely, 
if the weak coupling expansion was convergent,
then we should take $m$ as large as possible,
while if asymptotic,
then we should be more careful.

Let us gain some intuitions from experiences on one-point Pad\'e approximation
of small parameter expansion.
It is known that
one-point Pad\'e approximation including more terms
often give more precise approximation
even if the small parameter expansion is asymptotic.
For instance,
such behaviour appears in the series $\sum_n (-1)^n n! g^n$.
A sufficient condition for convergence to exact result 
has been found in \cite{Samuel:1995jc}.
Similar results are obtained in two-points approximation by FPR analysis
in 0d $\phi^4$ theory, 
average plaquette in 4d pure $SU(3)$ YM on lattice, and so on.
When we do not know about properties of expansions sufficiently,
we should conservatively choose $m$ to be close to
the optimized value\footnote{
When we have the series $F(g)=\sum_k c_k g^k$,
the optimized value of $k$ at $g=g_\ast$ is determined by
$\left. \frac{\del}{\del k}\log{c_k} \right|_{k=k_\ast} +\log{g_\ast} =0 $.
} $m_\ast (g)$ 
in a range which we would like to approximate.
So, independent of problems,
when we would like to have better approximation in the range $g\in [0,g_\ast ]$
we expect that
larger $m$ gives more precise approximation until $m\simeq m_\ast (g_\ast )$.

In our problem,
we expect that 
the weak coupling expansion is asymptotic and
behaves as $\sim m!$ at $m$-loop for large $m$ as in typical of field theory\footnote{
Since the $\mathcal{N}=4$ SYM is the special case of 4d $\mathcal{N}=2$ theories,
we also expect that
the weak coupling expansion is Borel summable 
from the previous studies \cite{Argyres:2012vv,Argyres:2012ka,Russo:2012kj,Aniceto:2014hoa,Honda:2016mvg}.
}.
We do not know 
whether we should take $m$ to be as large as possible or not.
If this is the case,
then we should take $m$ to be our maximal value, namely $m=4$ for spin-0, 2 and $m=3$ for spin-4.
If not,
then we should think of optimization for the weak coupling expansion
in the range $g\in [0,1]$
since S-duality relates this region to the other region.
Ideally, we would like to know the optimized value $m_\ast (g)$ at $g=1$
but our current information is not sufficient to estimate the optimized value.
However,
from many examples with factorial behaviour,
we expect that 
$m_\ast (g=1)$ is larger than $4$.
Thus we shall take $m=4$ for spin-0, 2 and $m=3$ for spin-4.

\subsubsection{Choice of $s$}
\label{sec:s}
\subsubsection*{Constraints from weak coupling perturbation theory}
The anomalous dimension $\gamma_M (g)$ has the small-$g$ expansion
\begin{\eq}
\gamma_M (g) = \sum_{k=1} s_k g^k ,
\end{\eq}
where only positive integer powers of $g$ appear.
On the other hand, the perturbative part of the interpolating function takes the form
\begin{\eq}
\left. F_m^{(s,\alpha )} (\tau ) \right|_{\rm pert.}
= \Biggl[ \frac{\sum_{k=1}^p c_k g^{-k}\bigl( \zeta (2s+2k) +\frac{\sqrt{\pi}\Gamma (s+k-1/2)}{\Gamma (s+k)} \zeta (2s+2k-1)g^{2s+2k-1} \bigr)}
{\sum_{k=1}^q d_k g^{-k}\bigl( \zeta (2s+2k) +\frac{\sqrt{\pi}\Gamma (s+k-1/2)}{\Gamma (s+k)} \zeta (2s+2k-1)g^{2s+2k-1} \bigr) } \Biggr]^\alpha ,
\end{\eq}
whose small-$g$ expansion contains fractional powers of $g$ for general $s$.
In order to guarantee absence of such fractional powers, we should take
\begin{\eq}
2s\in\mathbb{Z} .
\end{\eq}

\subsubsection*{Constraints from $1/N$ expansion and holography}
In the 't Hooft limit $\lambda =gN={\rm fixed}, N\gg 1$, 
$\gamma_M $ has the following $1/N$-expansion
\begin{\eq}
\gamma_M (\lambda ,N) =\sum_{\ell =0}^\infty \frac{a_\ell (\lambda )}{N^{2\ell}} ,
\end{\eq}
up to instanton corrections.
We have the following two expectations for this expansion.
\begin{enumerate}
\item Since the $\mathcal{N}=4$ SYM has only adjoint fields,
we do not have $\mathcal{O}(1/N^{2\ell +1})$ corrections.

\item  Since the leading twist operators are dual to multi particle states
appearing in the supergravity with $G_N \sim 1/N^2$,
large-$\lambda$ expansion of $a_\ell (\lambda )$ 
can be regarded as $\alpha'$-expansion, 
where $\alpha' \sim 1/\sqrt{\lambda}$.
Hence, 
we expect
that the $\alpha'$-expansion of $a_\ell (\lambda )$
begins with some non-negative integer powers,
namely $\mathcal{O}(\alpha'^0 )$ at lowest\footnote{
If this started with negative powers,
then higher derivative corrections to the SUGRA became very large
in the $\alpha' \rightarrow 0$ limit.
}.
\end{enumerate}
These points can be used 
for constraining interpolating functions
because the interpolating functions may not satisfy these conditions in general.

Indeed we find that
the interpolating functions with $s\in\mathbb{Z}$ have 
odd powers of $1/N$ in the large $N$ expansion
while the interpolating functions with half-odd $s$ do not have this problem.
For example,
the interpolating functions for the spin-0 operator 
with $(m,\alpha )=(4,1/4 )$ and $s\in\mathbb{Z}$
has $\mathcal{O}(1/N^{2s+1})$ corrections.
This means that 
we should take $s$ to be as large as possible
to make ``wrong $1/N$-corrections" as small as possible.

Regarding the second point,
we find that
interpolating functions with half-odd $s$ have strange $\alpha'$ corrections
while those with integer $s$ are completely fine.
For example,
the interpolating function for the spin-0 operator with $(m,\alpha  ,s)=(4,1/4 ,(2\ell -1)/2)$
has $\mathcal{O}(\lambda^{2\ell -1} )$ in large-$\lambda$ expansion of $a_\ell (\lambda)$.
Thus we should take $s$ to be large as possible for $2s\in\mathbb{Z}$. 

\begin{figure}[t]
\begin{center}
\includegraphics[width=7.4cm]{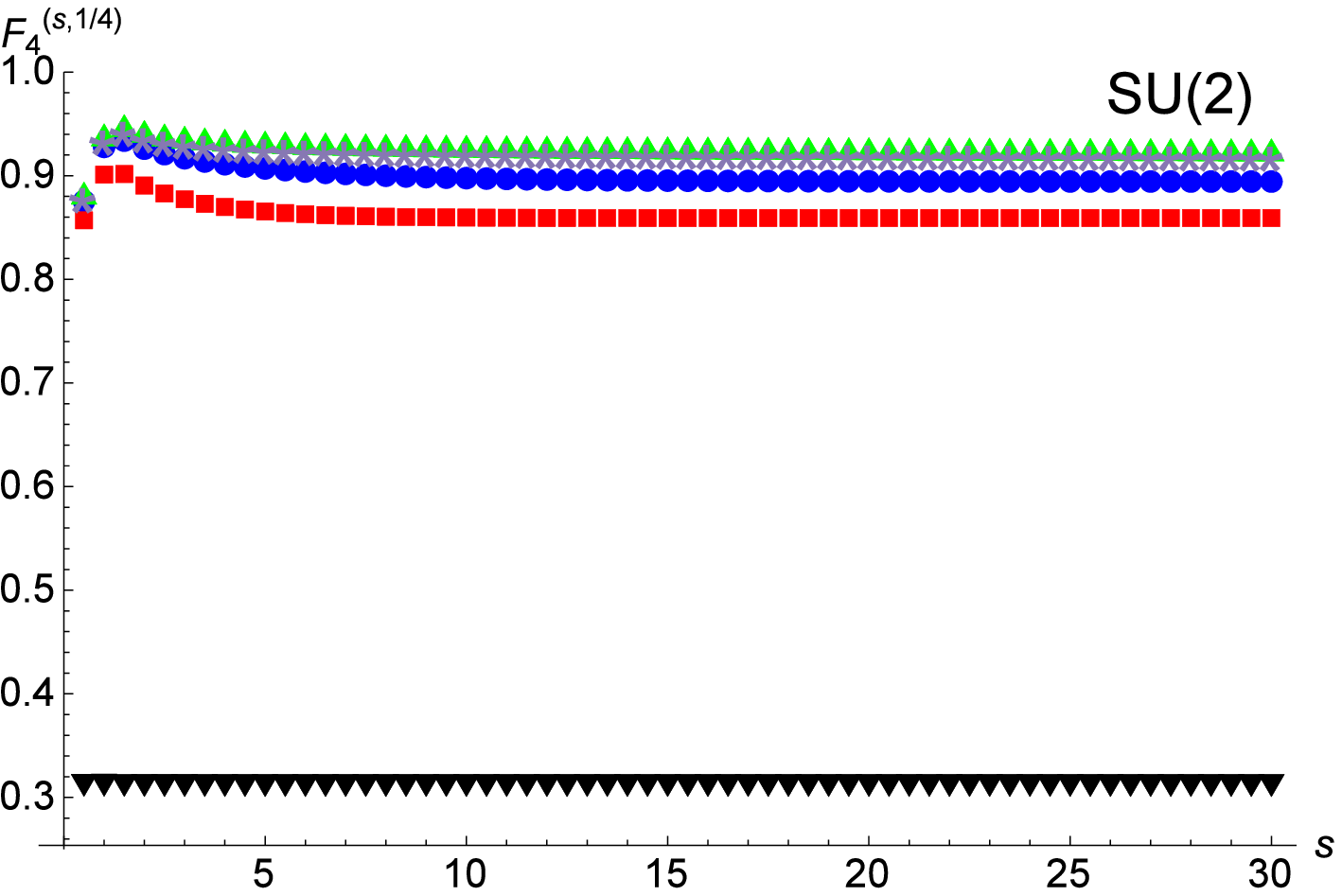}
\includegraphics[width=7.4cm]{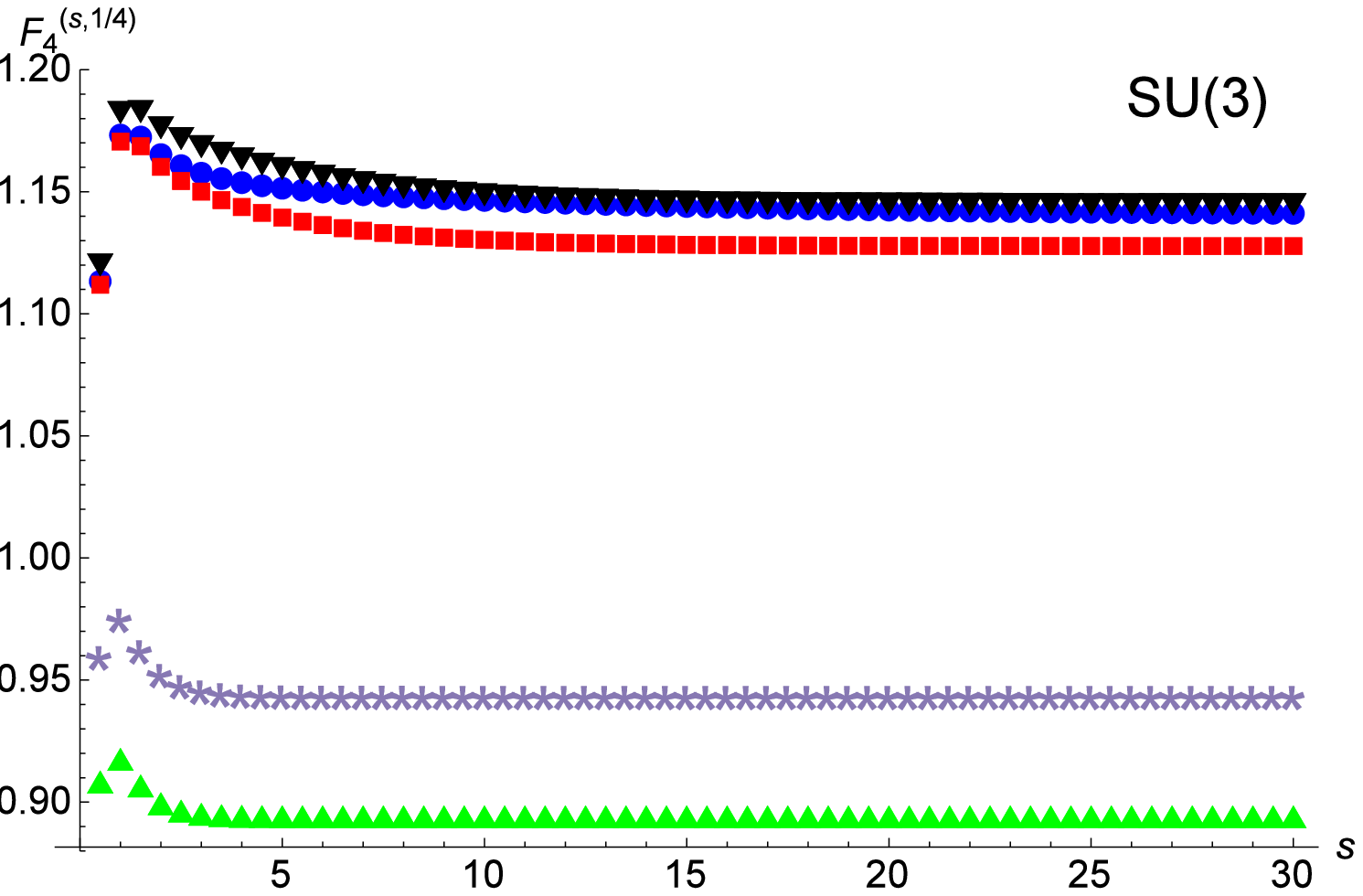}\\
\includegraphics[width=7.4cm]{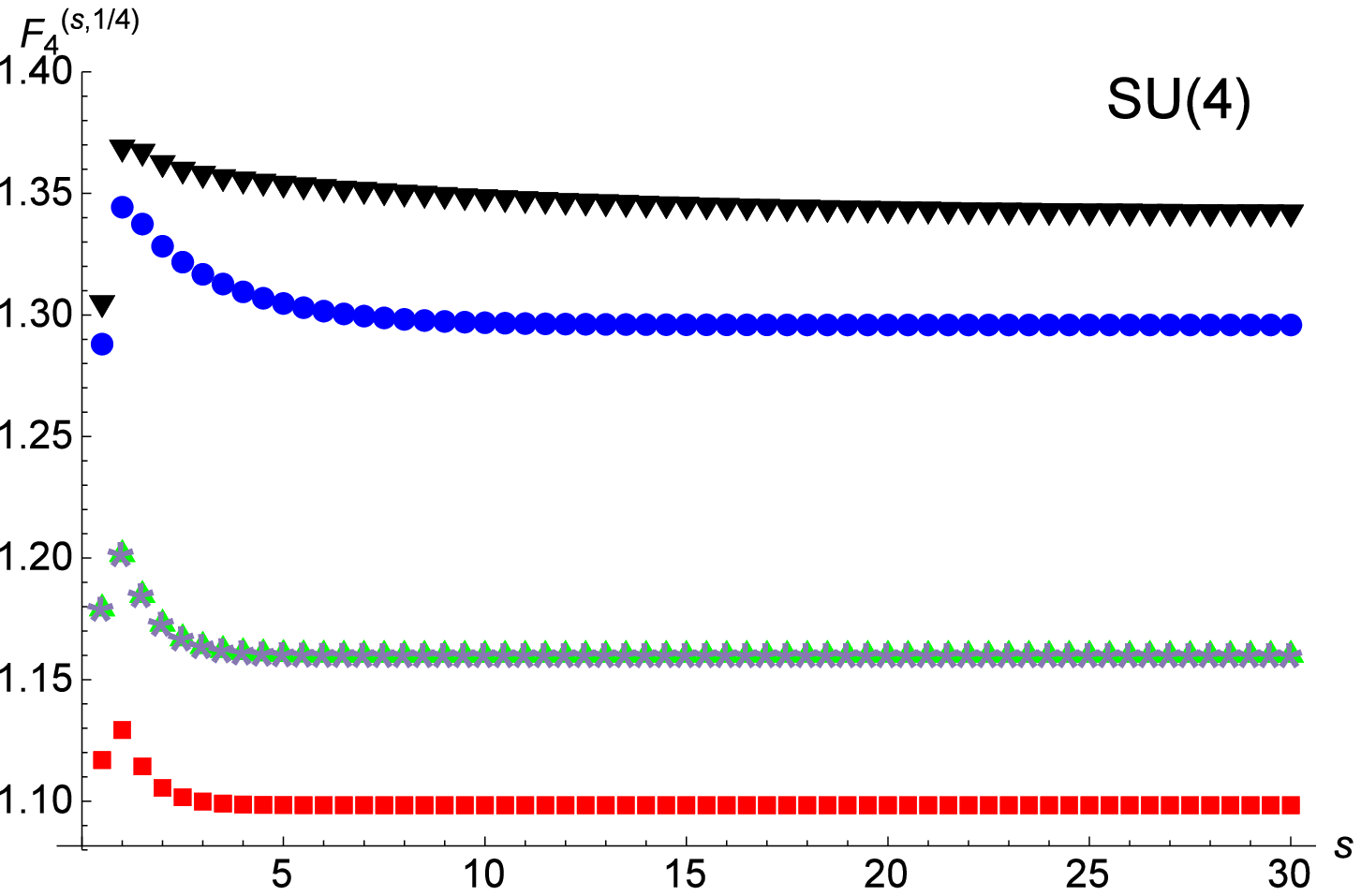}
\includegraphics[width=7.4cm]{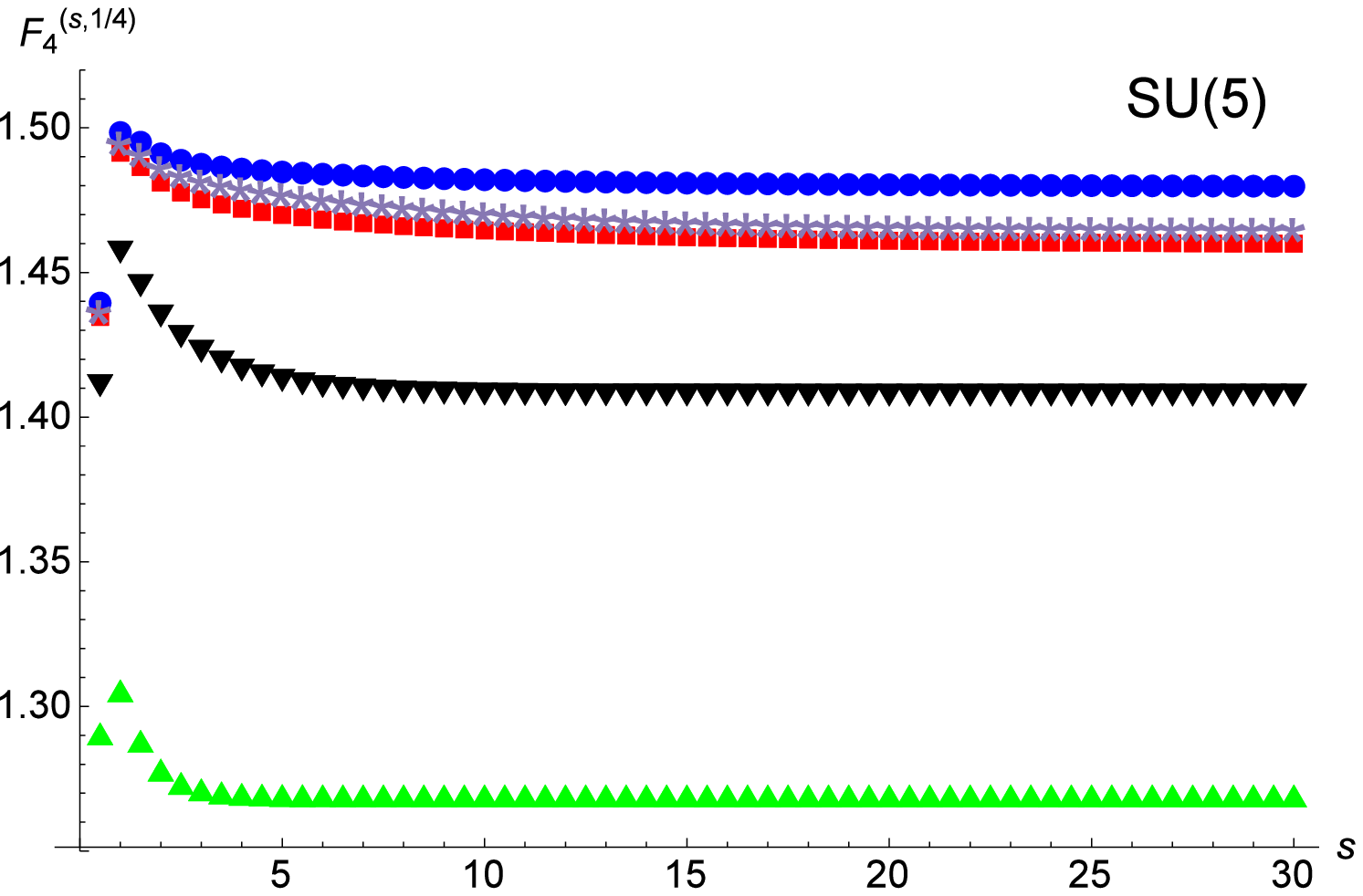}
\end{center}
\caption{$s$-dependence of 
the interpolating function $F_4^{(s, 1/4)}(\tau )$
for the spin-0 leading twist operator
at randomly chosen five points $\tau =(r_1 ,\cdots ,r_5 )$.
($r_1$: blue circle, $r_2$: red square, $r_3$: green triangle,
$r_4$: black inverse triangle, $r_5$: purple asterisk)
[Left-Top] The SU(2) case, $(r_1 ,r_2 ,r_3 ,r_4 ,r_5 )$ $\simeq$ 
$(0.5371 + 0.223i ,0.3408 +0.6288i ,0.6924 + 0.9223i ,
0.9543 +0.1698i ,0.4806 +0.6612i)$.
[Right-Top] The SU(3) case, $(r_1 ,r_2 ,r_3 ,r_4 ,r_5 )$ $\simeq$
$(0.04144 + 0.9375i, 0.5572 + 0.3810i, 0.01432 + 0.4833i,
0.6221 + 0.7688i, 0.7378 + 0.2039i )$.
[Left-Bottom] The SU(4) case, $(r_1 ,r_2 ,r_3 ,r_4 ,r_5 )$ $\simeq$
$(0.3239 + 0.7137i, 0.7915 +0.3959i, 0.7869 + 0.4904i,
0.4507 +0.7671i, 0.4965 +0.597i)$.
[Right-Bottom] The SU(5) case, $(r_1 ,r_2 ,r_3 ,r_4 ,r_5 )$ $\simeq$
$(0.6138 +0.9069i, 0.4550 + 0.6598i, 0.05669 + 0.5260i,
0.4407 + 0.1766i, 0.4346 + 0.7571i)$.
}
\label{fig:s_dependence}
\end{figure}

From the above discussion it is clear that we should look at large-$s$ behaviours of the interpolating functions.
So let us see $s$-dependences of our interpolating functions.
Fig.~\ref{fig:s_dependence} shows 
$s$-dependence of the interpolating function $F_4^{(s, 1/4)}(\tau )$ of the spin-0 leading twist operator
for various values of $\tau$ and $N$.
We can easily see that
the values of the interpolating functions for all the cases
become constant for large-$s$ regime\footnote{
It is worth to mention that
Alday-Bissi's interpolating function \eqref{eq:AB} does not show this behaviour.
They have a strong dependence on $s$. 
}.
These behaviours are not only for this particular interpolating function
but also for all other interpolating functions 
as long as we use interpolating functions of the type \eqref{eq:main_interpolation}.
See Appendix \ref{app:s-dep} 
for similar results on the other interpolating functions.
Furthermore 
we can analytically show the saturation for large-$s$
in weak coupling regime and
at the duality invariant points $\tau =\tau_S$ and $\tau =\tau_{TS}$.
For details, see Appendix \ref{app:s-dep_fp}.
Thus we should pick $s$ from the region having the plateau behaviour.
These plots indicate that
we can regard $s=30$ as sufficiently large $s$. 

\subsubsection{Choice of $\alpha$}
The parameter $\alpha$ determines 
the type of branch cuts of the interpolating functions.
In \cite{Honda:2015ewa} 
it was discussed for
the standard FPR \eqref{eq:FPR} that
correct values of $\alpha$ would be
related to analytic properties of exact results.
For example,
if observables under consideration had square type of branch cuts,
then interpolating functions with $\alpha =1/2$ 
would tend to be better approximations\footnote{
Also note that
interpolating functions with many poles may describe
different type of branch cuts.
For example,
it is known that
Pad\'e approximant often describes branch cuts
by bunch of poles.
}.
Since we do not know analytic properties of the anomalous dimensions,
we do not know what should be the correct value of $\alpha$
from this viewpoint.

However,
we now see that
the upper bounds obtained by the bootstrap
are useful to find ``wrong" choices of $\alpha$.
In the large-$N$ limit\footnote{
The large-$N$ limit taken in the context of the bootstrap so far
seems $g={\rm fixed},\ N\rightarrow\infty$ rather than the planar limit.
We expect that the results include the planar limit
because the limit $g={\rm fixed},\ N\rightarrow\infty$ would be equivalent 
to very strong 't Hooft coupling limit for this case \cite{Azeyanagi:2012xj,Azeyanagi:2013fla}.
},
the upper bounds on the anomalous dimensions are \cite{Beem:2013qxa,Alday:2014tsa} 
\begin{\eq}
\lim_{N\rightarrow\infty}\gamma_{0,2,4} \leq 2 .
\label{eq:bound_largeN}
\end{\eq}
If a planar limit of an interpolating function breaks this bound considerably,
then we can regard
this interpolating function as the wrong choice.

\begin{figure}[t]
\begin{center}
\includegraphics[width=7.4cm]{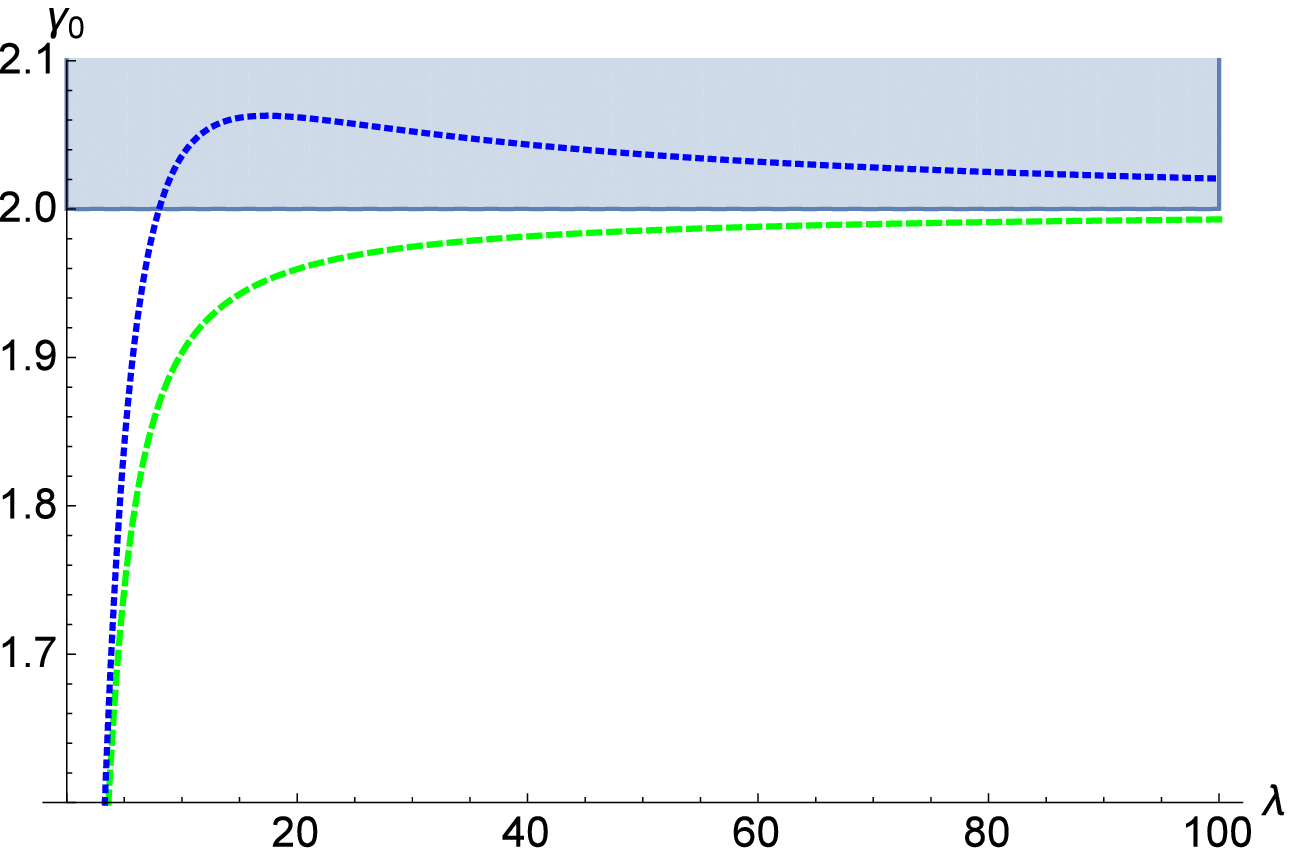}
\includegraphics[width=7.4cm]{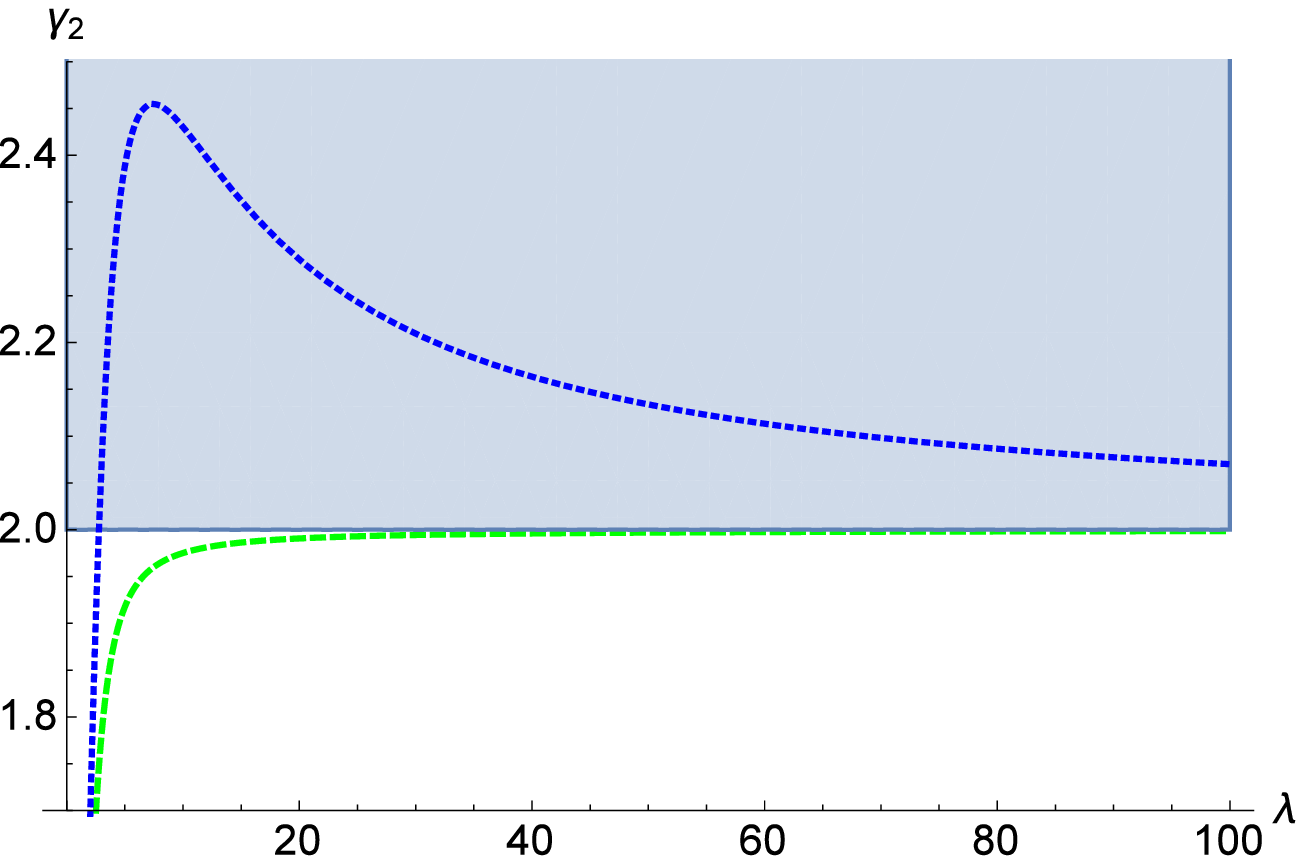}
\end{center}
\caption{
The planar limits of 
the interpolating functions $F_4^{(30,1/2)}$ (blue dotted line)
and $F_4^{(30,1/4)}$ (green dashed line)
with the upper bound by the bootstrap in the large-$N$ limit (shaded region).
[Left] The spin-0 case.
[Right] The spin-2 case.
}
\label{fig:planar_bound}
\end{figure}

Let us consider the spin-0 and spin-2 cases.
For these cases,
the maximum of $m$ is\footnote{
For the spin-4 case, 
$\alpha$ of the interpolating function with $m=3$
is uniquely determined as $\alpha =1/3$.
} $m=4$.
By the above arguments in this subsection,
we expect that
the best approximation among our interpolating function is
either $F_4^{(s,1/2)}(\tau )$ or $F_4^{(s,1/4)}(\tau )$ 
with sufficiently large $s$.
As we discussed,
we can regard $s=30$ as sufficiently large $s$.
Therefore
$F_4^{(30,1/2)}(\tau )$ or $F_4^{(30,1/4)}(\tau )$
would give the best approximation
(Their explicit forms are written in appendix \ref{app:explicit}).
In fig.~\ref{fig:planar_bound}
we plot the planar limits of 
the interpolating functions 
$F_4^{(30,1/2)}(\tau )$ and $F_4^{(30,1/4)}(\tau )$,
whose expressions are\footnote{
Note that the planar limit is described only by the perturbative part.
}
\begin{\eqa}
\left. F_4^{(30,1/2)}(\tau ) \right|_{\rm spin0,planar }
&=& 2\lambda \sqrt{\frac{ \lambda +4.14281}
{ \lambda ^3+1.8719 \lambda ^2+15.9554 \lambda +18.1724}} ,\NN\\
\left. F_4^{(30,1/4)}(\tau ) \right|_{\rm spin0,planar }
&=& \frac{6\lambda}
 {\left( 81 \lambda ^4 +109.116 \lambda ^3 +473.741 \lambda ^2 +1984.4 \lambda +1558.55 \right)^{1/4} } ,\NN\\
\left. F_4^{(30,1/2)}(\tau ) \right|_{\rm spin2,planar }
&=& 2\lambda \sqrt{\frac{\lambda +7.79869}
{\lambda ^3+0.49032 \lambda ^2+13.8773 \lambda +17.7339}}  ,\NN\\
\left. F_4^{(30,1/4)}(\tau ) \right|_{\rm spin2,planar }
&=&  \frac{50\lambda }
{\left( 390625\lambda ^4 +108254\lambda ^3+637497. \lambda ^2
  +2643220\lambda +2019870 \right)^{1/4}} , \NN\\
\end{\eqa}
together with the upper bounds \eqref{eq:bound_largeN} 
by the bootstrap.
We see that
the interpolating functions $F_4^{(30,1/2)}(\tau )$
both for spin-0 and spin-2 cases
break the upper bounds by the conformal bootstrap considerably
while $F_4^{(30,1/4)}(\tau )$ does not.
This indicates that
the interpolating function $F_4^{(30,1/2)}(\tau )$ is the wrong choice
although its a priori reason is unclear.
Thus we expect that
$F_4^{(30,1/4)}(\tau )$ gives the best approximation 
and
uses the interpolating function $F_4^{(30,1/4)}(\tau )$ 
for comparison with the $\mathcal{N}=4$ superconformal bootstrap.
It would be interesting 
if one can relate this to analytic property of the dimension
in the spirit of \cite{Honda:2015ewa}.

\section{Results on the leading twist operators for finite $N$}
\label{sec:results_finite}
In this section we present our result on the leading twist operators and 
compare this with the $\mathcal{N}=4$ superconformal bootstrap.
We also discuss the image of the conformal manifold
in the space of the dimensions.
Finally We study 
the dimension of the sub-leading twist operator and
the level crossing phenomenon with the leading twist operator.

\subsection{Comparison with the $\mathcal{N}=4$ superconformal bootstrap}
In this subsection
we compare our interpolating functions 
with the conjecture \cite{Beem:2013qxa} 
by the conformal bootstrap
that the upper bounds on the dimensions of the leading twist operators are saturated 
at one of the duality invariant points $\tau =\tau_S =i$ and $\tau =\tau_{TS} =e^{i\pi /3}$.

\subsubsection{Spin-0}
\begin{figure}[t]
\begin{center}
\includegraphics[width=7.4cm]{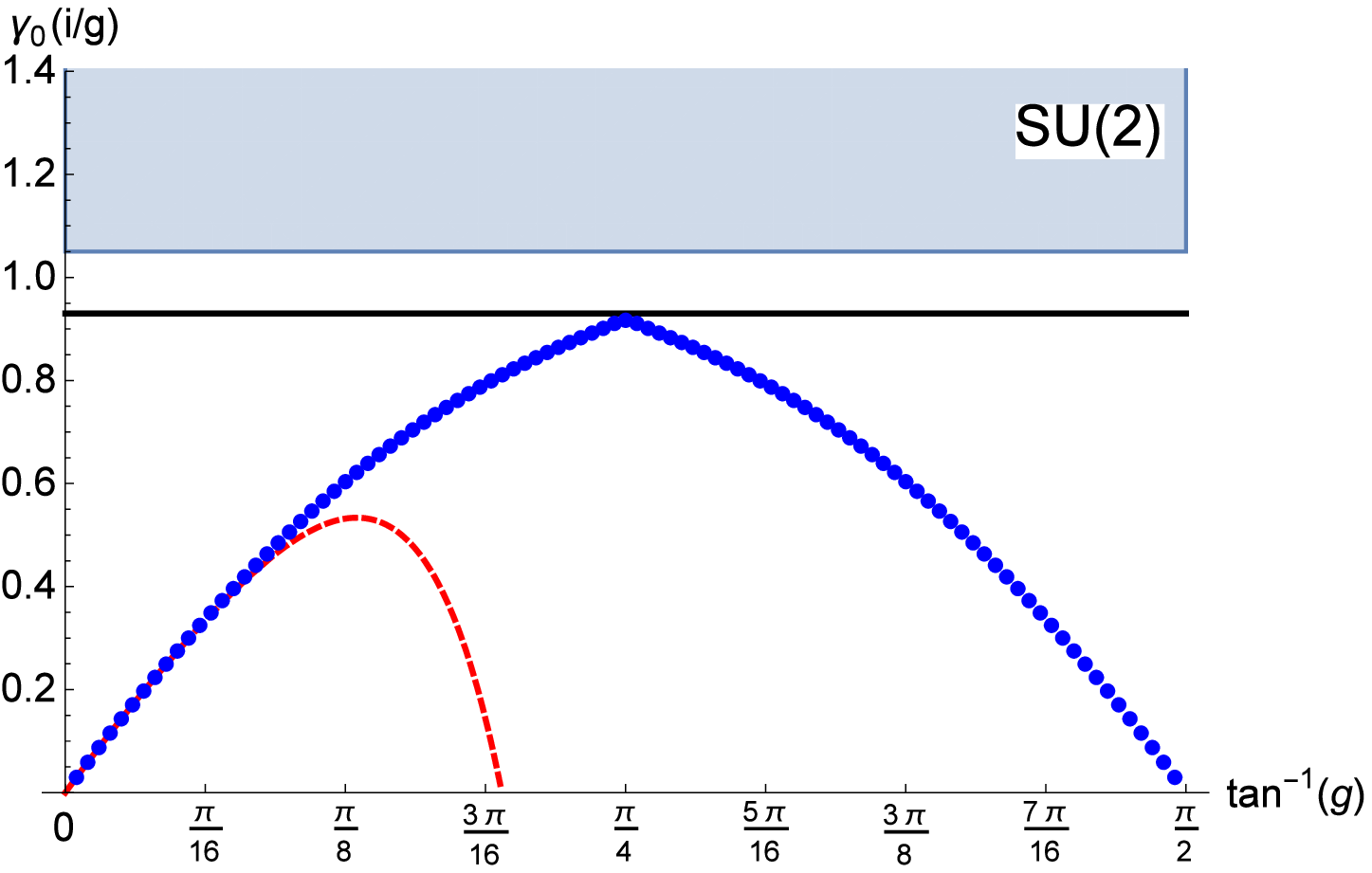}
\includegraphics[width=7.4cm]{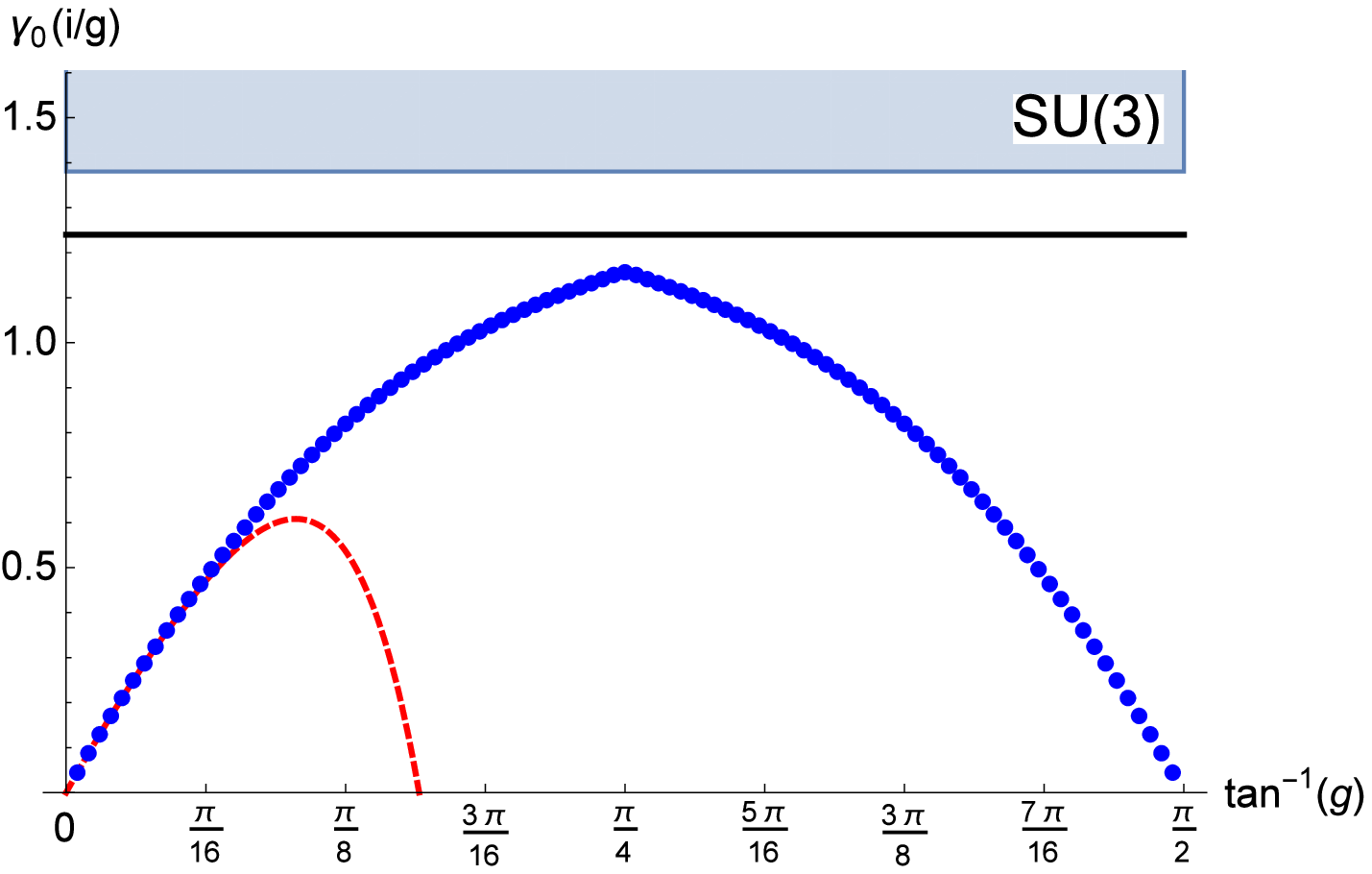}\\
\includegraphics[width=7.4cm]{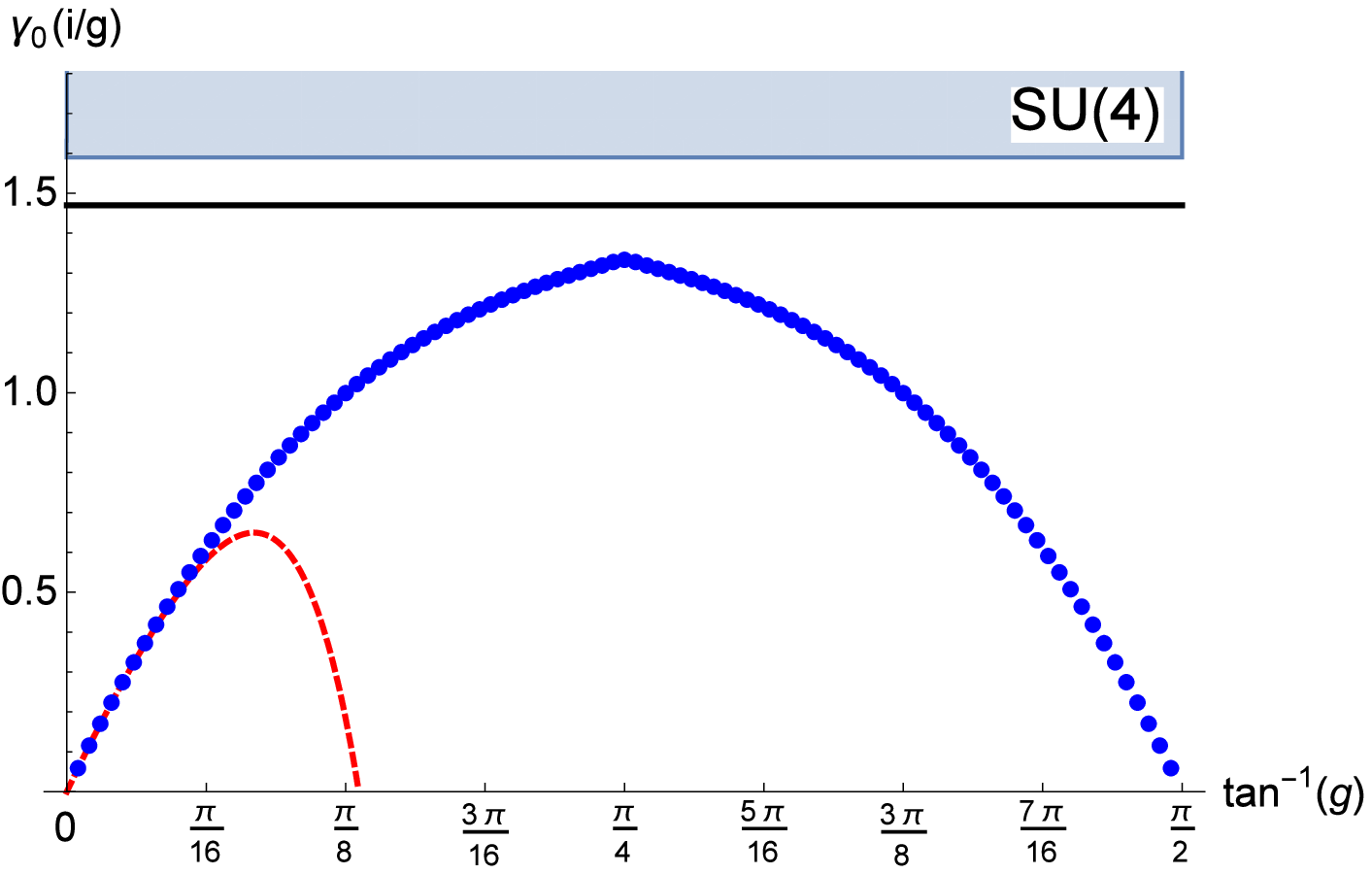}
\includegraphics[width=7.4cm]{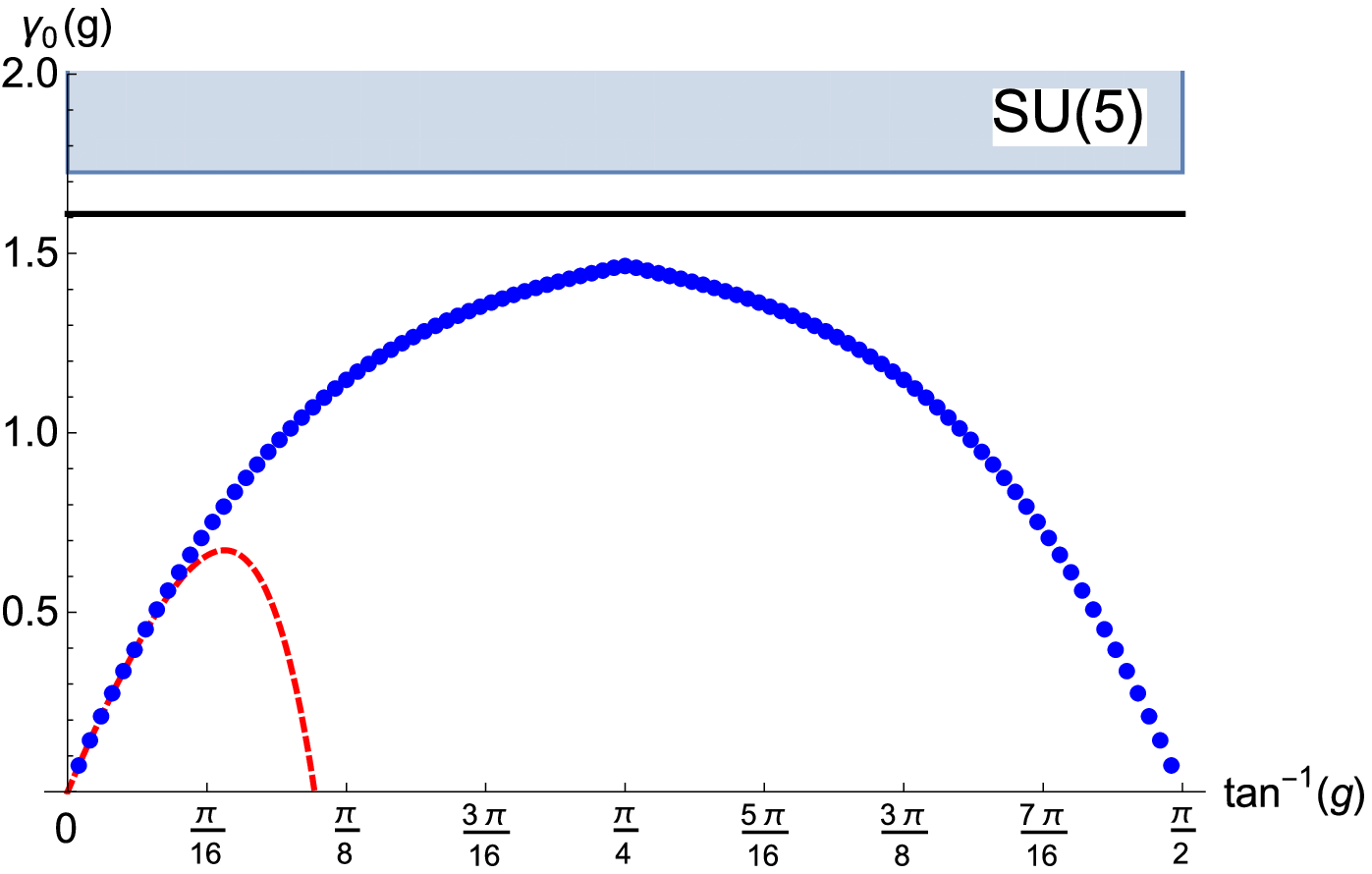}
\end{center}
\caption{
The interpolating function 
$F_4^{(30,1/4)}(\tau =i/g)$
for the spin-0 operator
is plotted against the gauge coupling $g$ for $\theta =0$ (blue dots).
The red dashed line denotes the weak coupling expansion up to four loop.
The shaded region and horizontal black solid line are 
the upper bounds and corner values obtained by the $\mathcal{N}=4$ superconformal bootstrap, respectively.
}
\label{fig:spin0_coupling}
\end{figure}
\begin{figure}[t]
\begin{center}
\includegraphics[width=7.4cm]{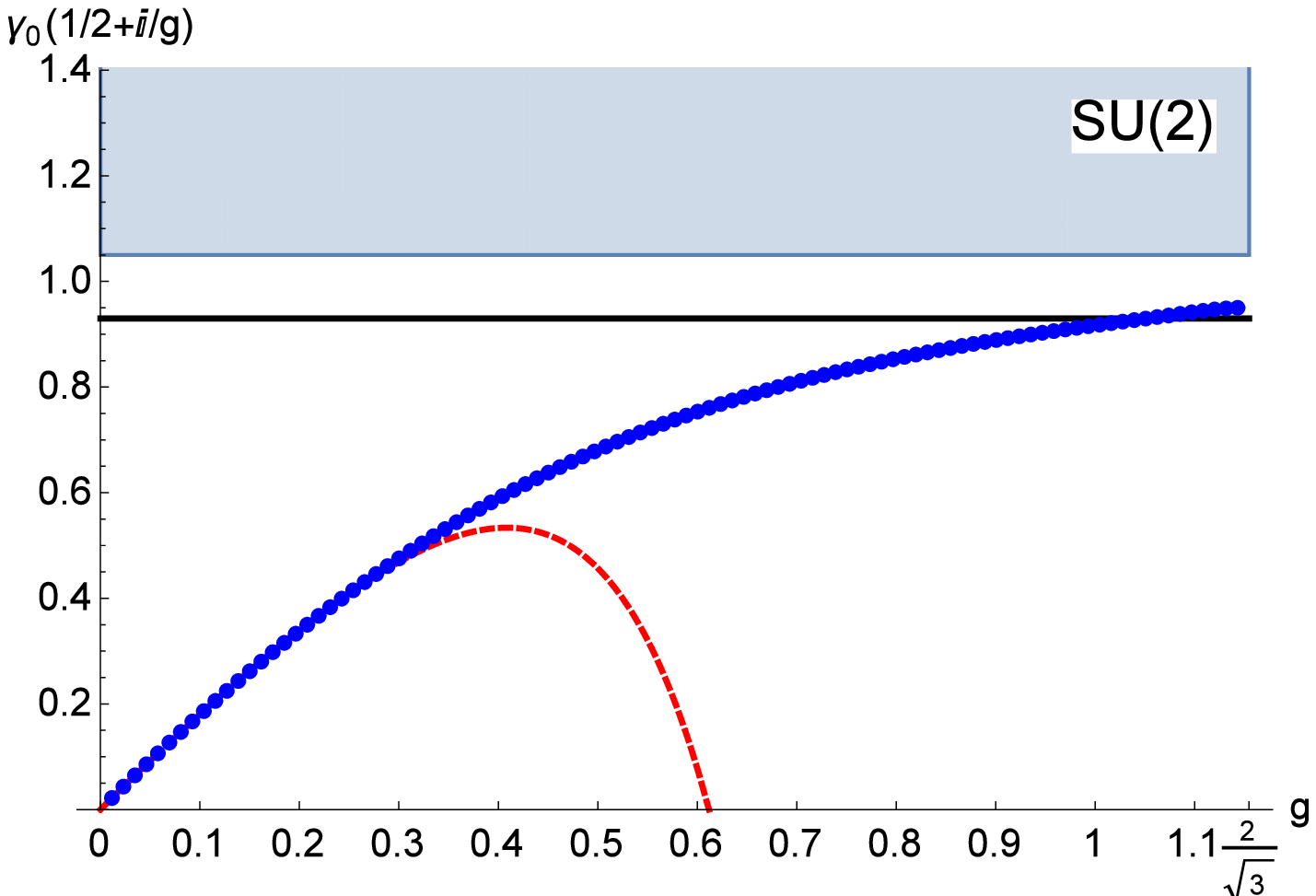}
\includegraphics[width=7.4cm]{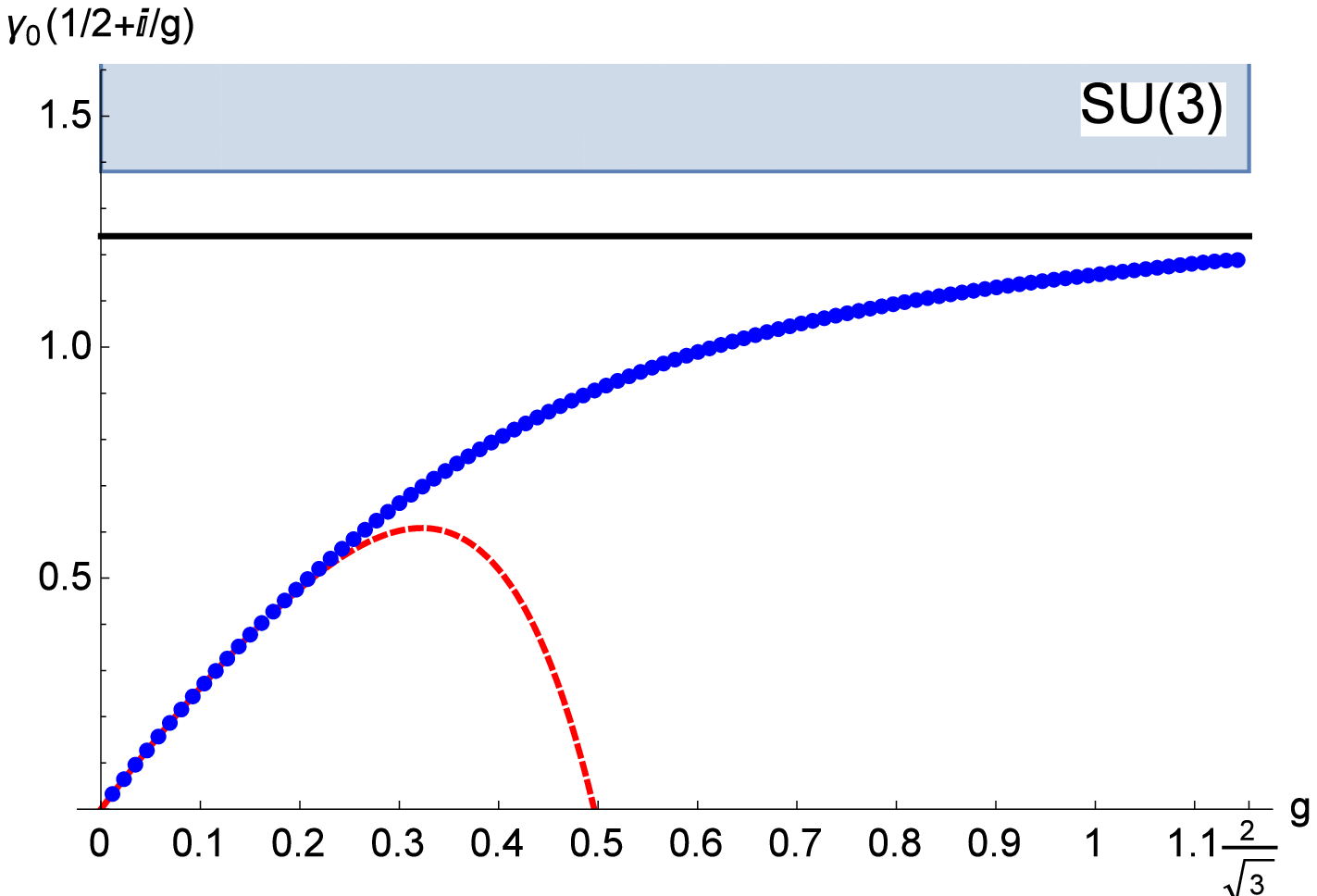}\\
\includegraphics[width=7.4cm]{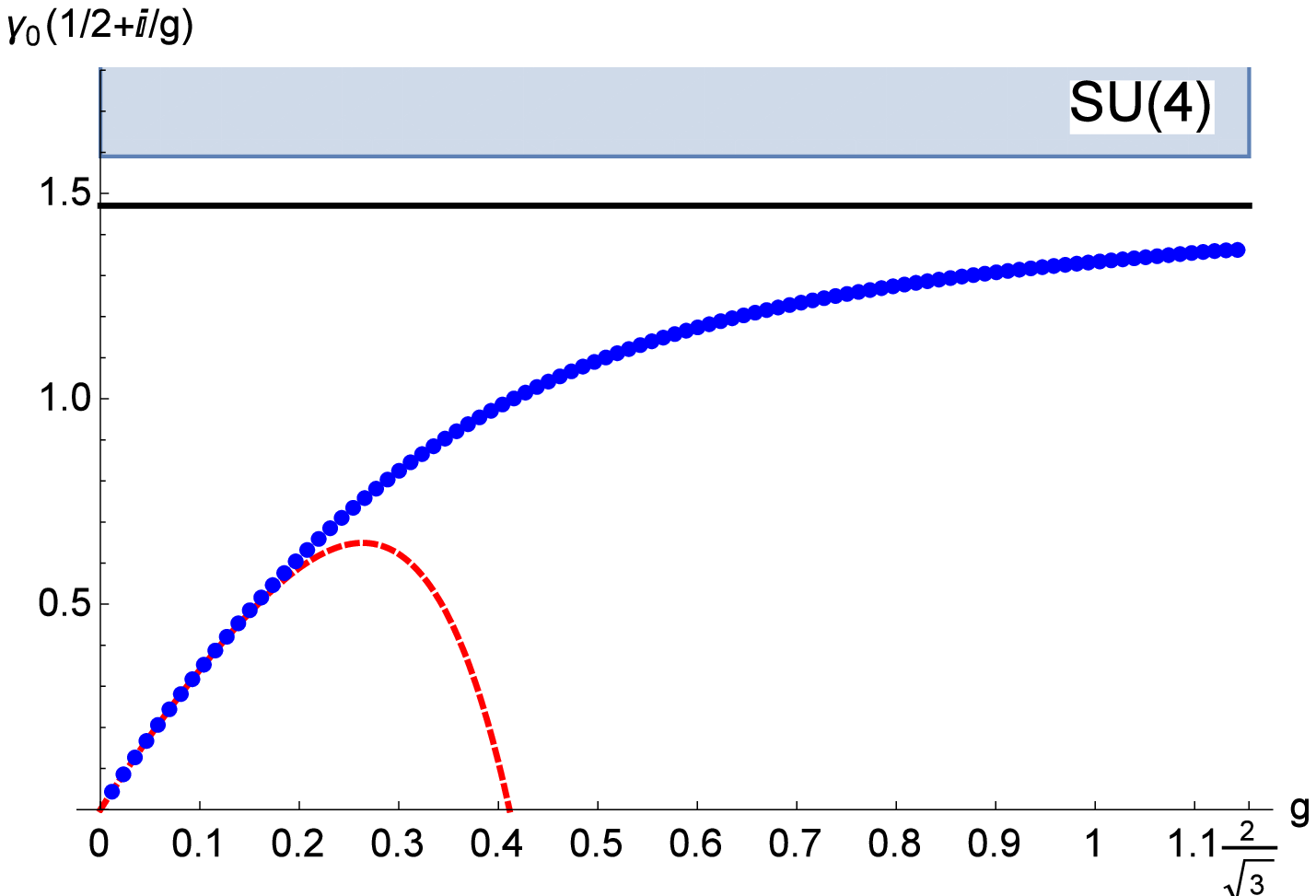}
\includegraphics[width=7.4cm]{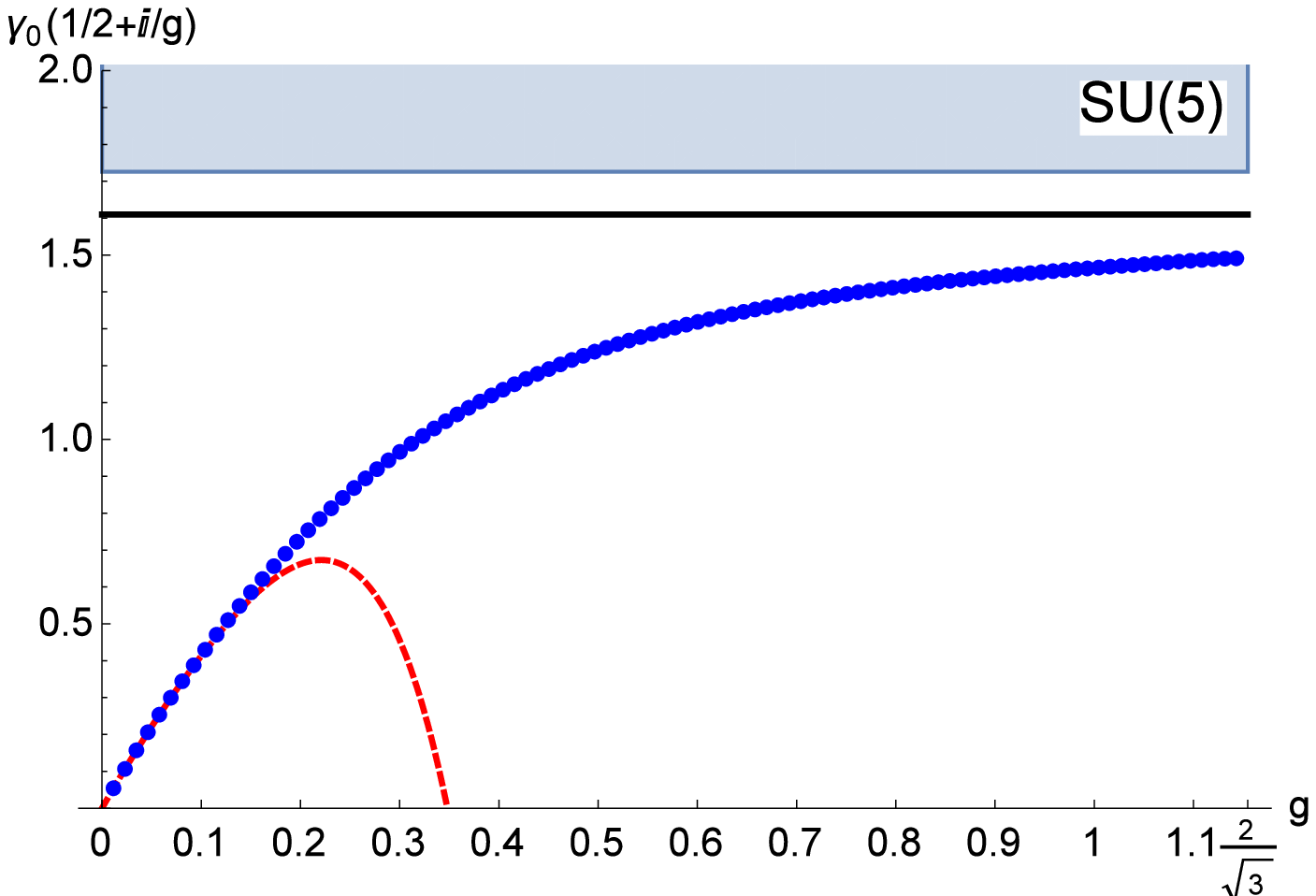}
\end{center}
\caption{
A similar plot as fig.~\ref{fig:spin0_coupling} for $\theta =\pi$.
The interpolating function $F_4^{(30,1/4)}(\tau =1/2 +i/g)$
for the spin-0 operator
is plotted as a function of $g$
together with the 4-loop result.
Note that $g=2/\sqrt{3}$ corresponds to $\tau=\tau_{TS}=e^{i\pi /3}$.
}
\label{fig:spin0_coupling_pi}
\end{figure}
We begin with the spin-0 leading twist operator.
By the arguments in sec.~\ref{sec:constraints},
we expect that
the best approximation among our interpolating functions is
$F_4^{(s,1/4)}(\tau )$ 
with sufficiently large $s$ and
$s=30$ can be regarded as sufficiently large $s$.
Therefore
we use 
the interpolating function $F_4^{(30,1/4)}(\tau )$ for comparison with the conformal bootstrap,
whose explicit forms are written in appendix \ref{app:explicit}.
This gives 
our predictions of the dimension for arbitrary values of the gauge coupling $\tau$ and $N$.

In order to compare our interpolating function with the $\mathcal{N}=4$ superconformal bootstrap,
we shall ask where the interpolating function takes its maximal value 
as a function of $\tau$.
We expect that
the maximal value is given 
at either of the duality invariant points $\tau =\tau_S$ or $\tau =\tau_{TS}$
if the interpolating function reasonably approximates the dimension.
Note that
the duality invariant points are also quite special for our interpolating functions. By construction our interpolating functions always have local extremum at $\tau =\tau_S$ and $\tau =\tau_{TS}$
because the building block $E_s (\tau )$ of the interpolating functions has local minimum\footnote{
Global minimum of $E_s (\tau )$ is given by $\tau =\tau_{TS}$.  
} at these points for arbitrary $s$.
What is nontrivial here is 
whether one of the extremum of the interpolating function at $\tau =\tau_S ,\tau_{TS}$ is global maximum or not.
We will see soon that 
the global maximum is given by\footnote{
Interpolating functions used in \cite{Alday:2013bha}  also have global maximum at $\tau =\tau_{TS}$
while those used in \cite{Beem:2013hha} have global maximum at $\tau =\tau_S$.
} $\tau =\tau_{TS}$.

In fig.~\ref{fig:spin0_coupling}
we plot coupling dependence of the interpolating function $F_4^{(30,1/4)}$ 
for $\theta =0$.
Note\footnote{
One might wonder that
the interpolating function has a cusp at $\tau =\tau_S$.
But we can prove analytically that
the interpolating function is differentiable at $\tau =\tau_S$.
} that
$\tan^{-1}{g}=\pi /4$ corresponds to
the duality invariant point $\tau =\tau_S =i$ 
under the $\mathbf{S}$-transformation.
First we easily see that
the interpolating function is consistent
with the upper bounds for all the values of $N$. 
Next we observe that
the interpolating function has 
the peak at $g=1$, namely, $\tau =\tau_S$.
This indicates that
$\tau =\tau_S$ gives the local maximum of the interpolating function.
In fig.~\ref{fig:spin0_coupling_pi},
we give similar plots for $\theta =\pi$ as fig.~\ref{fig:spin0_coupling},
whose right end $g=2/\sqrt{3}$ corresponds to
the duality invariant point $\tau=\tau_{TS}$.
We again see that
the interpolating functions have the local maximum at $\tau =\tau_{TS}$.

\begin{table}[t]
\begin{center}
  \begin{tabular}{|c||c | c| c| c| c| c|  }
  \hline  & $SU(2)$ & $SU(3)$ & $SU(4)$  & $SU(5)$ & $SU(6)$ & $SU(7)$ \\
\hline $\sqrt{a}$  
          & 0.86603& 1.4142  & 1.9365 & 2.4495 & 2.9580 & 3.4641 \\
\hline\hline $F_4^{(30,1/4)}(\tau_S )$   
& 0.916879    &   1.15649     &  1.33316 & 1.46501 & 1.56427 & 1.63975    \\
\hline $F_4^{(30,1/4)}(\tau_{TS})$   
&   0.950352    &  1.18875     &   1.36267 & 1.49133 & 1.58747 & 1.66015    \\
\hline \hline Corner value    &   0.93    &   1.24    &    1.47  
& 1.61  & 1.7  & 1.78 \\ 
\hline Strict upper bound  &   1.05    &   1.38     &    1.59   
& 1.726 & 1.816 & 1.878 \\
\hline
  \end{tabular}
\end{center}
\caption{The interpolating function for spin-0 at the duality invariant points
and data from the $\mathcal{N}=4$ superconformal bootstrap.}
\label{tab:fp_spin0}
\end{table}
Which of the duality invariant points does give the global maximum?
In table \ref{tab:fp_spin0},
we explicitly write down 
the values of the interpolating functions at $\tau =\tau_S$ and $\tau =\tau_{TS}$
for various $N$.
The table tells us that
the interpolating function has the larger values at $\tau =\tau_{TS}$
than the one at $\tau =\tau_S$.
Actually we have checked that this is true for many other values of $N$.
Thus we conclude that
the interpolating function has the global maximum at $\tau =\tau_{TS}$.

\begin{figure}[t]
\begin{center}
\includegraphics[width=7.4cm]{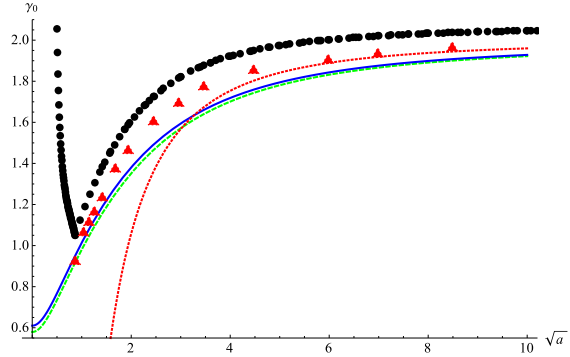}
\end{center}
\caption{
Comparison of the interpolating function $F_4^{(30,1/4)}(\tau)$ at the duality invariant points
with 
the $\mathcal{N}=4$ superconformal bootstrap for the spin-0 case.
The horizontal axis denotes the square root of the central charge: $\sqrt{a}$. 
The black circle symbols are 
the upper bounds of the anomalous dimension 
by the $\mathcal{N}=4$ superconformal bootstrap
while the red triangles are the corner values. 
The green dashed and blue solid lines are  
$F_4^{(30,1/4)}(\tau_S )$ and $F_4^{(30,1/4)}(\tau_{TS})$, respectively.
The red dashed line shows $2-16/N^2$,
which is obtained by numerical fitting of the corner values in the large-$a$ regime \cite{Beem:2013qxa}.
}
\label{fig:bound0}
\end{figure}
In fig.~\ref{fig:bound0},
we plot the values of our interpolating function at the duality invariant points
and the data of 
the $\mathcal{N}=4$ superconformal bootstrap.
The horizontal axis\footnote{
Note that
we have constructed the interpolating functions 
for $SU(N)$ gauge group
and therefore the smallest value of $a$,
which we can compare with the bootstrap,
is $a=3/4$ corresponding to the $SU(2)$ case. 
} 
is $\sqrt{a}$, 
where $a$ is the central charge\footnote{
$a$ is defined as $\langle T_\mu^\mu \rangle = \frac{c}{16\pi^2}W^2 -\frac{a}{16\pi^2}E_{\rm 4d}$,
where $W$ is the Weyl tensor and $E_{\rm 4d}$ is the 4d Euler density.
For superconformal case, 
$a$ is related to $U(1)_R^3$ and $U(1)_R$-gravity$^2$ anomalies as
$a=\frac{3}{32}(3{\rm tr}_{\rm fermion}R^3 -{\rm tr}_{\rm fermion}R)$.
} 
given by\footnote{
For gauge group $G$, $a={\rm dim}(G)/4$.}
\begin{\eq}
a=\frac{N^2 -1}{4} .
\end{\eq}
Note that
information on the gauge group in the $\mathcal{N}=4$ superconformal bootstrap
is packaged into the central charge $a$.

As discussed above
the interpolating function $F_4^{(30,1/4)}(\tau )$
has the greater values at $\tau =\tau_{TS}$ 
than the one at $\tau =\tau_S $.
Thus ``prediction" 
for the maximum of the dimension
from our interpolating functions is $F_4^{(30,1/4)}(\tau_{TS})$, 
whose formula is explicitly given by\footnote{
$
F_4^{(30,1/4)}(\tau_{S})
=2N \left(  N^4 +1.34711N^3 +37.8487N^2 +82.4118N +19.2413 \right)^{-1/4} 
$.
}
\begin{\eq}
F_4^{(30,1/4)}(\tau_{TS})
= \frac{2N}{\left(  N^4 +1.16663 N^3 +36.3865N^2 +66.0665N +10.8232 \right)^{1/4}} .
\end{\eq}
From fig.~\ref{fig:bound0}
we see that
our result is close to the corner values
in the small-$a$ regime and large-$a$ regime
but 
there are about $10\%$ discrepancies in the intermediate regime.
To interpret this,
note that the 
accuracy of the interpolating function
should depend on $a$ (or equivalently $N$) for the following two reasons.
First of all,
we have imposed 
matching with the holographic computation 
in the supergravity limit: $\lambda \gg 1,N\gg 1$
and
consistency with the upper bound \eqref{eq:bound_largeN} in the planar limit.
Since the maximal value of the interpolating functions
satisfying these conditions is two,
our interpolating function matches with the corner values 
in the large-$a$ regime almost\footnote{
Although we have imposed matching with the holographic computation
also at $\mathcal{O}(1/N^2 )$,
we have not imposed 
anything on the maximal value of the interpolating functions 
at $\mathcal{O}(1/N^2 )$.
} by construction.
Second,
recalling that the effective coupling constant is $gN$
rather than $g$,
we know that the weak coupling expansion is more precise 
for the small-$a$ regime.
Thus the intermediate region is harder to be approximated
by our interpolating functions
compared to the other regime.
Hence,
the interpolating function
would give relatively worst approximation 
in the intermediate regime.
Thus we interpret
the discrepancies as the lack of the accuracy of the interpolating function.

As a result,
when we expect good approximations by the interpolating function,
we have found that
the interpolating function at $\tau =\tau_{TS}$ is very close to the corner values.
This supports the conjecture of the $\mathcal{N}=4$ superconformal bootstrap \cite{Beem:2013qxa}
that the upper bounds on the dimensions are saturated at one of 
the duality-invariant points $\tau =\tau_S $ and $\tau =\tau_{TS}$.
We certainly expect that
if we could include more higher order terms of the perturbation theory,
then the interpolating function would have a better  approach to the corner values. 
In the rest of this subsection
we will see that
similar results hold also for the spin-2 and spin-4 operators.

\subsubsection{Spin-2}
\begin{figure}[t]
\begin{center}
\includegraphics[width=7.4cm]{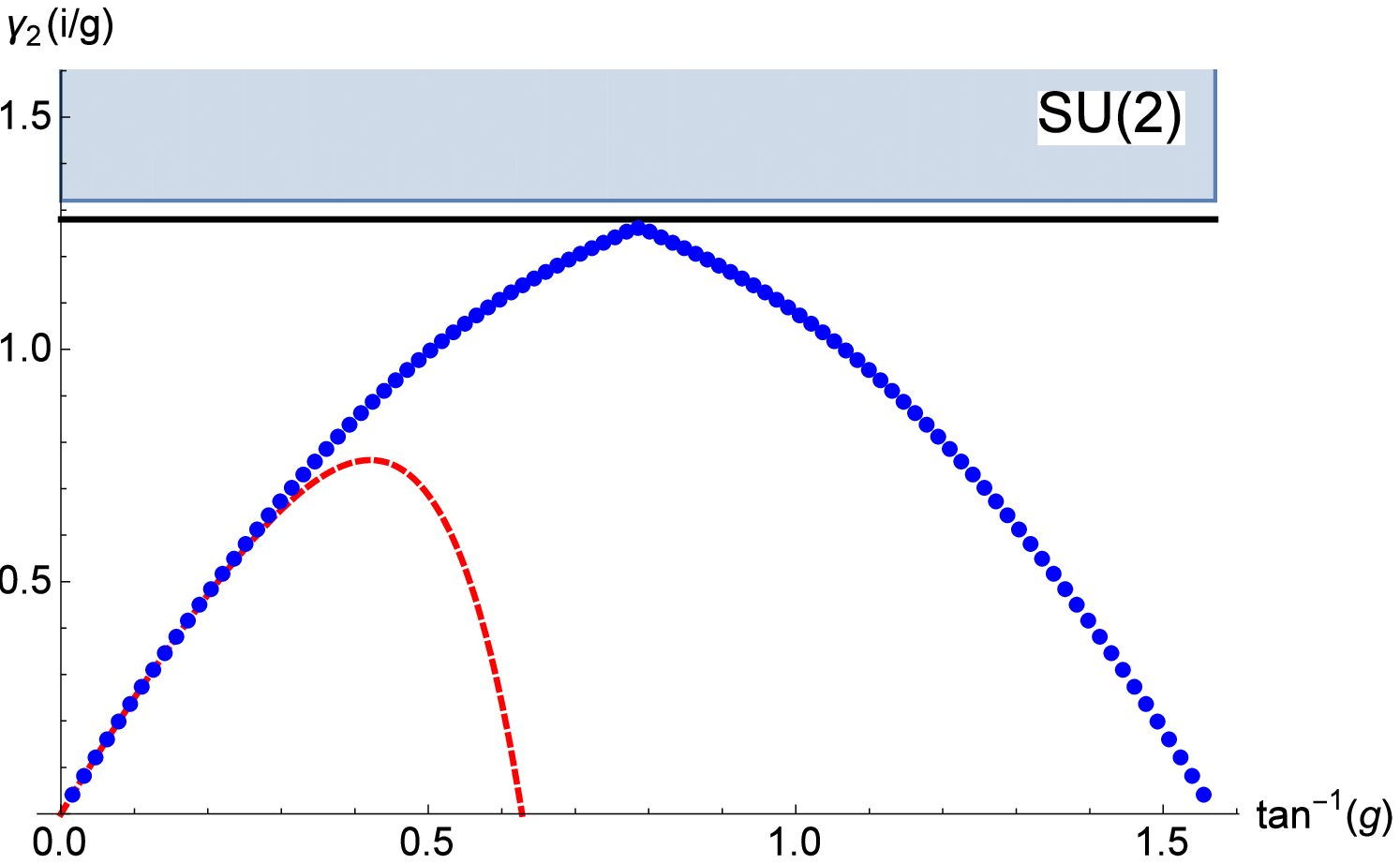}
\includegraphics[width=7.4cm]{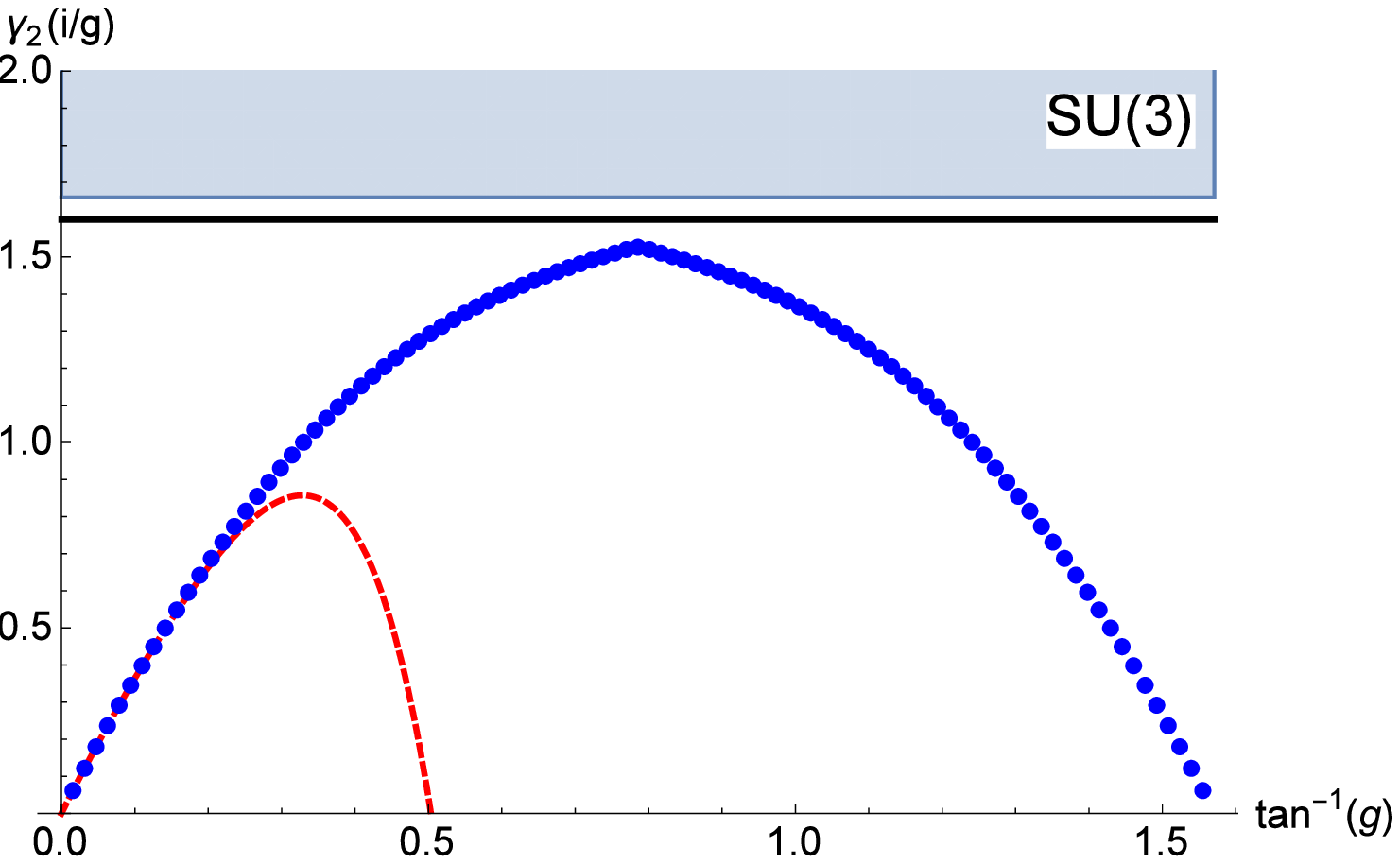}\\
\includegraphics[width=7.4cm]{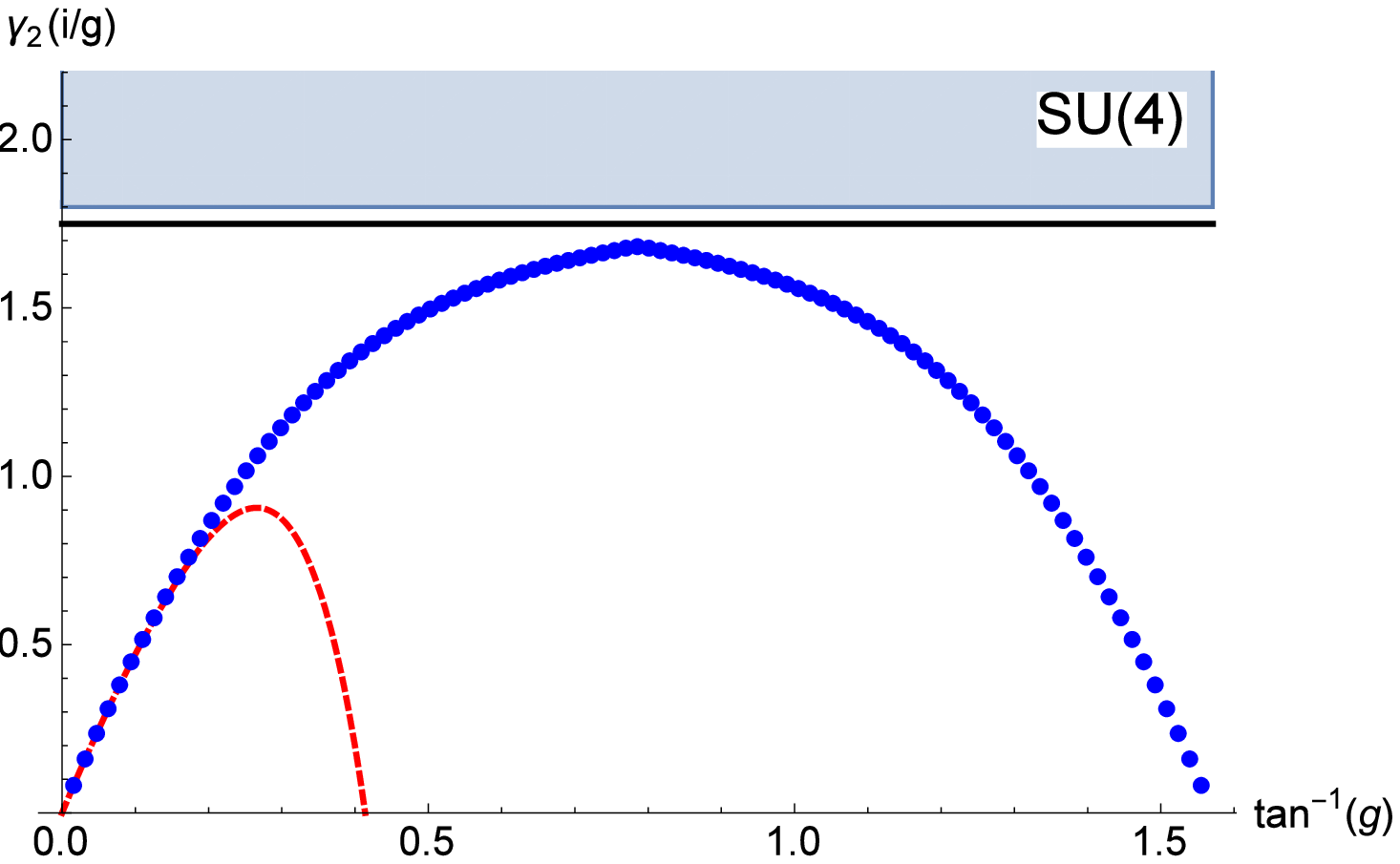}
\includegraphics[width=7.4cm]{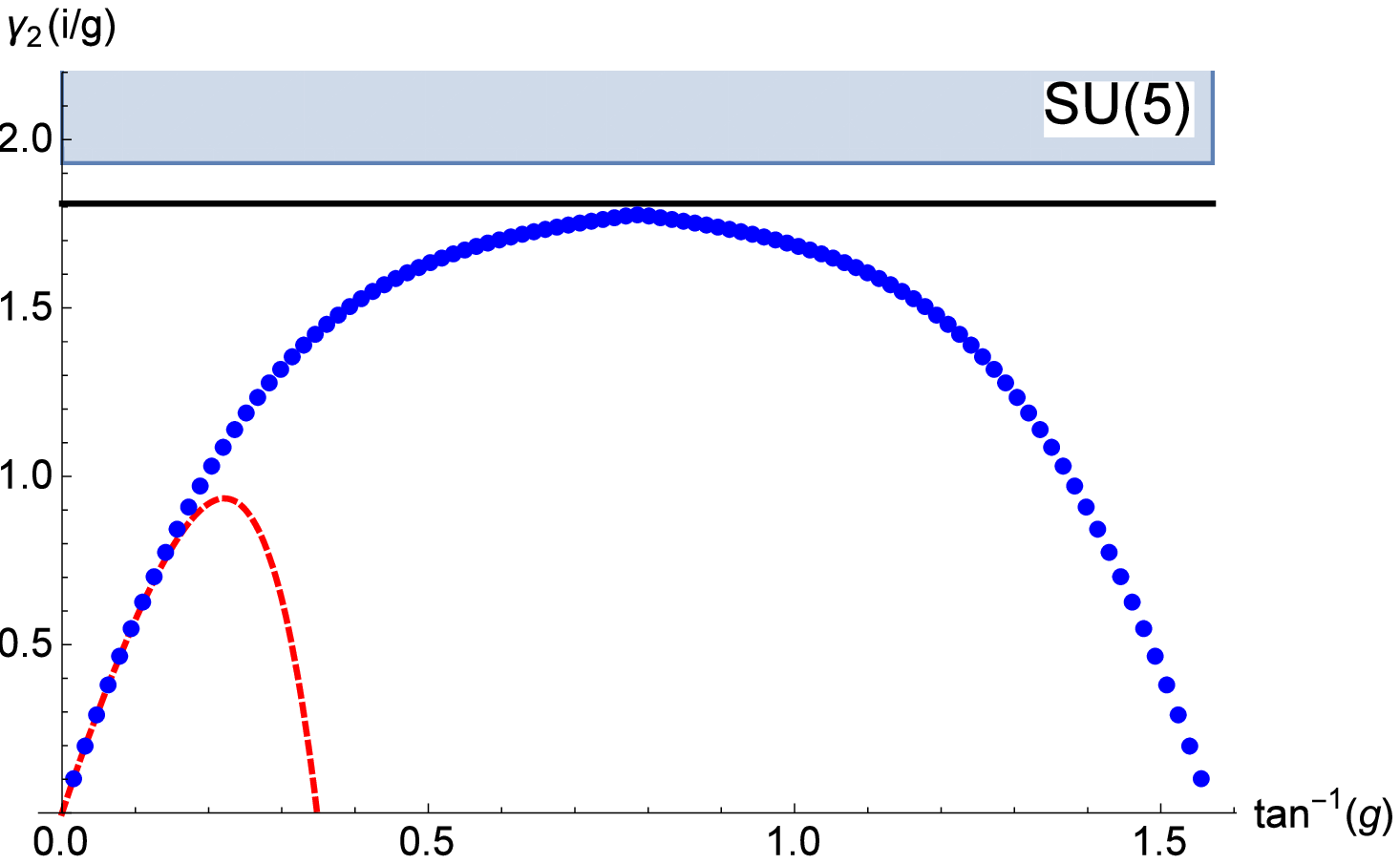}
\end{center}
\caption{
Similar plots for the spin-2 operator as fig.~\ref{fig:spin0_coupling}.
The interpolating function $F_4^{(30,1/4)}(\tau )$
for $\tau =i/g$ is plotted as the function of $g$.
}
\label{fig:spin2_coupling}
\end{figure}
\begin{figure}[t]
\begin{center}
\includegraphics[width=7.4cm]{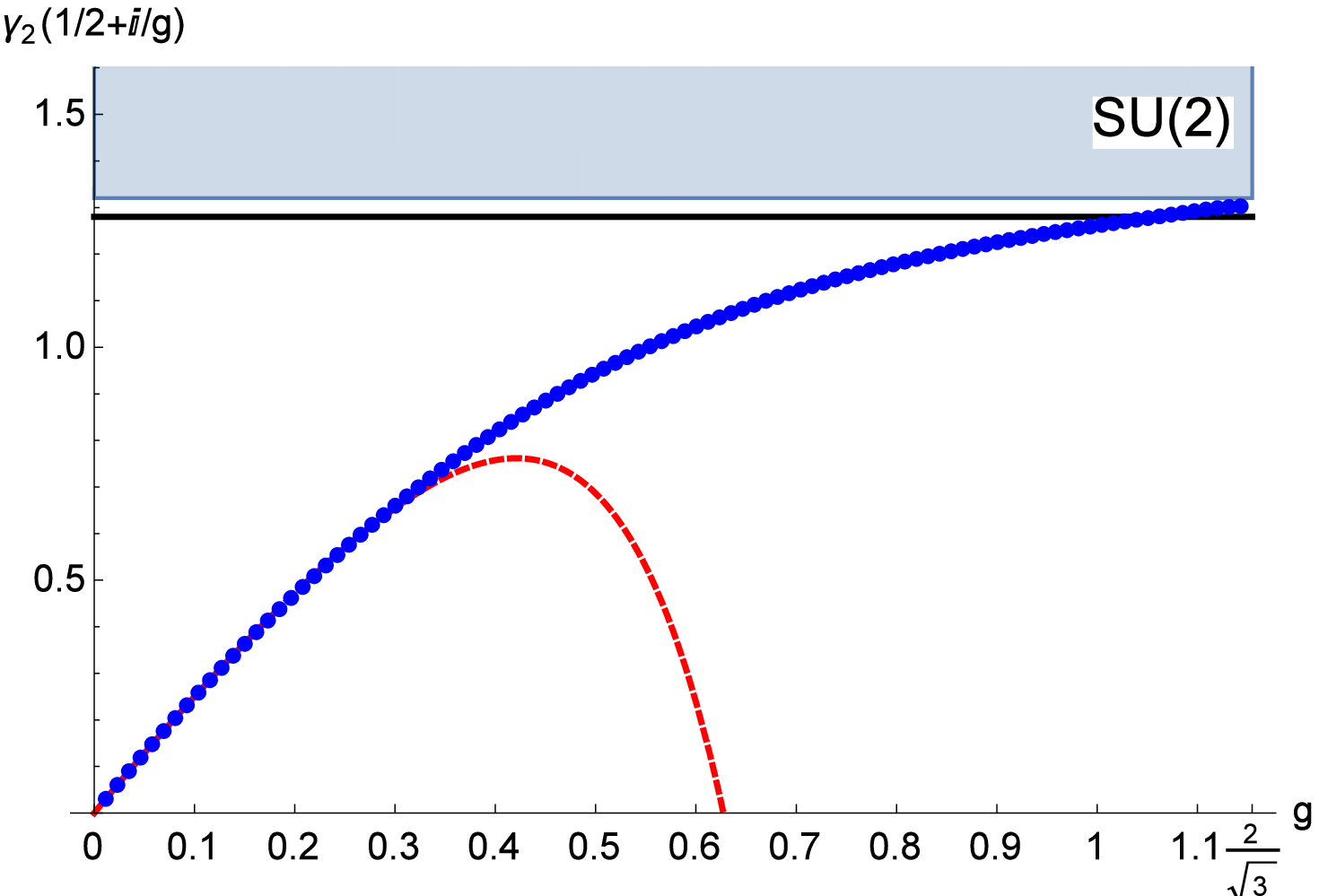}
\includegraphics[width=7.4cm]{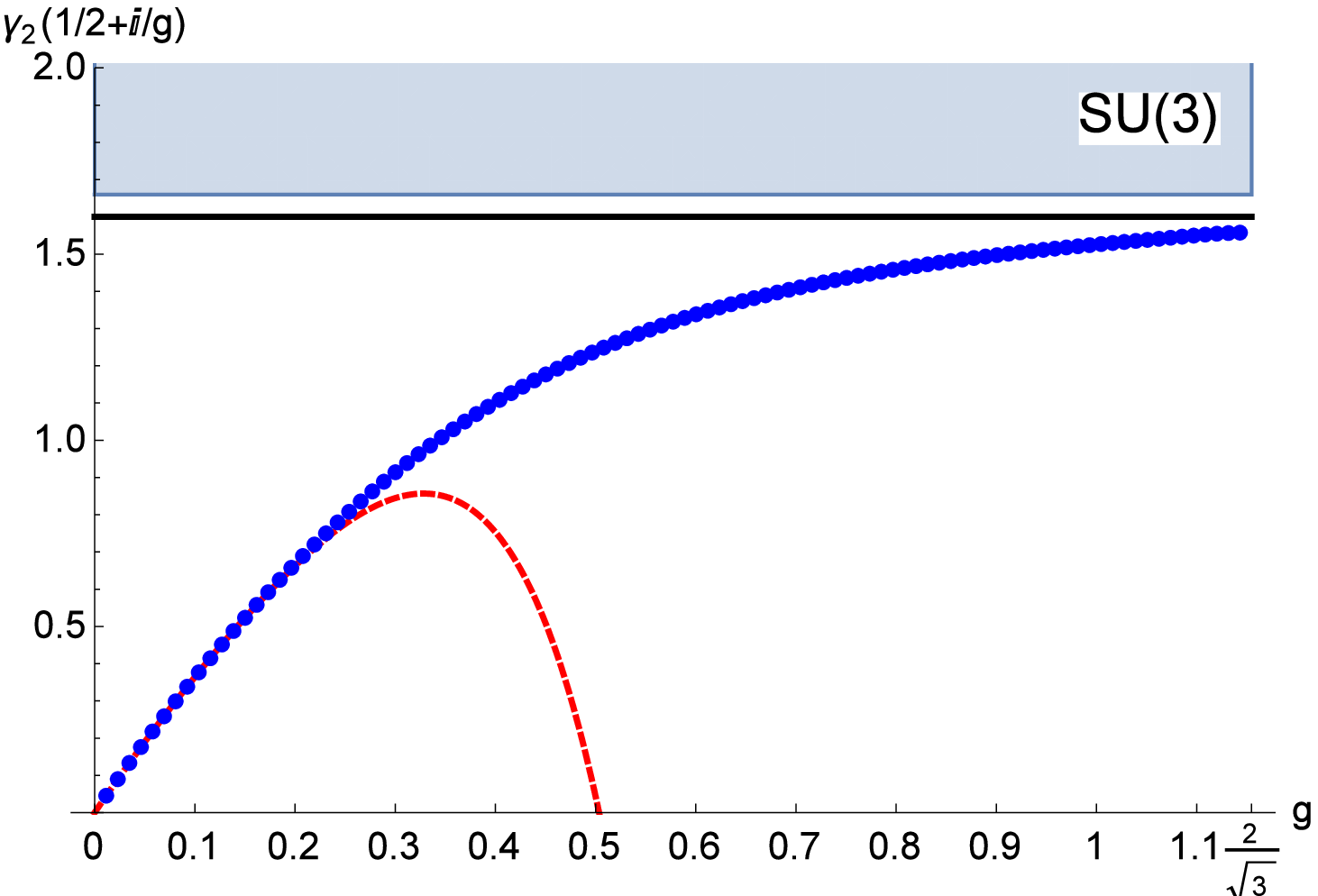}\\
\includegraphics[width=7.4cm]{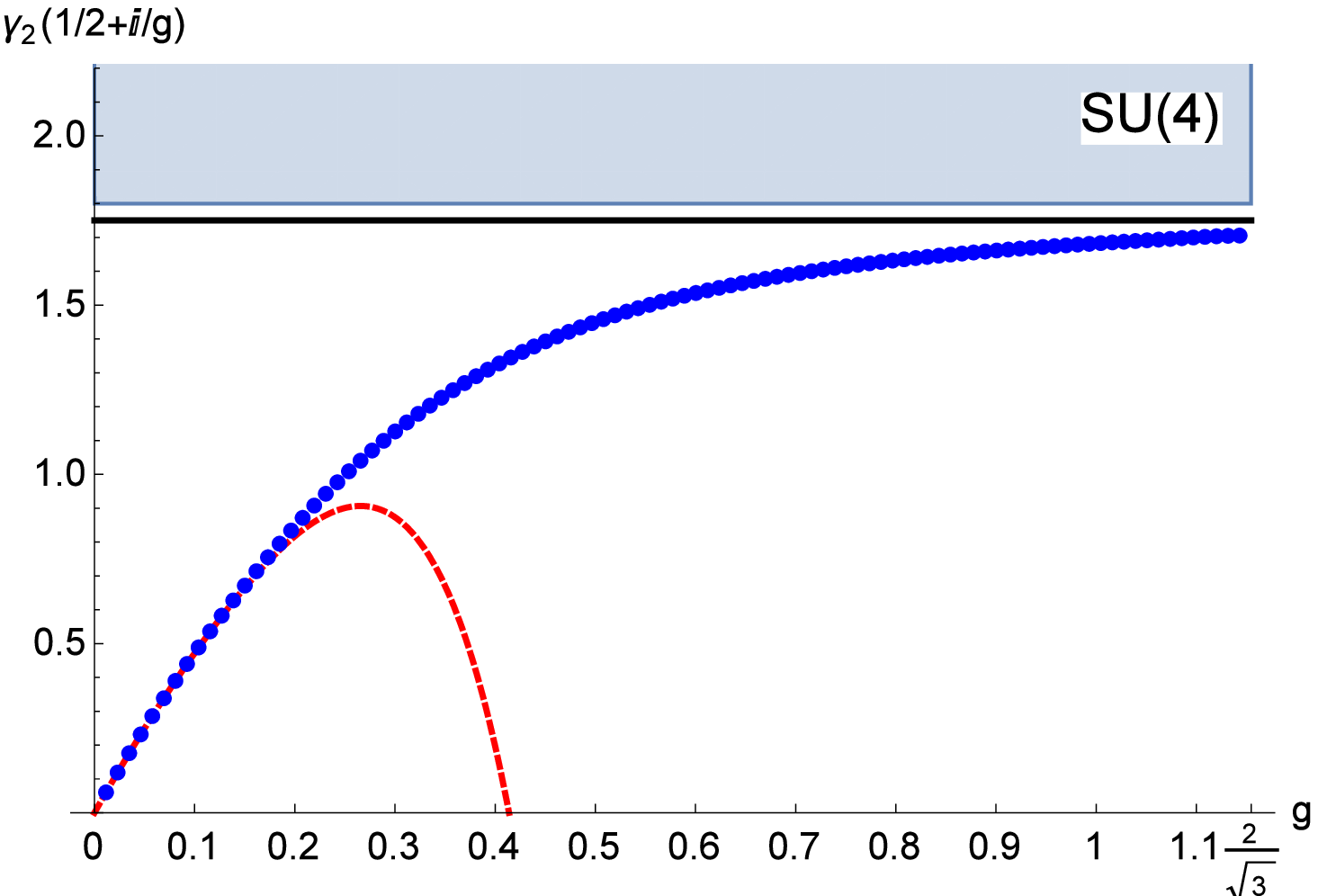}
\includegraphics[width=7.4cm]{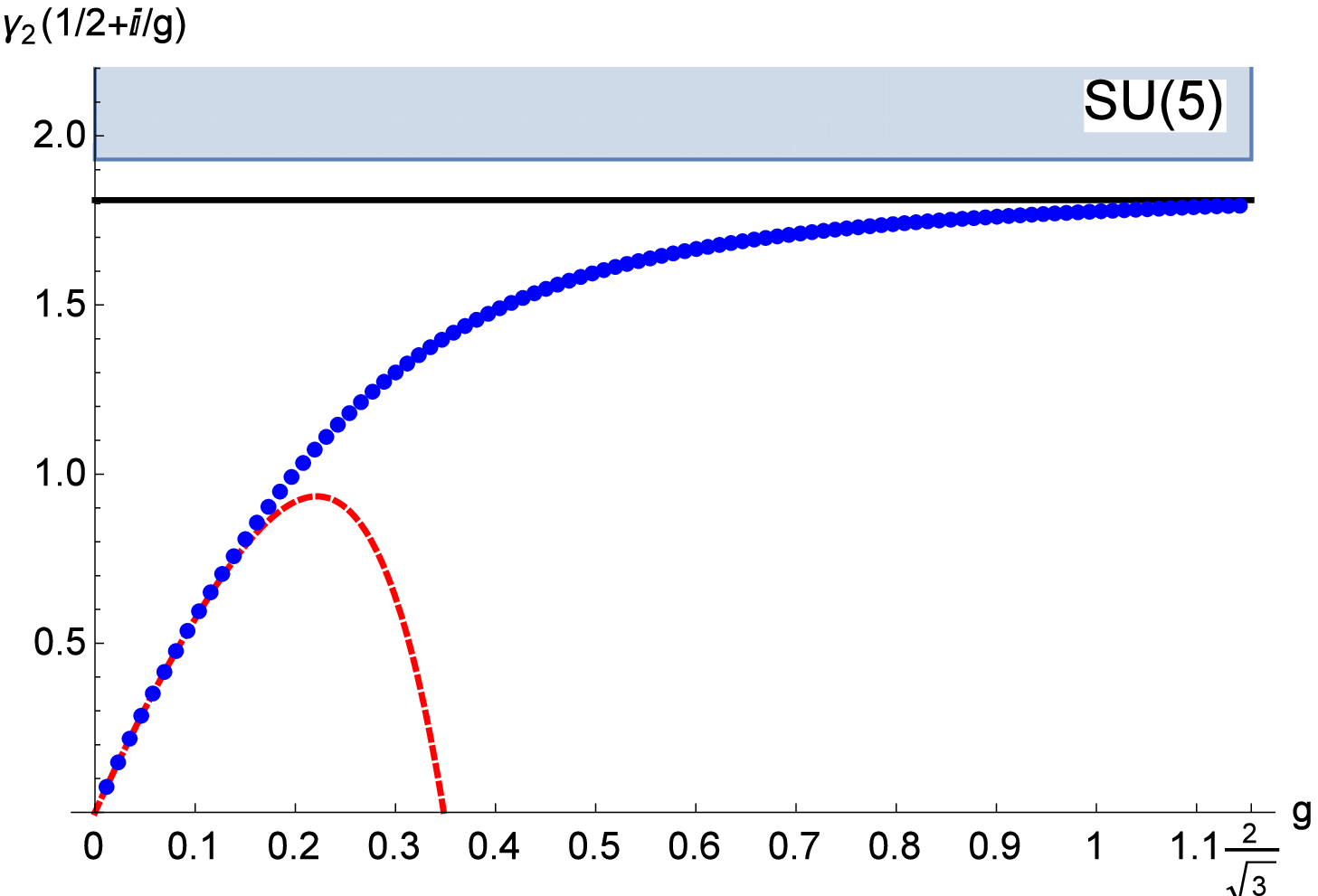}
\end{center}
\caption{
Similar plots for the spin-2 case
as fig.~\ref{fig:spin2_coupling} for $\theta =\pi$.
The interpolating function 
$F_4^{(30,1/4)}(1/2 +i/g)$
is plotted as a function of $g$.
}
\label{fig:spin2_coupling_pi}
\end{figure}
Next we consider the spin-2 leading twist operator.
According to 
sec.~\ref{sec:constraints},
we expect that
the best approximation is given by
$F_4^{(30,1/4)}(\tau )$ as in the spin-0 case.
In fig.~\ref{fig:spin2_coupling} and \ref{fig:spin2_coupling_pi},
we plot coupling dependence of the interpolating function 
for $\theta =0$ and $\pi$
as in fig.~\ref{fig:spin0_coupling} and \ref{fig:spin0_coupling_pi},
respectively.
From these figures,
we see that
$F_4^{(30,1/4)}(\tau )$ has 
the local maximum at the duality invariant points $\tau =\tau_S$ and $\tau =\tau_{TS}$.
In table \ref{tab:fp_spin2}
we compare $F_4^{(30,1/4)}(\tau_{S})$ with $F_4^{(30,1/4)}(\tau_{TS})$
as in table \ref{tab:fp_spin0}.
This indicates that
$F_4^{(30,1/4)}(\tau_{TS})$ is always larger than $F_4^{(30,1/4)}(\tau_{S})$
and hence we conclude that
the interpolating function has the global maximum at $\tau =\tau_{TS}$. 
For any $N$, 
the global maximum $F_4^{(30,1/4)}(\tau_{TS})$ takes the form\footnote{
$
F_4^{(30,1/4)}(\tau_{S})
=  2N \left( N^4 +0.27713 N^3 +9.63199N^2 +19.6152N+5.17088 \right)^{-1/4} .
$
}
\begin{\eq}
F_4^{(30,1/4)}(\tau_{TS})
= \frac{2N}{\left(  N^4 +0.240002N^3 +9.224N^2 +15.5222N +2.90862 \right)^{1/4}} .
\end{\eq}

\begin{table}[h]
\begin{center}
  \begin{tabular}{|c||c | c| c| c| c| c|  }
  \hline  & $SU(2)$ & $SU(3)$ & $SU(4)$  & $SU(5)$ & $SU(6)$ & $SU(7)$ \\
\hline $\sqrt{a}$  
          & 0.86603& 1.4142  & 1.9365 & 2.4495 & 2.9580 & 3.4641 \\
\hline\hline $F_4^{(30,1/4)}(\tau_S )$   
& 1.26131  & 1.52569   & 1.68222 & 1.77664 & 1.83585 & 1.87465    \\
\hline $F_4^{(30,1/4)}(\tau_{TS})$   
& 1.30315  & 1.55797   & 1.70566 & 1.79367 & 1.84852 & 1.88436    \\
\hline \hline Corner value  &   1.28    &   1.60     &   1.75 
& 1.81  & 1.89  & 1.92 \\ 
\hline Strict upper bound   &   1.32    &   1.66     &   1.80 
& 1.93  & 1.915 & 1.935  \\
\hline
  \end{tabular}
\end{center}
\caption{The interpolating function for spin-2 at the duality invariant points
and data from the $\mathcal{N}=4$ superconformal bootstrap.}
\label{tab:fp_spin2}
\end{table}

\begin{figure}[t]
\begin{center}
\includegraphics[width=7.4cm]{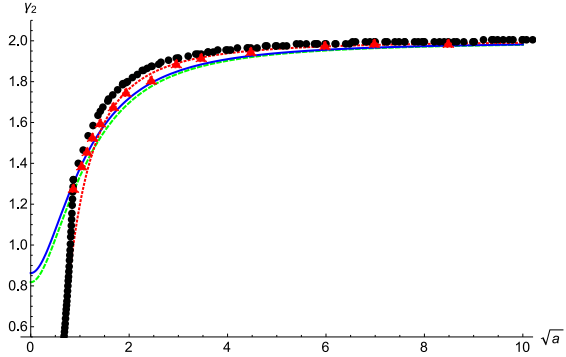}
\end{center}
\caption{
A similar plot for the spin-2 operator as in  fig.~\ref{fig:bound0}.
The red dashed line shows $2-4/N^2$,
which is numerical fitting of the corner values for large-$a$ \cite{Beem:2013qxa}.
}
\label{fig:bound2}
\end{figure}

In fig.~\ref{fig:bound2},
we compare the interpolating function at the duality invariant points
with the $\mathcal{N}=4$ superconformal bootstrap.
We easily see that
the result of $F_4^{(30,1/4)}(\tau =\tau_{TS})$ is very close to the corner values
in the whole region.
This situation is different from the spin-0 case,
where we have about $10\%$ discrepancies in the intermediate regime.
We interpret this 
as large-$a$ regime being effectively broader
for larger spin case.
Indeed $\mathcal{O}(1/N^2 )$ correction to the holographic result for spin-$M$
behaves as $-96/(M+1)(M+6)$ in the supergravity limit \cite{Dolan:2001tt} and
therefore the  anomalous dimension with larger-$M$ converges to $\gamma =2$ faster with increasing $N$.
Since our interpolating function correctly approximates
the large-$a$ regime,
we expect that
the interpolating function for the spin-2 case 
gives reliable approximation in broader range of $a$ 
compared to the spin-0 case.
Thus, 
we expect that our interpolating function provides better approximation
in the whole range for the spin-2 case.
Since the interpolating function $F_4^{(30,1/4)}(\tau_{TS})$ 
is close to the corner values, 
our result supports the conjecture in \cite{Beem:2013qxa} also for the spin-2 case.

\subsubsection{Spin-4}
\begin{figure}[th]
\begin{center}
\includegraphics[width=7.4cm]{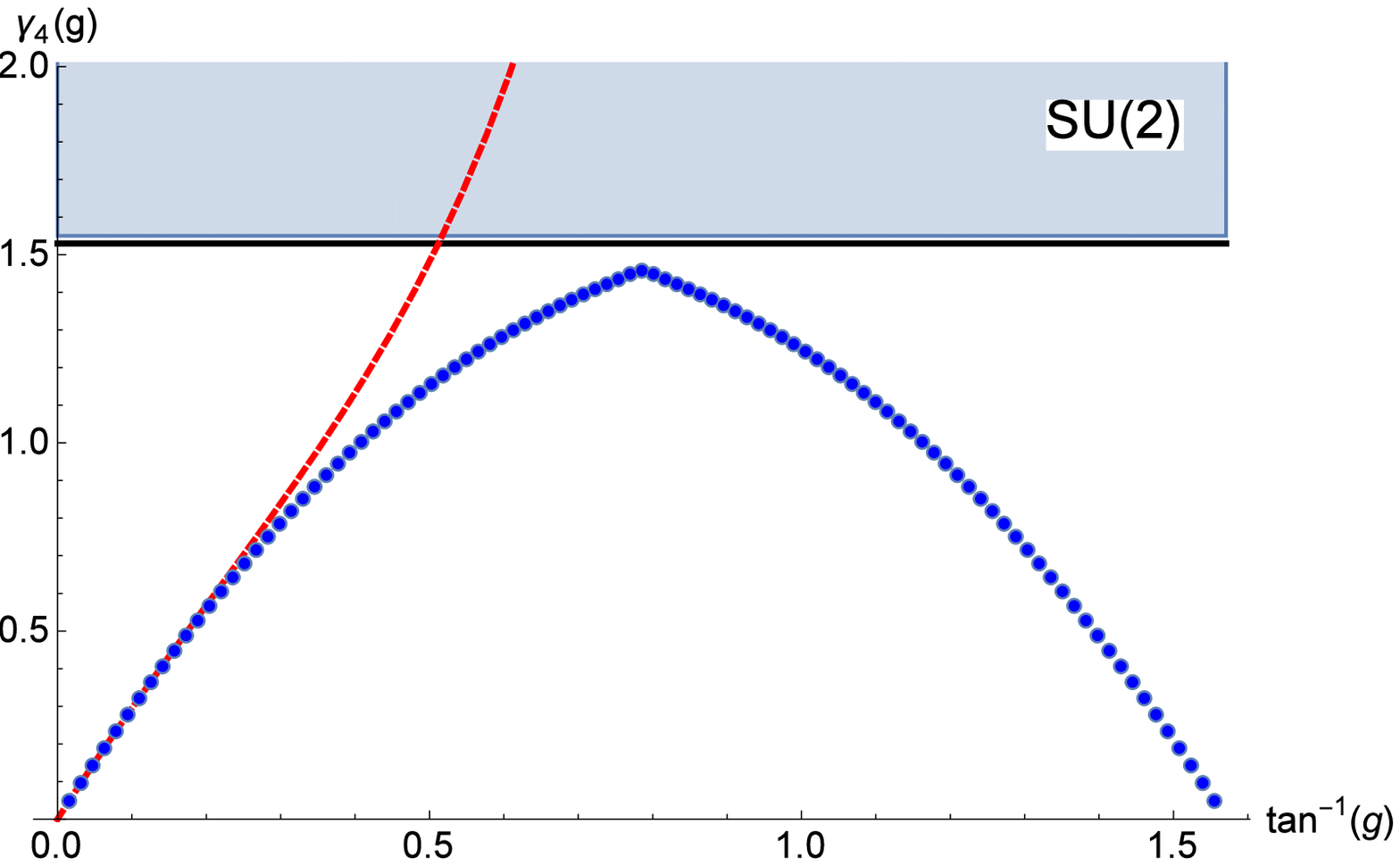}
\includegraphics[width=7.4cm]{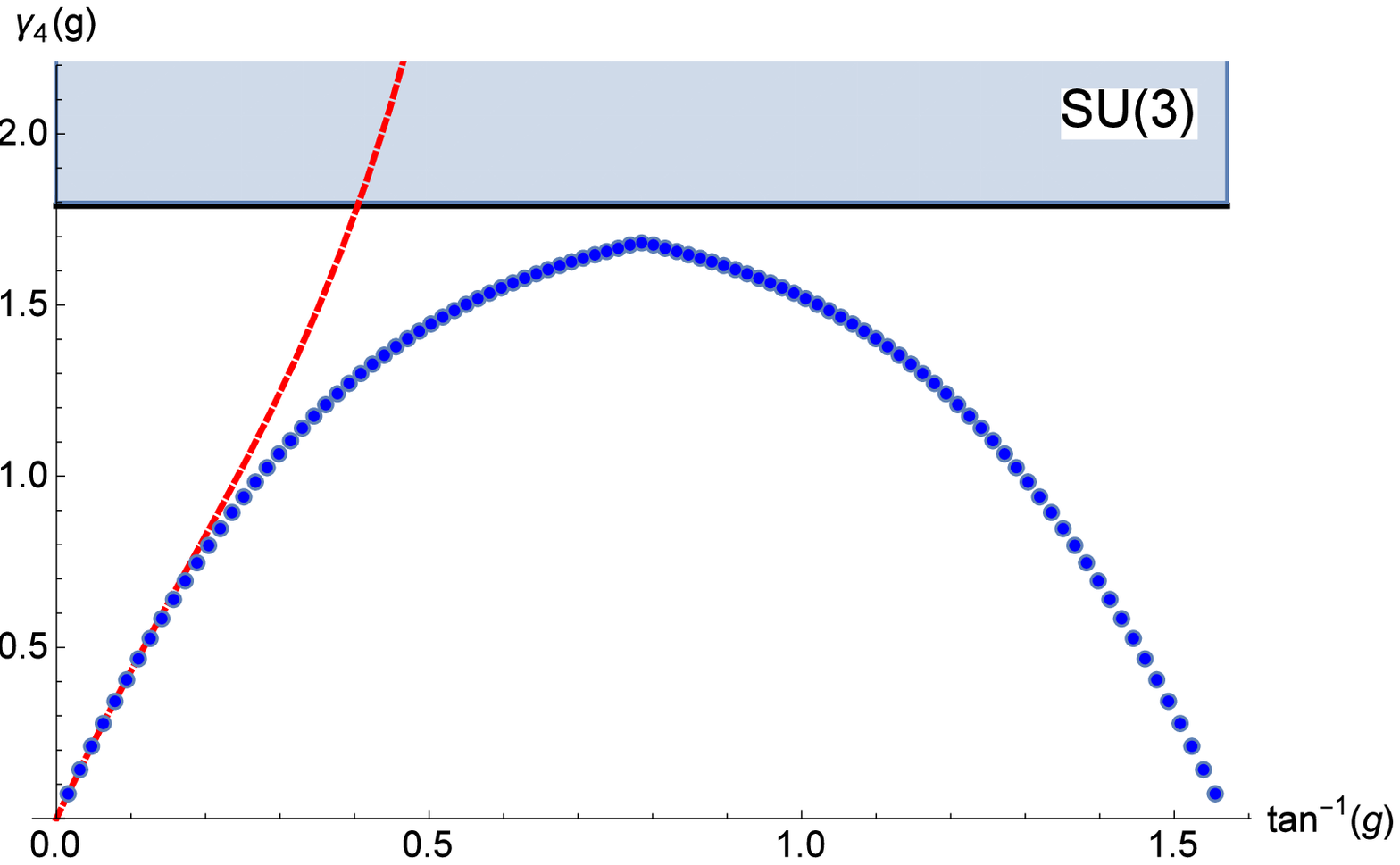}\\
\includegraphics[width=7.4cm]{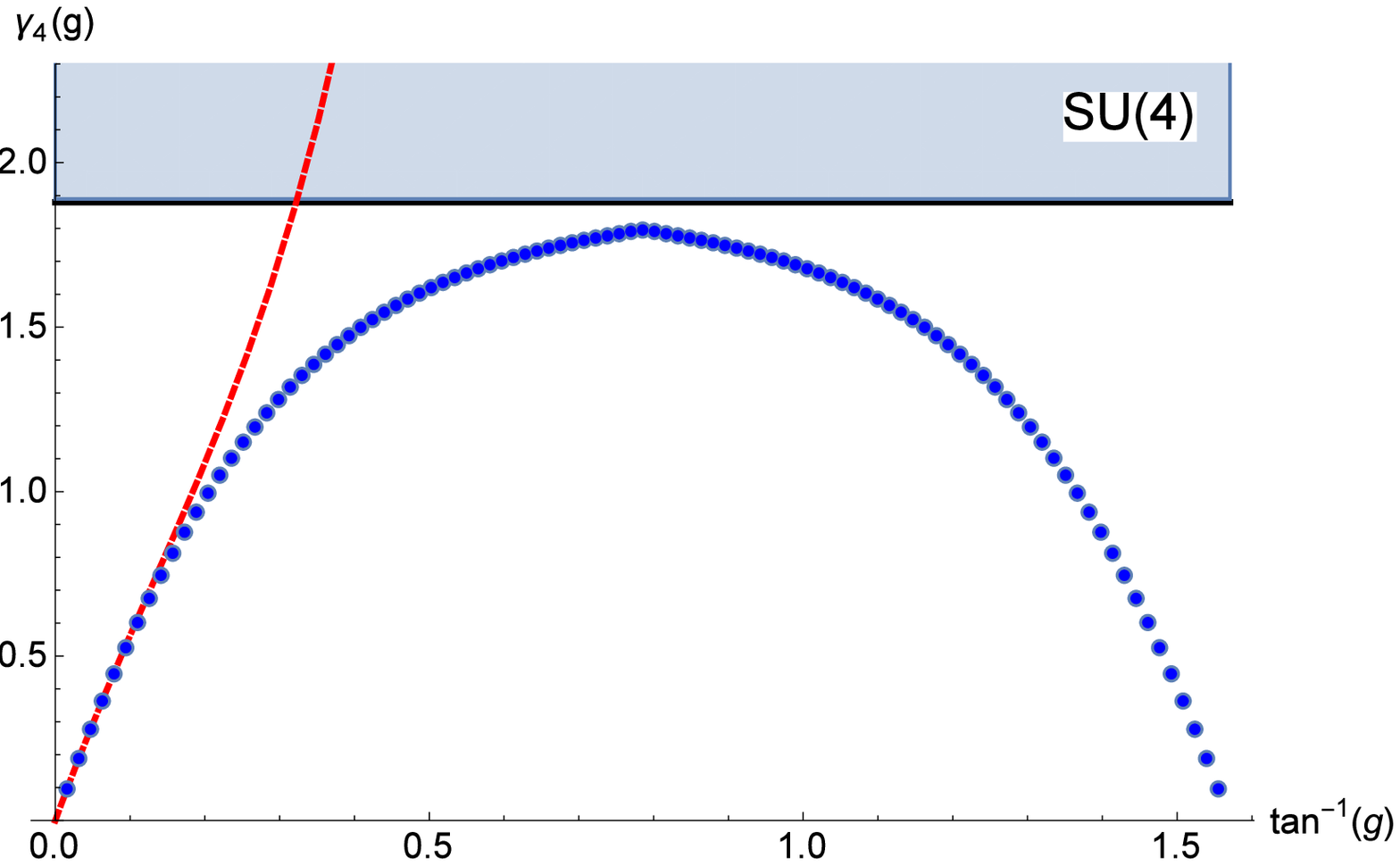}
\includegraphics[width=7.4cm]{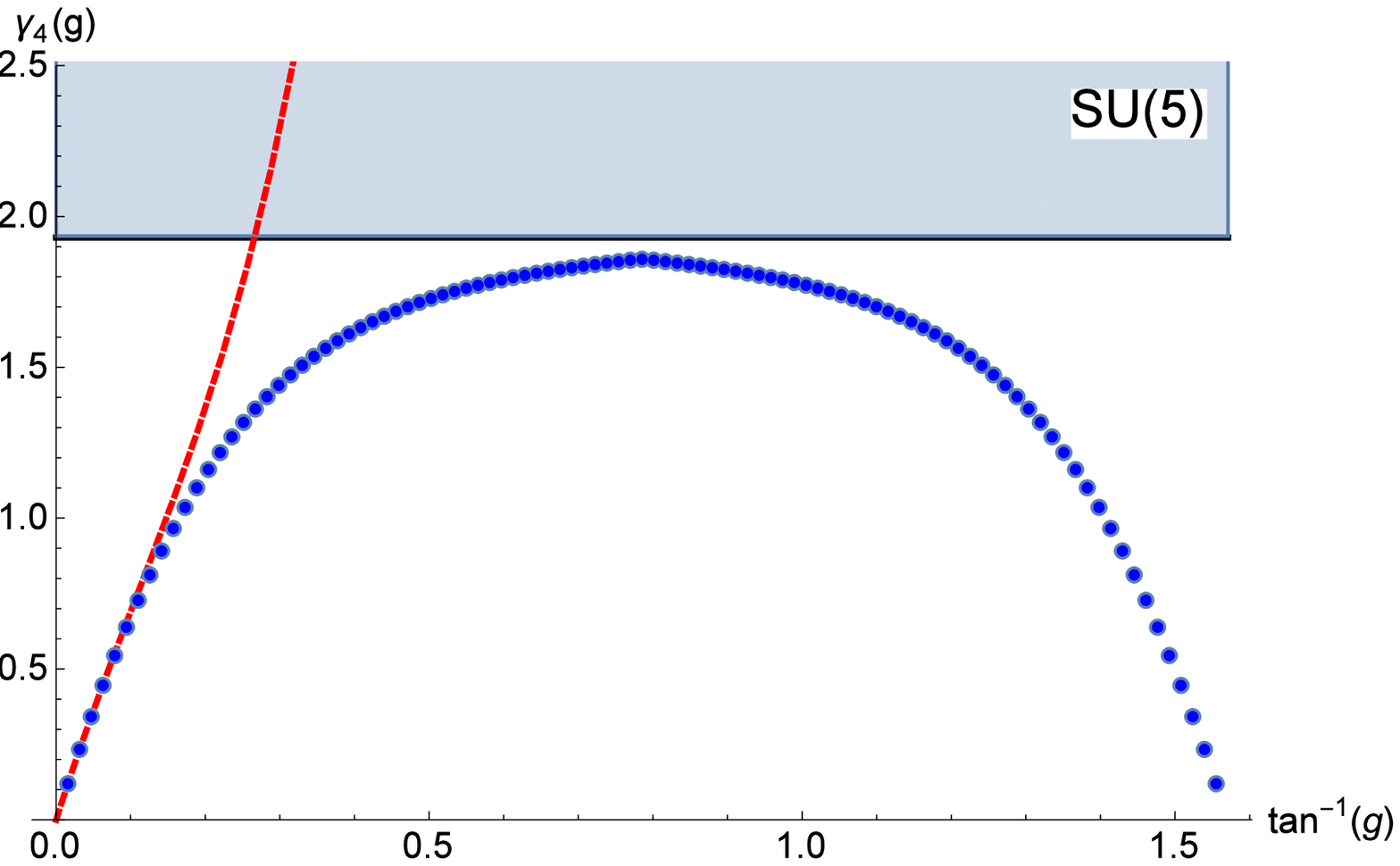}
\end{center}
\caption{
Similar plots for the spin-4 operator 
as fig.~\ref{fig:spin0_coupling} and \ref{fig:spin2_coupling}.
The interpolating function $F_3^{(30,1/3)}(\tau )$ for $\theta =0$
is plotted as the function of $g$.
The red dashed line denotes
the weak coupling expansion up to 3-loop.
}
\label{fig:spin4_coupling}
\end{figure}
\begin{figure}[th]
\begin{center}
\includegraphics[width=7.4cm]{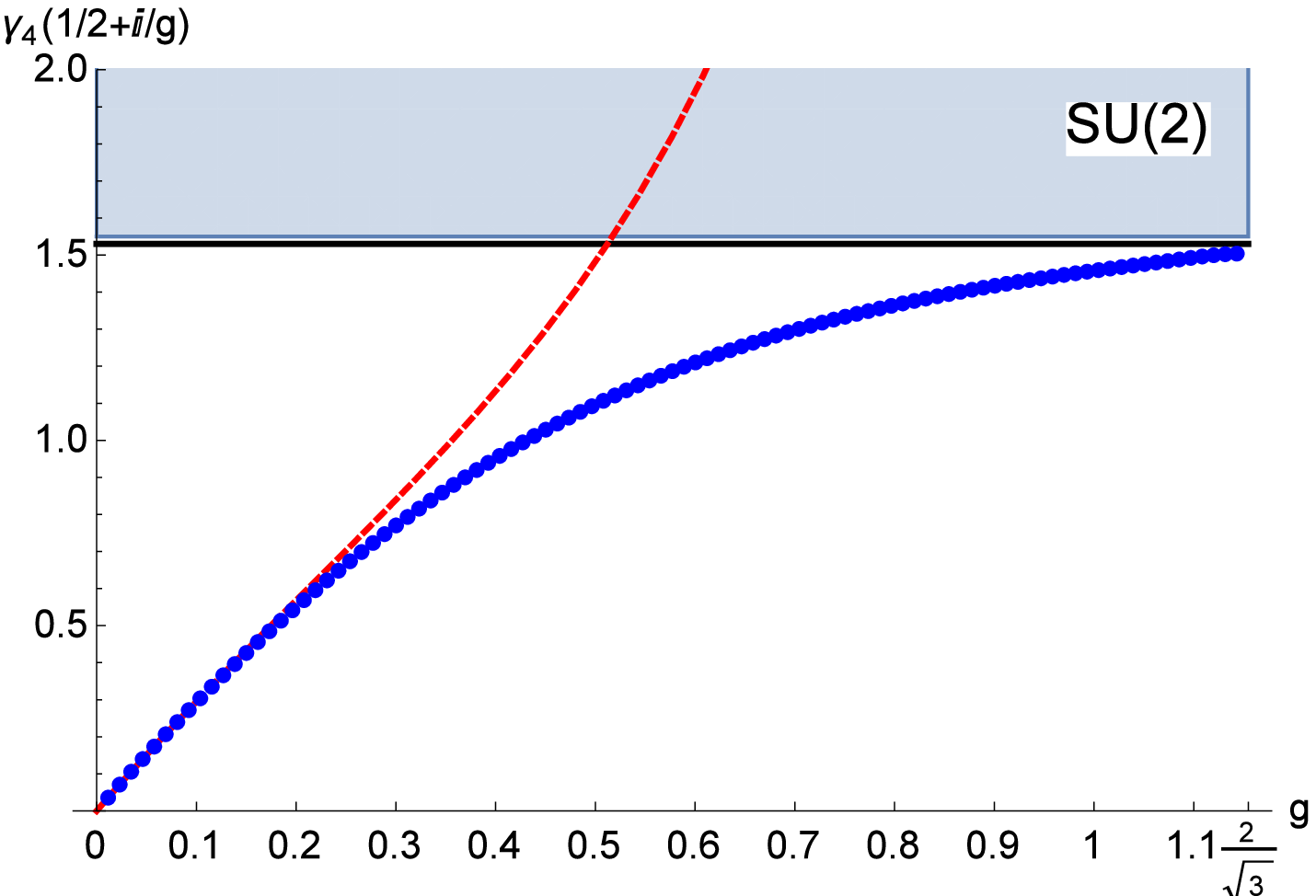}
\includegraphics[width=7.4cm]{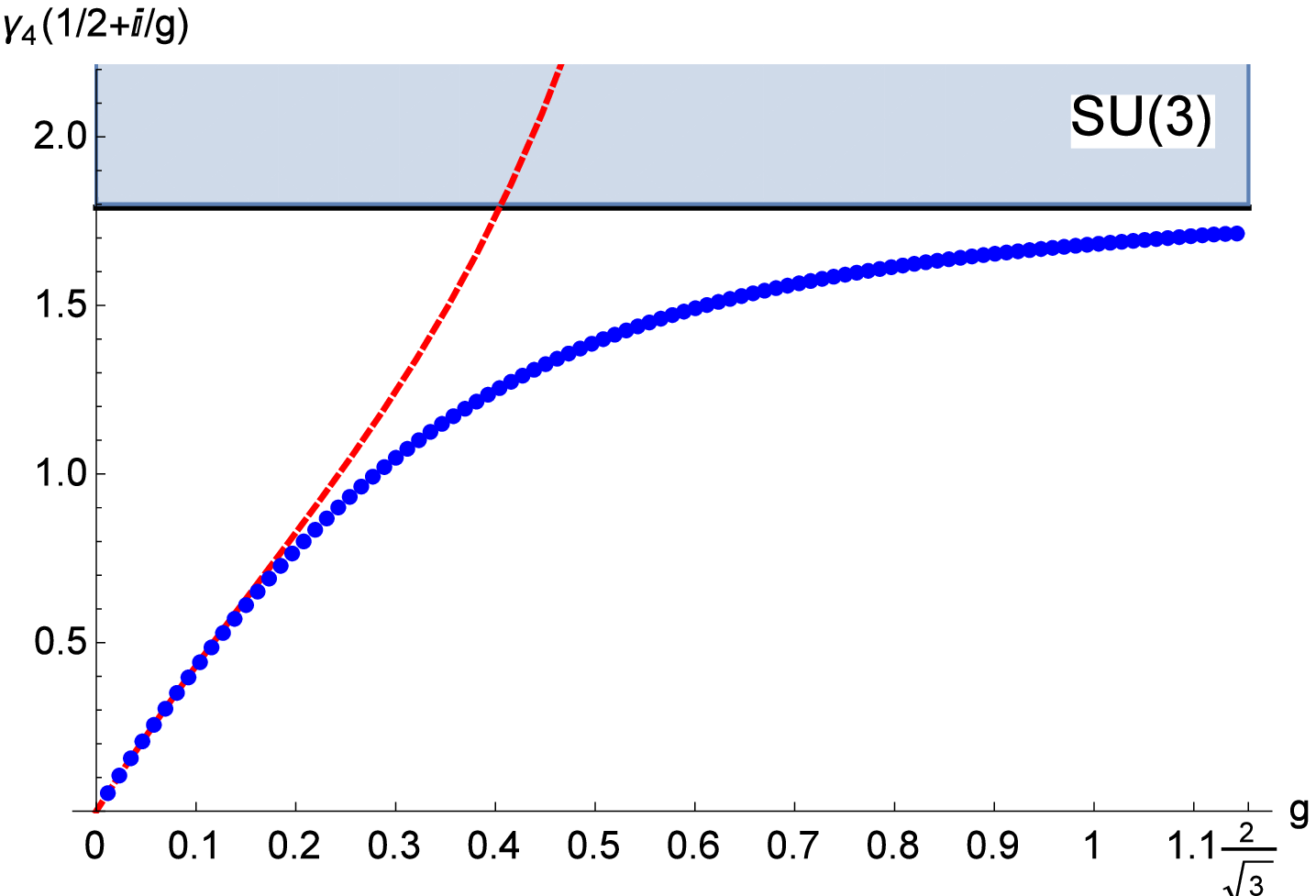}\\
\includegraphics[width=7.4cm]{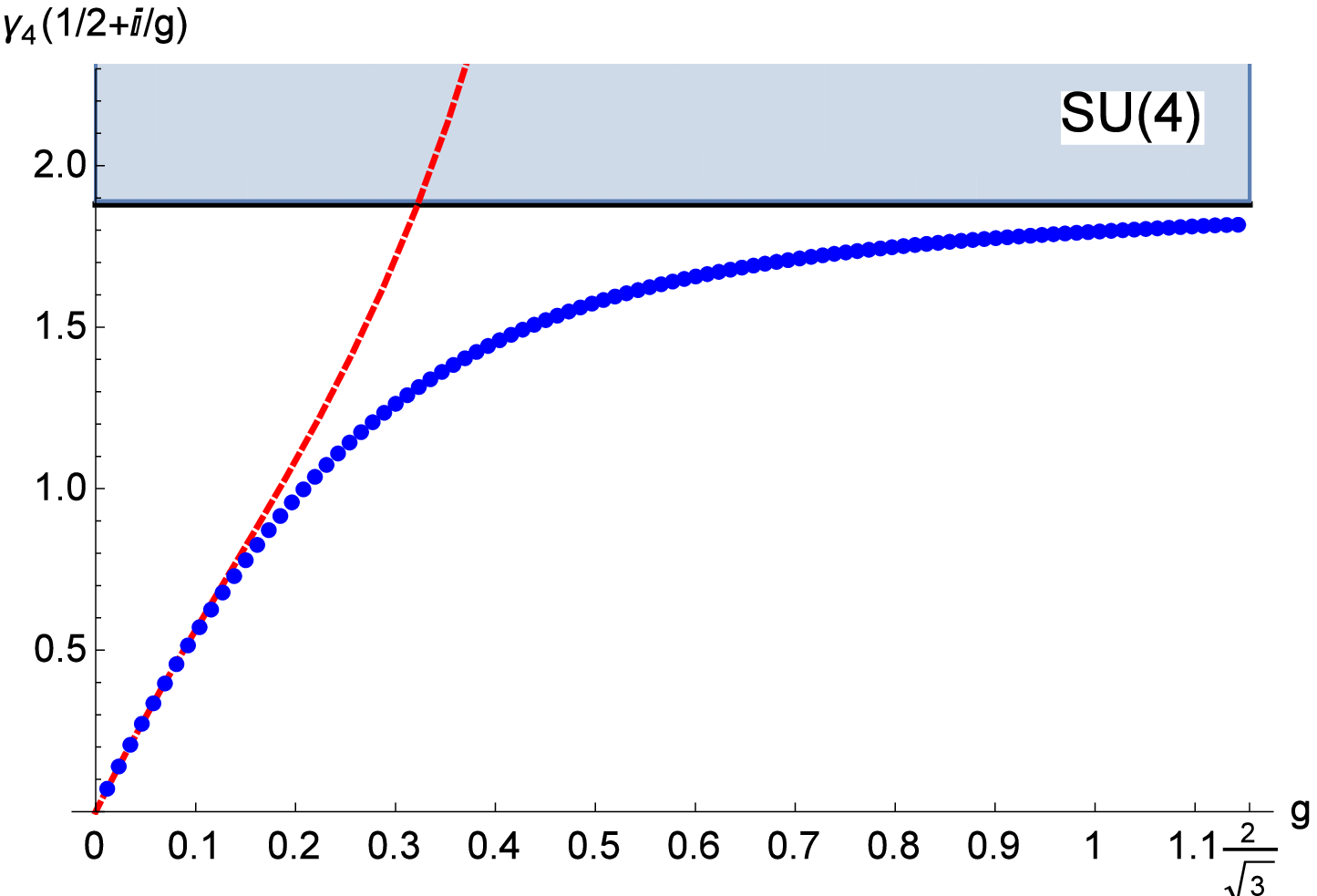}
\includegraphics[width=7.4cm]{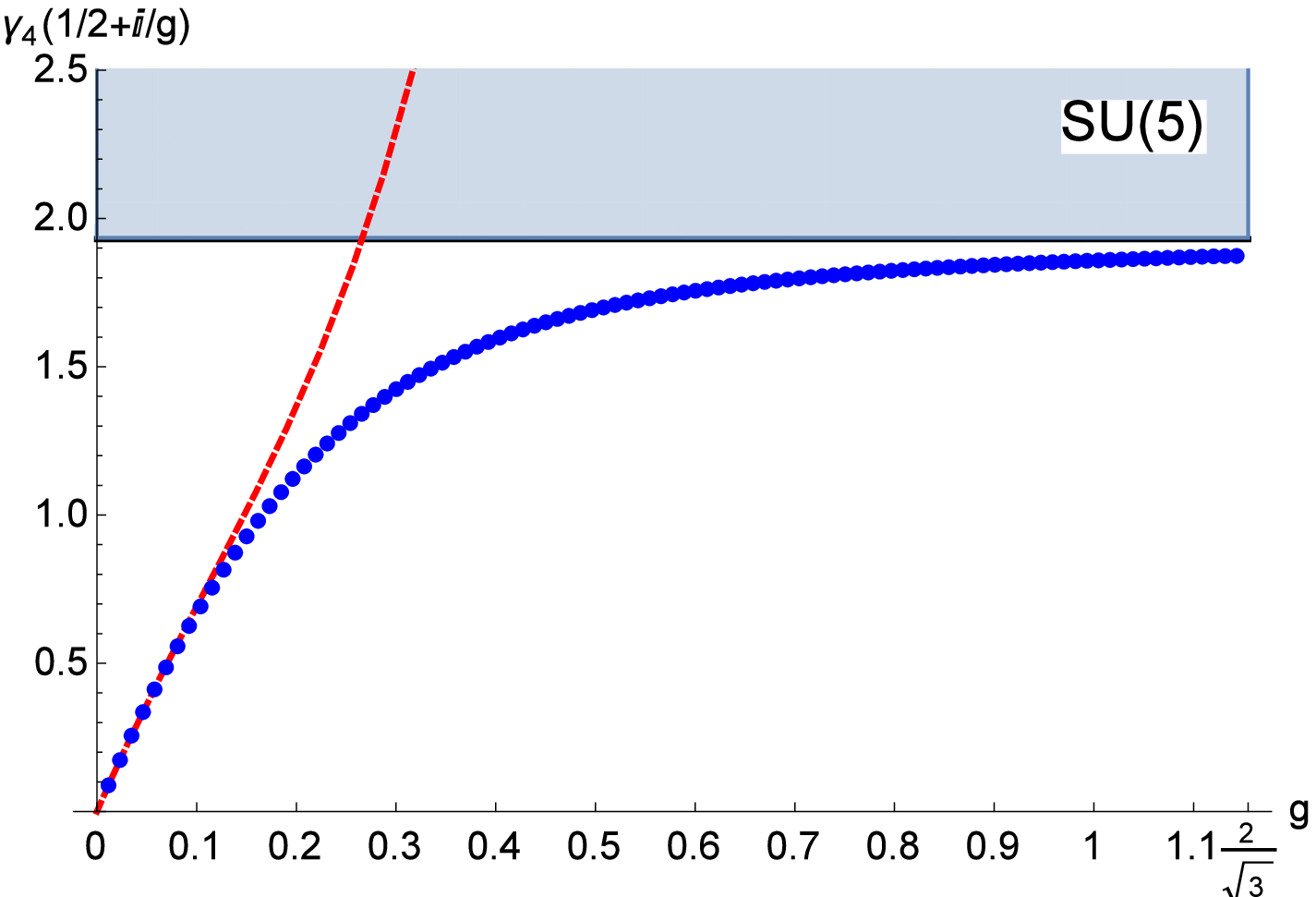}
\end{center}
\caption{
A similar plot as fig.~\ref{fig:spin4_coupling} for $\theta =\pi$.
The interpolating functions 
$F_3^{(30,1/3)}(1/2 +i/g)$
for the spin-4 operator
is plotted as a function of $g$.
}
\label{fig:spin4_coupling_pi}
\end{figure}
Finally let us consider the spin-4 case.
For this case 
we know the weak coupling expansion only up to three loop.
Hence according to sec.~\ref{sec:constraints} we shall consider interpolating functions with $m=3$. Contrary to the previous case, now $\alpha$ is uniquely determined as $\alpha =1/3$.
Thus
we expect that
the best approximation is given 
by the interpolating function $F_3^{(30,1/3)}(\tau )$. 

\begin{table}[h]
\begin{center}
  \begin{tabular}{|c||c | c| c| c| c| c|  }
  \hline  & $SU(2)$ & $SU(3)$ & $SU(4)$  & $SU(5)$ & $SU(6)$ & $SU(7)$ \\
  \hline $\sqrt{a}$  
            & 0.86603& 1.4142  & 1.9365 & 2.4495 & 2.9580 & 3.4641 \\  
\hline\hline $F_3^{(30,1/3)}(\tau_S )$   
& 1.45762 & 1.68167 & 1.79528 & 1.85814 & 1.8959  & 1.92016   \\
\hline $F_3^{(30,1/3)}(\tau_{TS})$   
& 1.5043  & 1.7136  & 1.81717 & 1.87378 & 1.90757 & 1.9292    \\
\hline \hline Corner value   &   1.53    &   1.79     &   1.88 
& 1.93  & 1.95  & 1.965     \\
\hline Strict upper bound  &    1.55    &   1.80     &   1.89   
& 1.935 & 1.955 & 1.965    \\
\hline
  \end{tabular}
\end{center}
\caption{The interpolating function for spin-4 at the duality invariant points
and data from the $\mathcal{N}=4$ superconformal bootstrap.}
\label{tab:fp_spin4}
\end{table}

As in the spin-0 and spin-2 cases,
we find 
from fig.~\ref{fig:spin4_coupling} and \ref{fig:spin4_coupling_pi} that
the interpolating function $F_3^{(30,1/3)}(\tau )$ 
has local maximum 
at the duality invariant points $\tau =\tau_S$ and $\tau =\tau_{TS}$.
From table \ref{tab:fp_spin4},
we see that
the interpolating function at $\tau =\tau_{TS}$
is greater than the one at $\tau =\tau_S$.
Thus our prediction for the maximum anomalous dimension 
is given by\footnote{
$
F_3^{(30,1/3)} (\tau_{S})
= 2N \left(  N^3 +0.157951N^2 +4.96272N +2.10839 \right)^{-1/3} .
$
}
\begin{\eq}
F_3^{(30,1/3)} (\tau_{TS})
= \frac{2N}{\left(   N^3 +0.136789 N^2 +4.44204N +1.36944 \right)^{1/3}} .
\end{\eq}

\begin{figure}[t]
\begin{center}
\includegraphics[width=7.4cm]{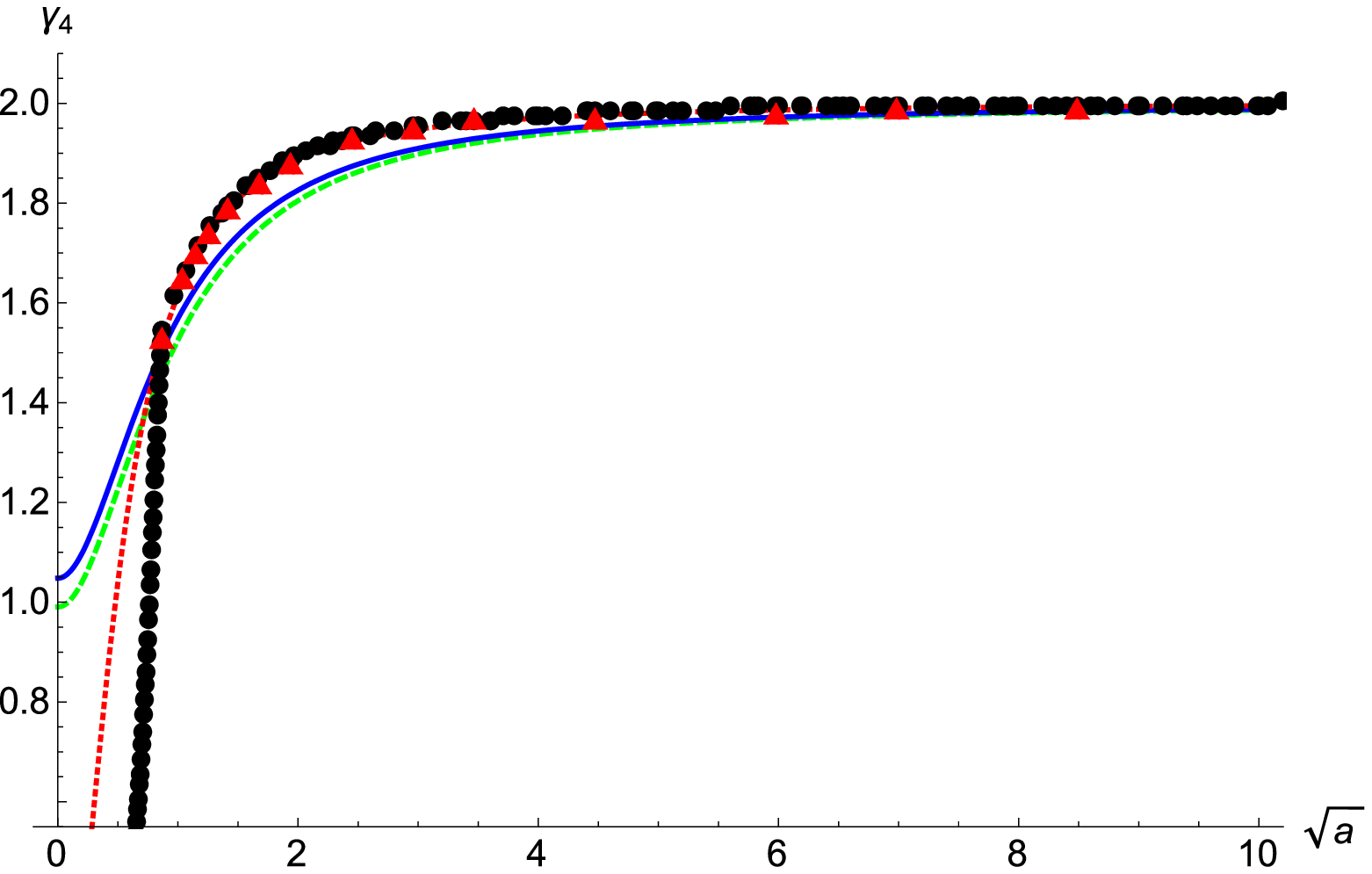}
\end{center}
\caption{
A similar plot for the spin-4 case 
as fig.~\ref{fig:bound0} and \ref{fig:bound2}.
The red dashed line shows $2-48/25N^2$,
which is numerical fitting of the corner values in the large-$a$ regime \cite{Beem:2013qxa}.
}
\label{fig:bound4}
\end{figure}

Fig.~\ref{fig:bound4}
compares our result with the $\mathcal{N}=4$ superconformal bootstrap.
In this figure we again see the agreement with the corner values
in the small-$a$ and large-$a$ regime
but there are about $10\%$ discrepancies in the intermediate regime.
To interpret this,
note that region, where the interpolating function would nicely approximate,
is different from the spin-0 and spin-2 cases.
First 
we expect that
accuracy of the interpolating function $F_3^{(30,1/3)}(\tau )$ 
is less than the spin-0 and spin-2 cases in the small-$a$ region.
This is 
because the result of the weak coupling expansion
is available only up to three loop for this case\footnote{
It is worthwhile to note that
the weak coupling expansion up to three loop is the same as the planar limit and
the fourth loop is the first order to deviate from the planar limit.
Thus we expect including the four-loop to be important.
}.
Secondly, as discussed in the spin-2 case,
large-$a$ regime is effectively broader for the spin-4 case.
We expect that 
the interpolating function is not reliable in the intermediate regime
but reliable in the small-$a$ and large-$a$ regime.

As a conclusion of this subsection,
we have seen that
when we expect reasonable approximation by the interpolating functions,
the maximal values of the interpolating functions
are close to the corner values of the bootstrap.
Thus we conclude that
our interpolating function approach strongly supports
the conjecture by the $\mathcal{N}=4$ superconformal bootstrap,
which states that 
the upper bounds on the dimensions are saturated at one of 
the duality-invariant points $\tau =\tau_S $ or $\tau =\tau_{TS}$.
Obviously,
if higher orders of the weak coupling perturbative series
become available,
then we can obtain more precise interpolating functions.
It would be nice 
if one can obtain the higher order results.

\subsubsection*{Comments on other gauge groups}
A priori our interpolating functions
are valid only for the $SU(N)$ gauge group
since we have used the weak coupling expansion and holographic computation
for the $SU(N)$ case. However, both 
the data of the bootstrap and the interpolating functions looks continuous for $a\geq 3/4$,
whereas the bootstrap data has a cusp at $a=3/4$ ($SU(2)$ case) which is not reproduced by the interpolating functions.
Thus we expect that
our interpolating functions would reasonably approximate
the dimension of the leading twist operator for other gauge groups as long as $a\geq 3/4$.
For $a<3/4$,
the only possible gauge group is $U(1)$, which is Abelian.
The bootstrap results have cusps at $a=3/4$ and
this implies 
a kind of transition from Abelian theory to non-Abelian theory,
or from free theory to interacting theory.
Contrary to the bootstrap,
our interpolating function is smooth across $a=3/4$.
Presumably the difference comes from the fact that
naive continuation of our interpolating function to general $a$
is a continuation suited to  interacting theory and
therefore we cannot apply the interpolating function 
for the $U(1)$ case.

\subsubsection*{Comments on instanton corrections}
The small-$g$ expansions of our interpolating functions
contain exponentially suppressed corrections,
whose weights are the same as the instanton actions.
This feature technically comes from imposing the full $S$-duality.
Recently Alday-Korchemsky \cite{Alday:2016tll} computed
instanton corrections to the dimension of the Konishi operator 
at $\mathcal{O}(g^2)$
by expanding around the instanton configuration.
They have found that
the instanton corrections start from $\mathcal{O}(g^2)$
and
the one-instanton correction for the $SU(2)$ case is given by
 \begin{\eq}
-\frac{9g^2}{20\pi^2} \left( e^{2\pi i\tau} +e^{-2\pi i\bar{\tau}}\right) .
\label{eq:1inst}
\end{\eq}
They also computed
$n$-instanton correction in the large-$N$ limit\footnote{
This is $g:{\rm fixed}$, $N\rightarrow\infty$ limit.
} as
\begin{\eq}
-\frac{27 g^2}{10\pi^{5/2} n^{3/2} N^{3/2}} 
\left( e^{2\pi in\tau} +e^{-2\pi in\bar{\tau}}\right) \sum_{d|n}\frac{1}{d^2},
\label{eq:ninst}
\end{\eq}
by using the technique of \cite{Dorey:1999pd}.
On the other hand,
the instanton corrections of our interpolating functions 
with general parameters
start from\footnote{
For the interpolating function $F_4^{(s,1/4)}(\tau )$,
it starts from $\mathcal{O}(g^{s+2})$.
} $\mathcal{O}(g^{s+1+{\rm min}(p,q)})$.
Since we would like to take $s$ to be sufficiently large
as discussed in sec.~\ref{sec:s},
the interpolating function $F_4^{(s,1/4 )}(\tau )$ with large-$s$
cannot reproduce the results \eqref{eq:1inst} and \eqref{eq:ninst}.
It is nice 
if one can construct a new class of interpolating functions,
which are consistent with \eqref{eq:1inst} and \eqref{eq:ninst}
in addition to the four-loop result, holographic result and full $S$-duality.

However, it is worthwhile to note that there is a subtlety in \cite{Alday:2016tll}.
We can also compute 
instanton corrections to the circular Wilson loop by using the same technique 
as in \cite{Bianchi:2002gz}.
However, the result of \cite{Bianchi:2002gz} states that
there are non-trivial instanton corrections to the circular Wilson loop.
This does not agrees with the results  
obtained by summing ladder diagrams \cite{Erickson:2000af,Drukker:2000rr} 
and the localization method \cite{Pestun:2007rz},
where instanton corrections are trivial.
Thus we should be careful on this point.

\subsubsection*{Comments on higher order corrections in the planar limit}
Although we have used the four-loop result of the weak coupling expansion
to construct the spin-0 interpolating functions,
there is a seven-loop result in the planar limit \cite{Bajnok:2012bz,Gromov:2014bva,Marboe:2014gma},
whose explicit form is given by \eqref{eq_Konishiw}.
We did not use the seven-loop result
because we also need non-planar corrections 
to completely fix the coefficients in interpolating functions
and the interpolating functions strongly depends 
on the values of the non-planar higher order corrections for small $N$.
Here we just compare
the higher order correction in the planar limit 
with the ones of the interpolating function.
The coefficients of the higher order small-$\lambda$ expansion of 
the interpolating function $F_4^{(30,1/4)}(\tau )$ in the planar limit are
\begin{\eqa}
 \left. F_4^{(30,1/4)}(\tau ) \right|_{{\rm spin-0,planar},\mathcal{O}(\lambda^5 ) }
&=& \frac{3 (-240 \zeta (3)+600 \zeta (5)+329)}{64 \pi ^5}
\simeq  0.101504 ,\NN\\
 \left. F_4^{(30,1/4)}(\tau ) \right|_{{\rm spin-0,planar},\mathcal{O}(\lambda^6 ) }
&=&\frac{9 (300 \zeta (3)-750 \zeta (5)-143)}{64 \pi ^6}
\simeq -0.0819242 ,\NN\\
 \left. F_4^{(30,1/4)}(\tau ) \right|_{{\rm spin-0,planar},\mathcal{O}(\lambda^7 ) }
&=& \frac{3 \left(360 \zeta (3)^2-120 \zeta (3) (15 \zeta (5)+98)+150 \zeta (5) (15 \zeta (5)+196)+871\right)}{256 \pi ^7} \NN\\
&\simeq &  0.0695153 ,
\end{\eqa}
while the correct values are
\begin{\eq}
\left. \Delta_{\rm Konishi} \right|_{{\rm planar},\mathcal{O}(\lambda^5 )} 
\simeq  0.119731 ,\quad
\left. \Delta_{\rm Konishi} \right|_{{\rm planar},\mathcal{O}(\lambda^6 )} 
\simeq  0.11623 ,\quad
\left. \Delta_{\rm Konishi} \right|_{{\rm planar},\mathcal{O}(\lambda^7 )} 
\simeq  0.117987  .
\end{\eq}
It is attractive
if one can construct interpolating functions,
which appropriately include
the higher order corrections in the planar limit.

\subsubsection*{Comments on $\alpha'$-corrections}
Recently the large-$N$ bootstrap 
for the $\mathcal{N}=4$ SCFT \cite{Alday:2014tsa}
studied structures of $\alpha'$-corrections
to the dimensions of the leading twist operators in the supergravity limit.
It has turned out that
the $\alpha'$-corrections (large-$\lambda$ expansion) 
at $\mathcal{O}(1/N^2)$ starts from $\mathcal{O}(1)$
and 
the next order is $\mathcal{O}(\alpha^{\prime 3})=\mathcal{O}(\lambda^{-3/2})$.
This feature is different from
our interpolating functions $F_4^{(s,1/4)}(\tau )$ for the spin-0,2 cases and
$F_3^{(s,1/3)}(\tau )$ for the spin-4 case.
Namely
their large-$\lambda$ expansions at $\mathcal{O}(1/N^2)$
start from $\mathcal{O}(1)$
but the next orders are $\mathcal{O}(\lambda^{-1})=\mathcal{O}(\alpha^{\prime 2})$.
Furthermore 
the large-$\lambda$ expansions have 
only non-negative integer powers of $\lambda^{-1}$.
It is illuminating 
if we can construct interpolating functions to be consistent with
the result of \cite{Alday:2014tsa}.

\subsection{Image of conformal manifold}
\label{sec:manifold}
The complex coupling $\tau$ is the exactly marginal parameter and
hence the coordinate of the conformal manifold in the $\mathcal{N}=4$ SYM.
At every point $\tau$ on the conformal manifold,
we have a set of dimensions of the leading twist operators:
$(\Delta_0 ,\Delta_2 ,\Delta_4 ,\cdots )$.
Since we have constructed the approximations of $(\Delta_0 ,\Delta_2 ,\Delta_4 )$
by the interpolating functions,
we can draw
an image of the conformal manifold projected 
to the $(\Delta_0 ,\Delta_2 ,\Delta_4 )$-space.

\subsubsection{$SU(2)$ case}
\begin{figure}[t]
\begin{center}
\includegraphics[width=7.4cm]{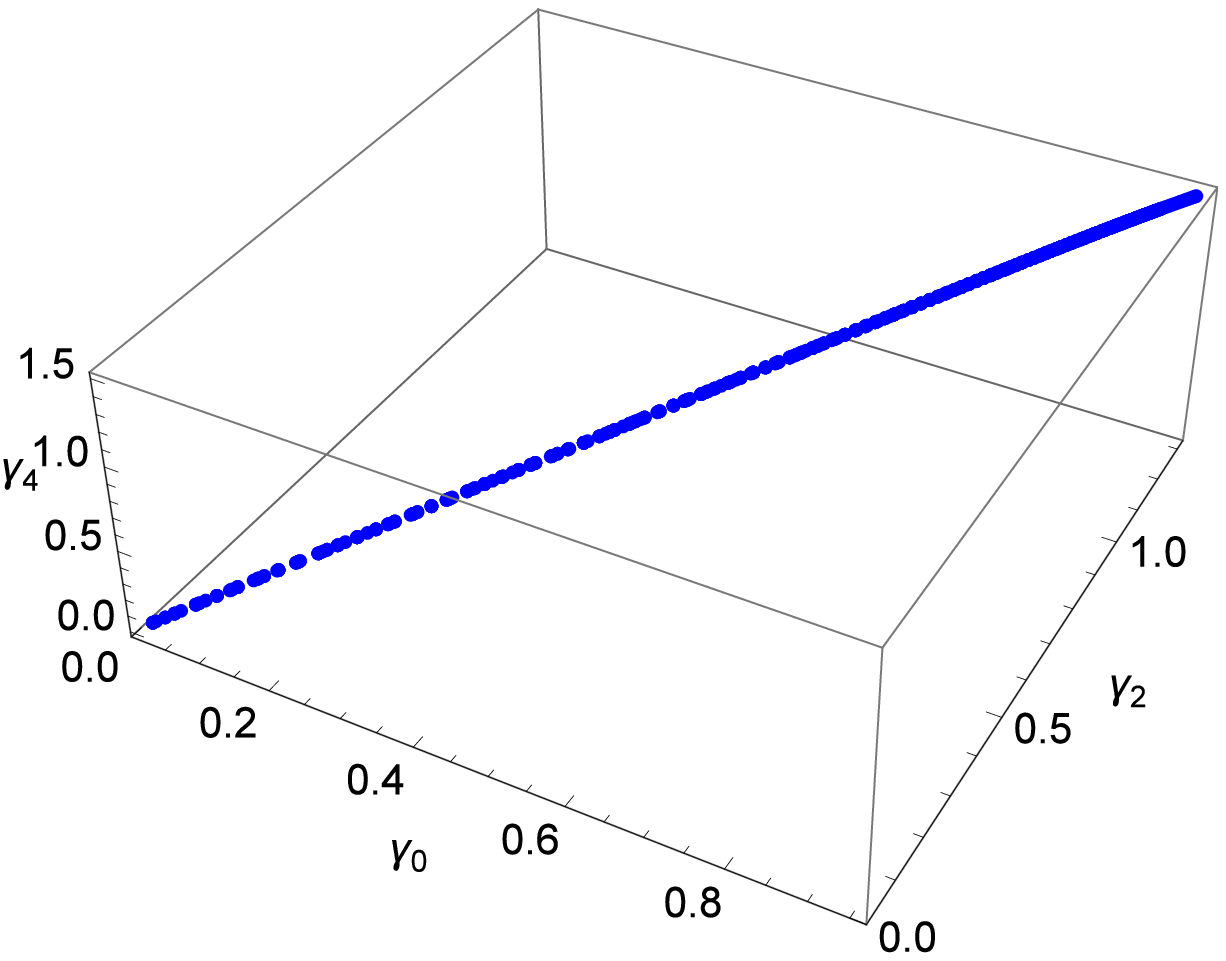}
\includegraphics[width=7.4cm]{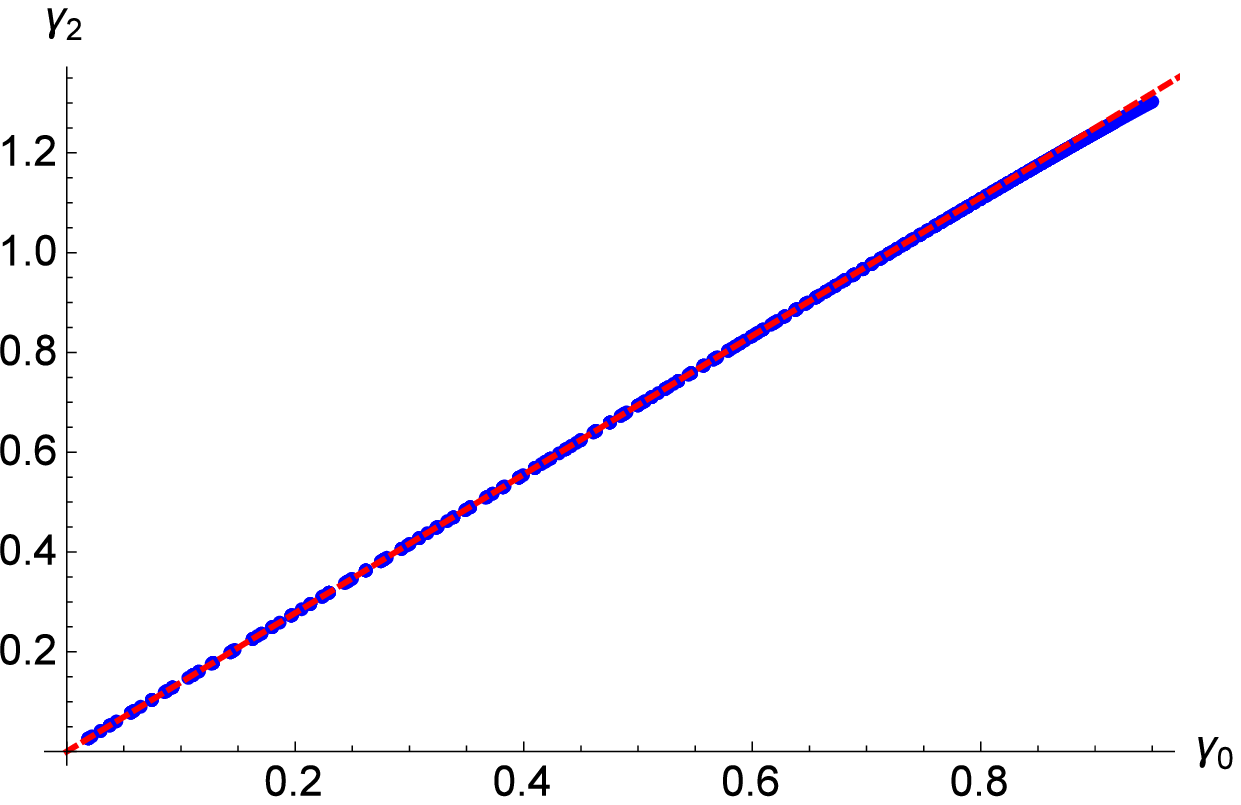}\\
\includegraphics[width=7.4cm]{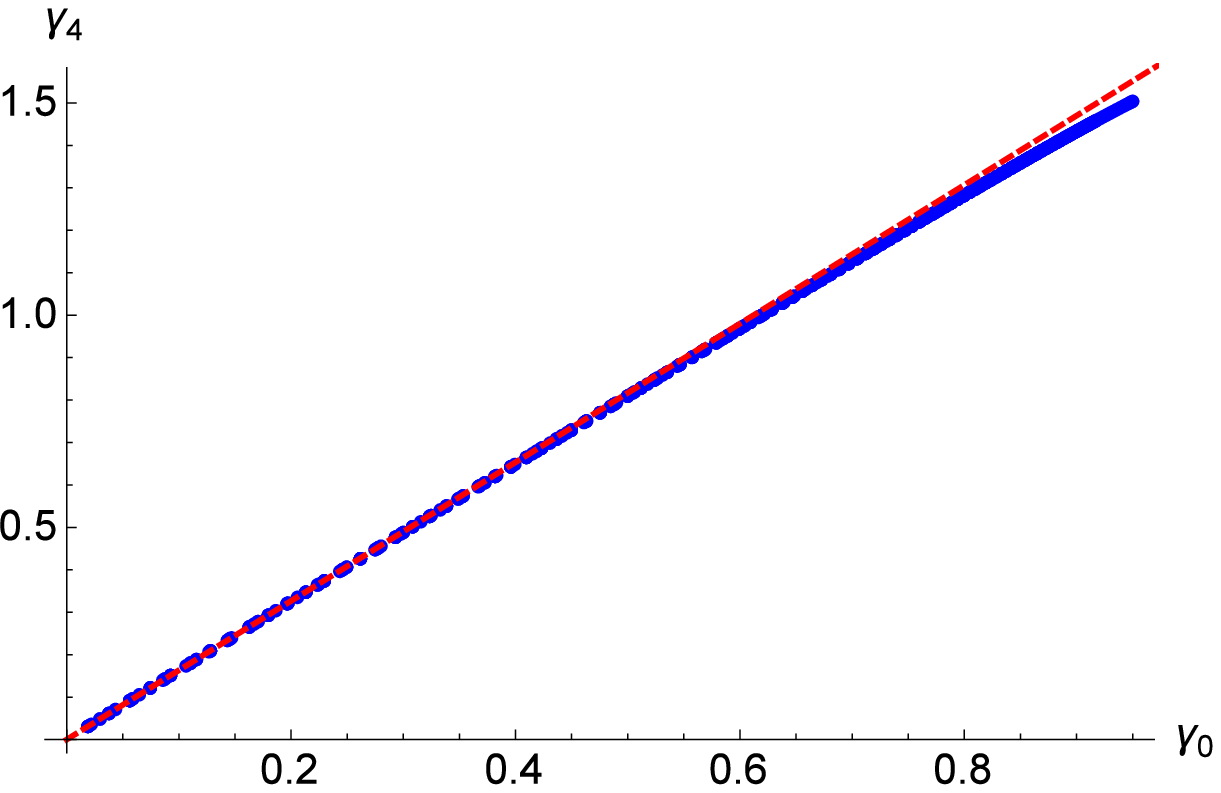}
\includegraphics[width=7.4cm]{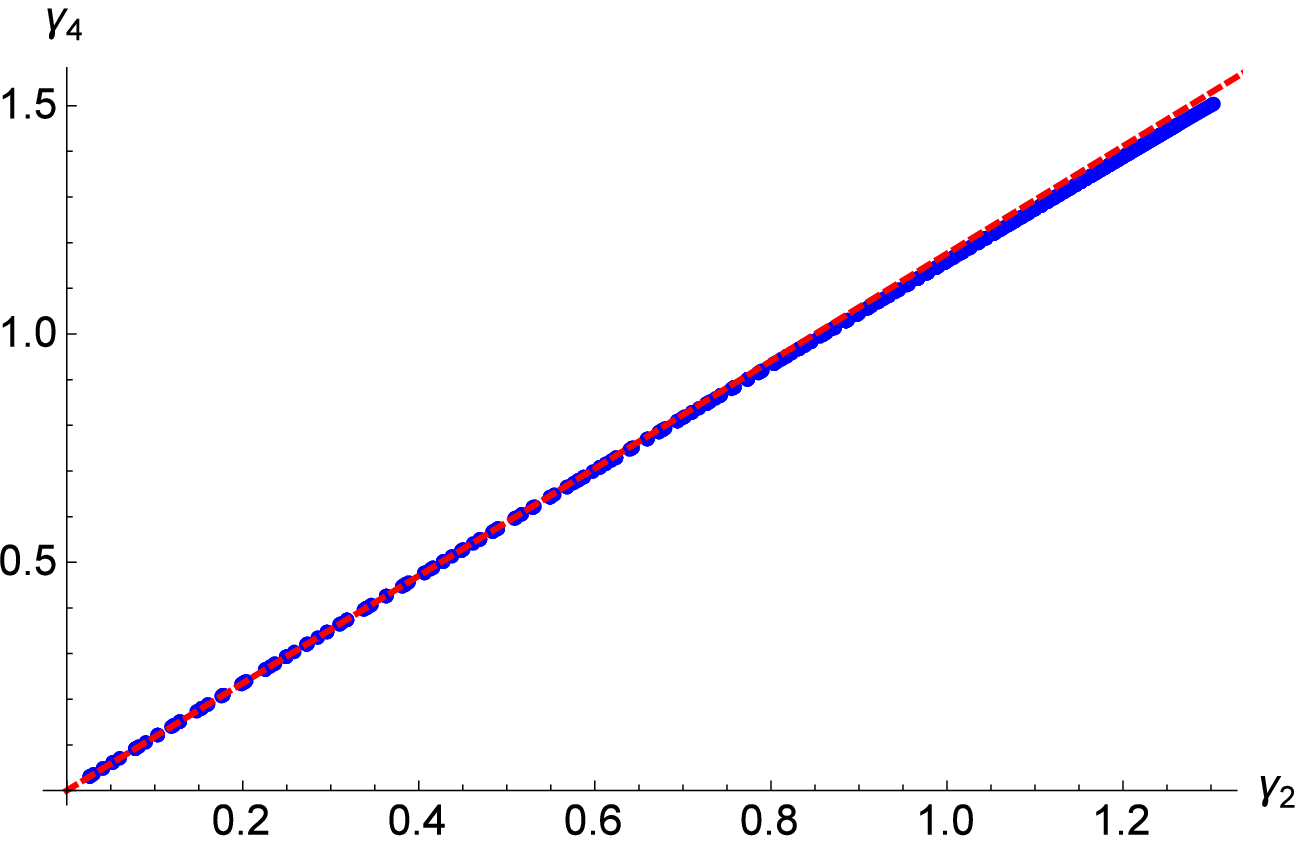}
\end{center}
\caption{The  figure shows the image of 
the conformal manifold for $N=2$
in the $(\gamma_0,\gamma_2 ,\gamma_4 )$-space.
We use the interpolating function $F_4^{(30,1/4)}(\tau )$ 
for $\gamma_0 ,\gamma_2$ and
$F_3^{(30,1/3)}(\tau )$ for $\gamma_4 $.
We have evaluated the interpolating functions
for $\theta =\pi n/5$ with $n=0,1,\cdots ,5$.
[Left-Top] The 3D plot in the $(\gamma_0,\gamma_2 ,\gamma_4 )$-space.
[Right-Top] The projection to the $(\gamma_0 ,\gamma_2 )$-plane.
The red dashed line shows $\gamma_2 =(25/18)\gamma_0$.
[Left-Bottom] The projection to the $(\gamma_0 ,\gamma_4 )$-plane.
The red dashed line shows $\gamma_4 =(49/30)\gamma_0$.
[Right-Bottom] The projection to the $(\gamma_2 ,\gamma_4 )$-plane.
The red dashed line shows $\gamma_4 =(147/125)\gamma_2$.
}
\label{fig:mfdN2}
\end{figure}
In fig.~\ref{fig:mfdN2} [Left-Top]
we plot the image of the conformal manifold for the $SU(2)$ case
projected\footnote{
We can get the result in the $(\Delta_0 ,\Delta_2 ,\Delta_4 )$-space
just by shifting the anomalous dimensions by two.
} to the $(\gamma_0 ,\gamma_2 ,\gamma_4 )$-space.
We have sampled the interpolating functions
for $\theta =\pi n/5$ with $n=0,1,\cdots ,5$ and many values of $g$.
The figure shows that
the image is almost one continuous straight line and
this is different from the following naive expectation.
Since we are considering the image of the two dimensional conformal manifold,
one naively expects that
the image was almost a straight line around the origin
but it starts spreading as the dimension increases.
This is 
because the anomalous dimensions in the weak coupling regime
are almost independent of $\theta$ 
but the $\theta$-dependence becomes important
in the strongly coupled regime,
where the anomalous dimensions become large.
Thus, if the above naive expectation was correct,
then we should observe six distinguishable lines
as we consider the six different values of $\theta$.

\begin{figure}[t]
\begin{center}
\includegraphics[width=7.4cm]{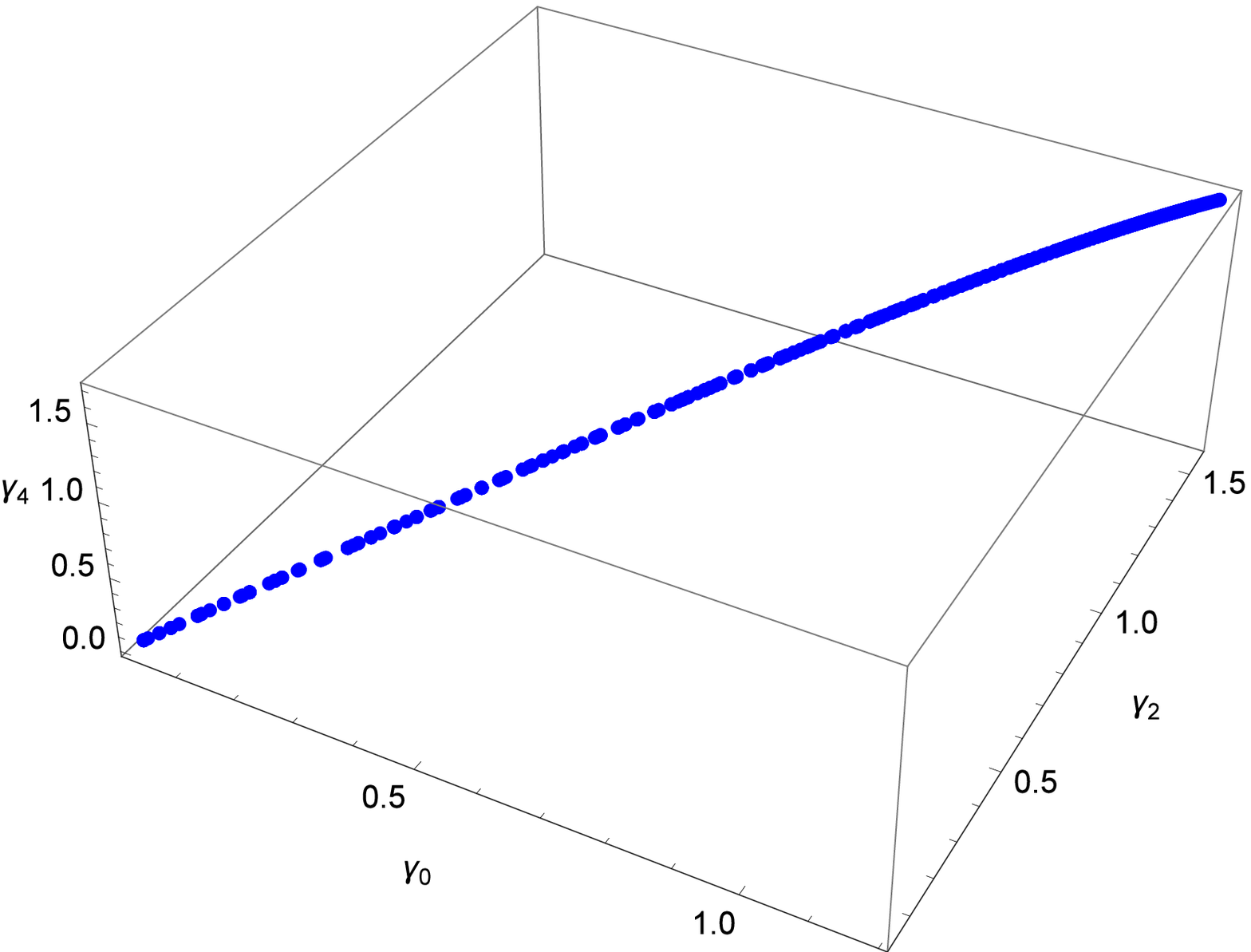}
\includegraphics[width=7.4cm]{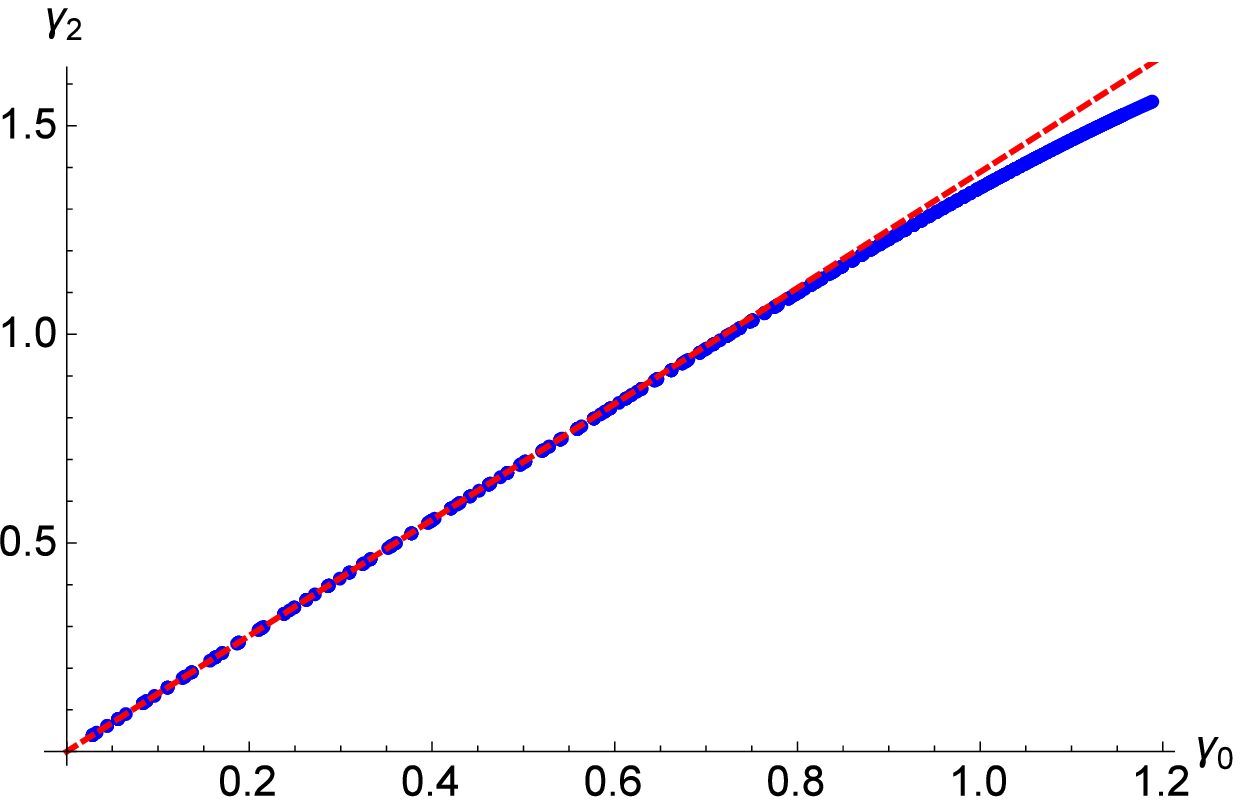}\\
\includegraphics[width=7.4cm]{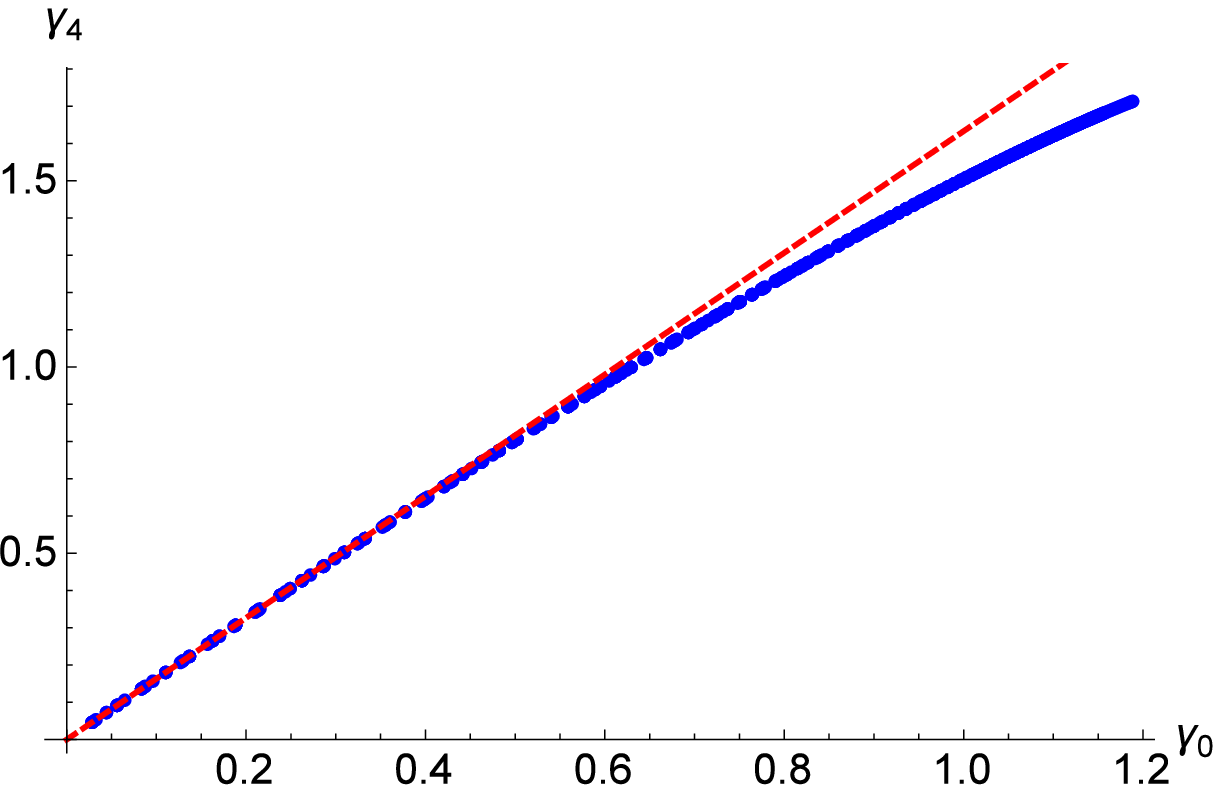}
\includegraphics[width=7.4cm]{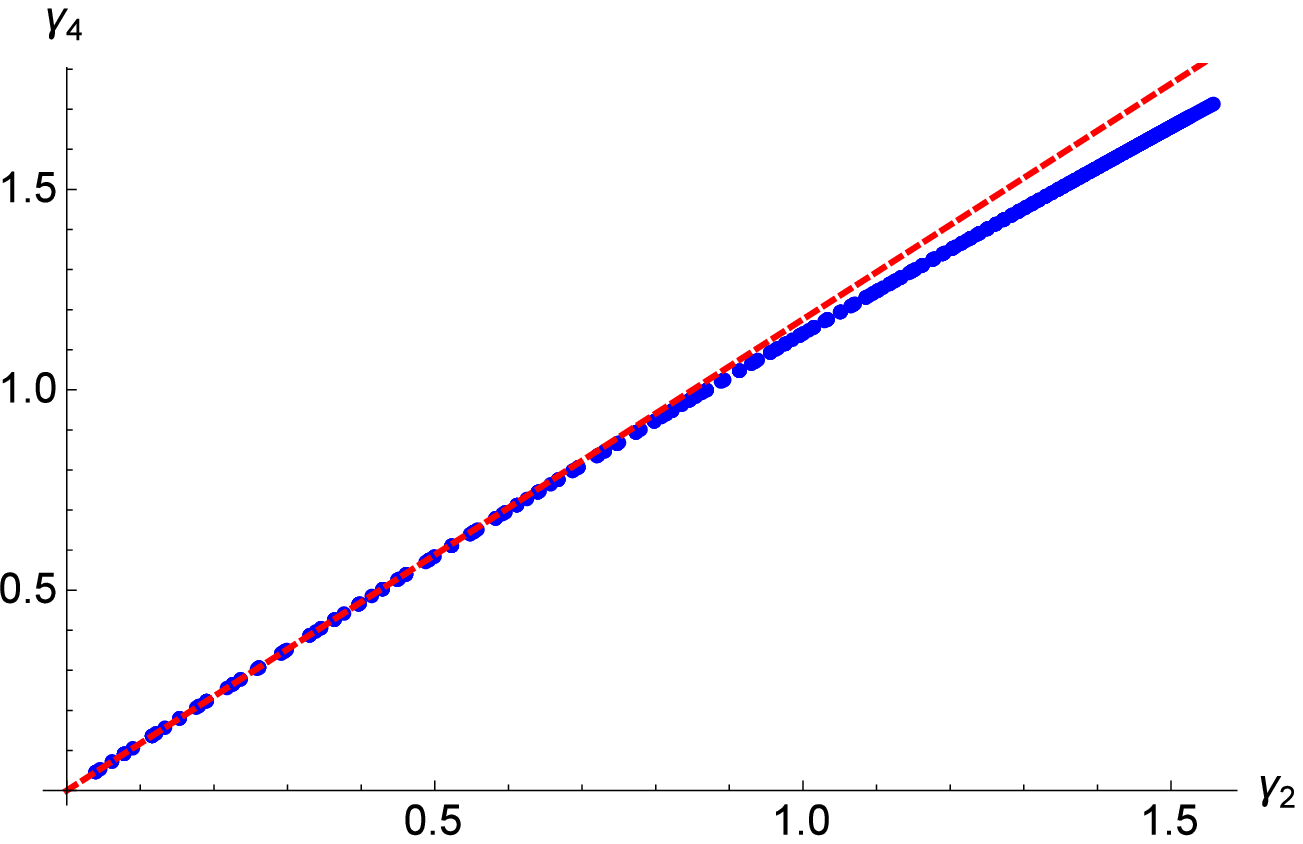}
\end{center}
\caption{
Similar plots as fig.~\ref{fig:mfdN2} for the $SU(3)$ case.
}
\label{fig:mfdN3}
\end{figure}

This property has been already observed 
for the $(\gamma_0 ,\gamma_2 )$-plane in \cite{Beem:2013hha}
by using the different interpolating functions
partially invariant 
under the $SL(2,\mathbb{Z})$ transformations.
In \cite{Beem:2013hha}
the authors found that
the slope of the almost straight line in the $(\gamma_0 ,\gamma_2 )$-plane 
is very close to $25/18\simeq 1.38889$,
which is the same as 
the ratio between the one-loop anomalous dimensions:
\begin{\eq}
 \frac{\left. \gamma_2 (\tau )\right|_{\mathcal{O}(g)} }
      {\left. \gamma_0 (\tau ) \right|_{\mathcal{O}(g)}} 
= \frac{25}{18} .
\end{\eq}
Fig.~\ref{fig:mfdN2} [Right-Top] 
shows that this is true also for our interpolating function.

In fig.~\ref{fig:mfdN2} [Left-Bottom] and \ref{fig:mfdN2} [Right-Bottom],
we show the similar plots 
in the $(\gamma_0 ,\gamma_4 )$ and $(\gamma_2 ,\gamma_4 )$-planes, 
respectively.
The straight lines show
$\gamma_4 =(49/30)\gamma_0$ and $\gamma_4 =(147/125)\gamma_2$,
whose slopes are the same as
\begin{\eq}
 \frac{\left. \gamma_4 (\tau ) \right|_{\mathcal{O}(g)}}
  {\left. \gamma_0 (\tau ) \right|_{\mathcal{O}(g)}}
= \frac{49}{30} ,\quad
\frac{\left. \gamma_4 (\tau ) \right|_{\mathcal{O}(g)}}
{\left. \gamma_2 (\tau ) \right|_{\mathcal{O}(g)}}
= \frac{147}{125} .
\end{\eq}
We find that
the images projected to the $(\gamma_0 ,\gamma_4 )$ and $(\gamma_2 ,\gamma_4 )$-planes
are very close to the straight lines as well.
This implies that
the observation in \cite{Beem:2013hha}
is also true for the $(\gamma_0 ,\gamma_4 )$ and $(\gamma_2 ,\gamma_4 )$-planes.
In the rest of this subsection
we see that
the situation is different for the higher $N$ cases.

\subsubsection{Higher $N$}
\begin{figure}[t]
\begin{center}
\includegraphics[width=7.4cm]{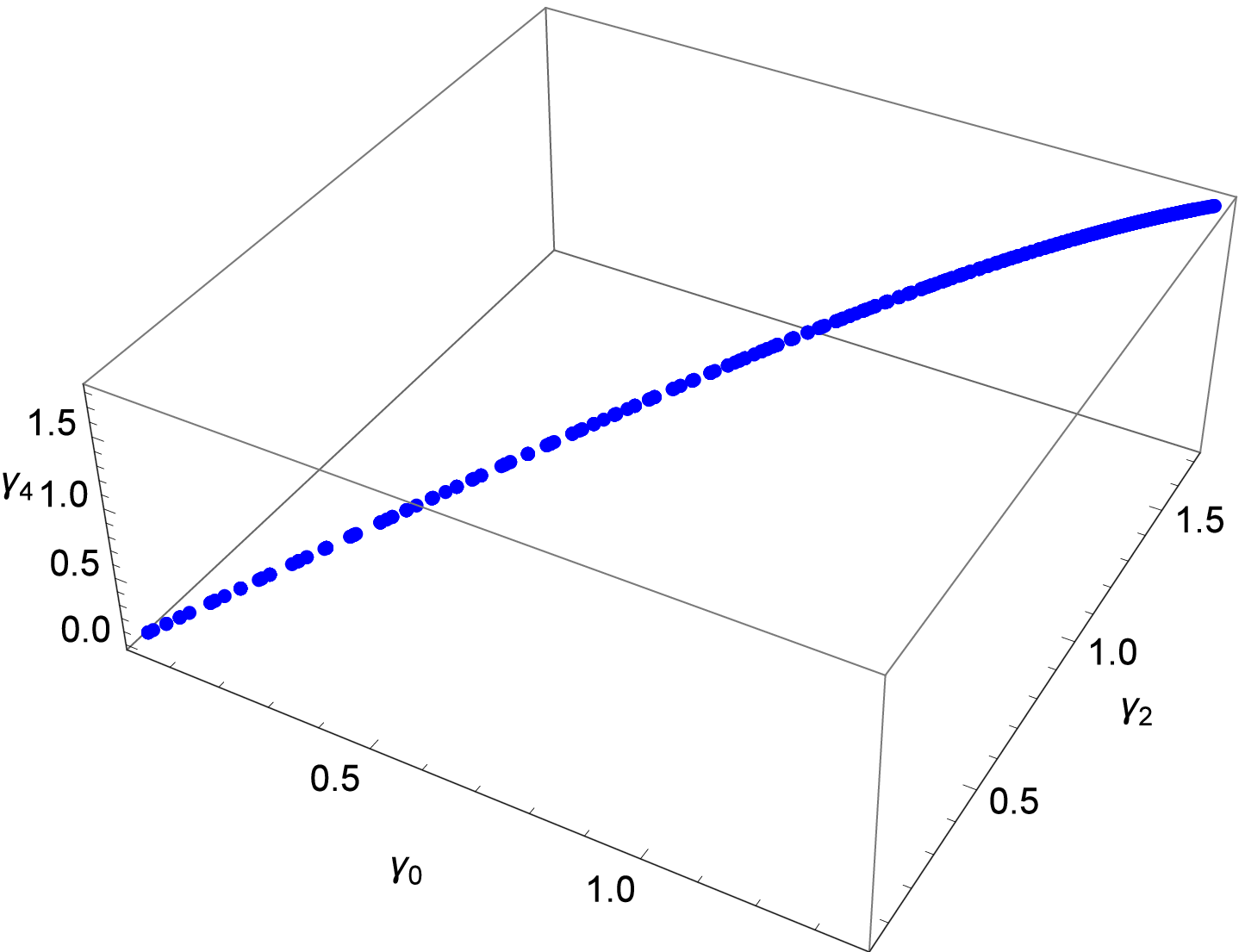}
\includegraphics[width=7.4cm]{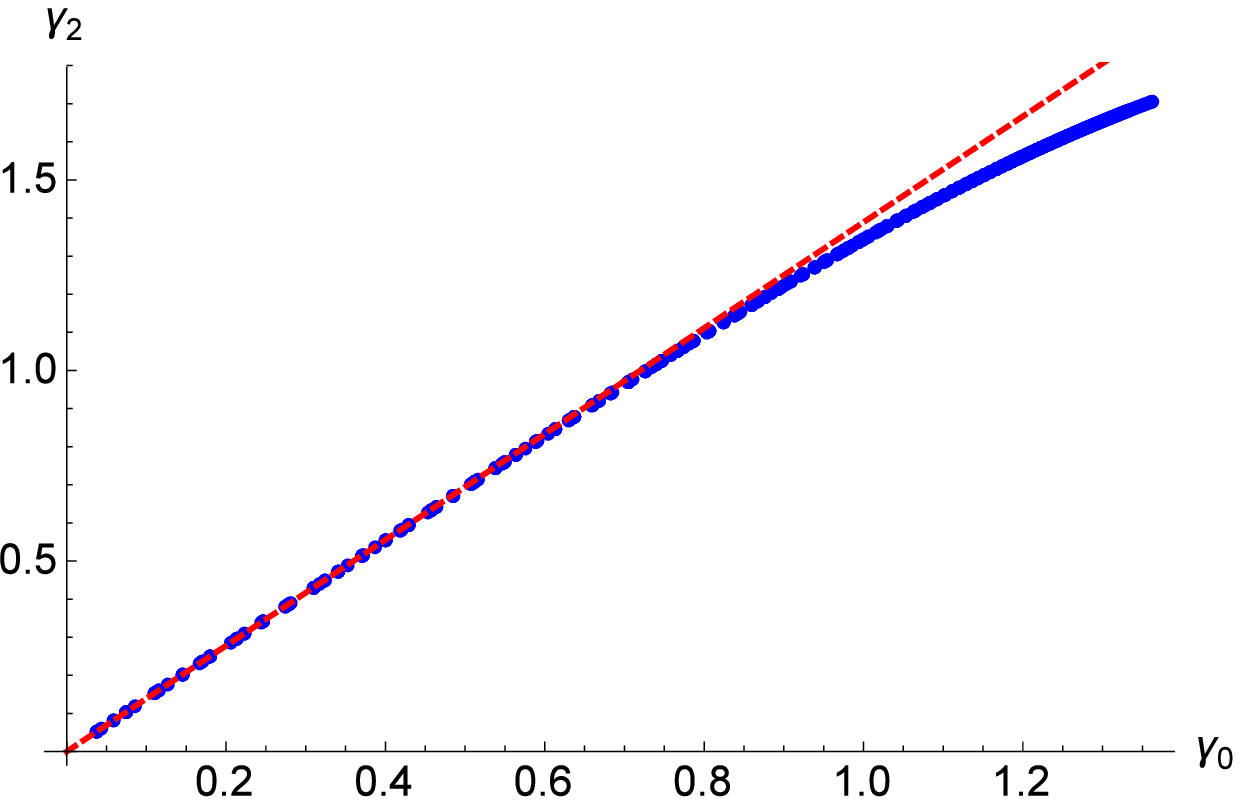}\\
\includegraphics[width=7.4cm]{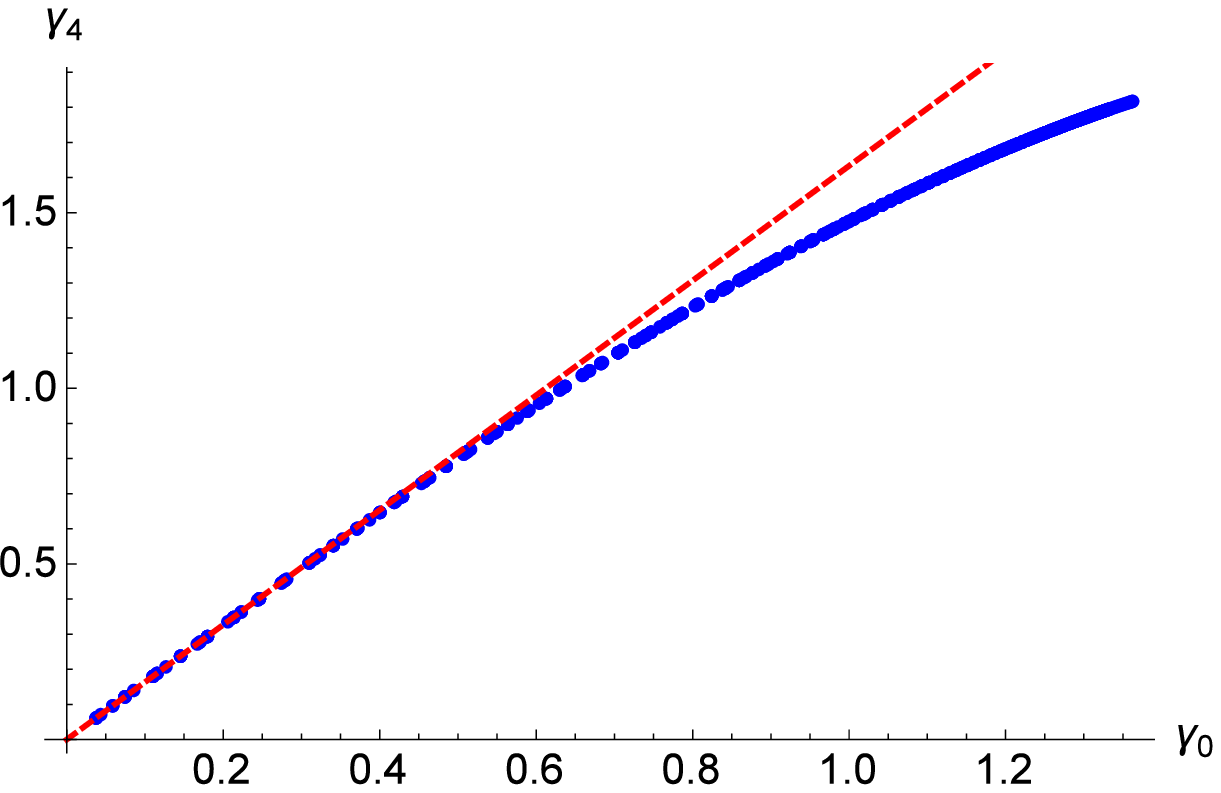}
\includegraphics[width=7.4cm]{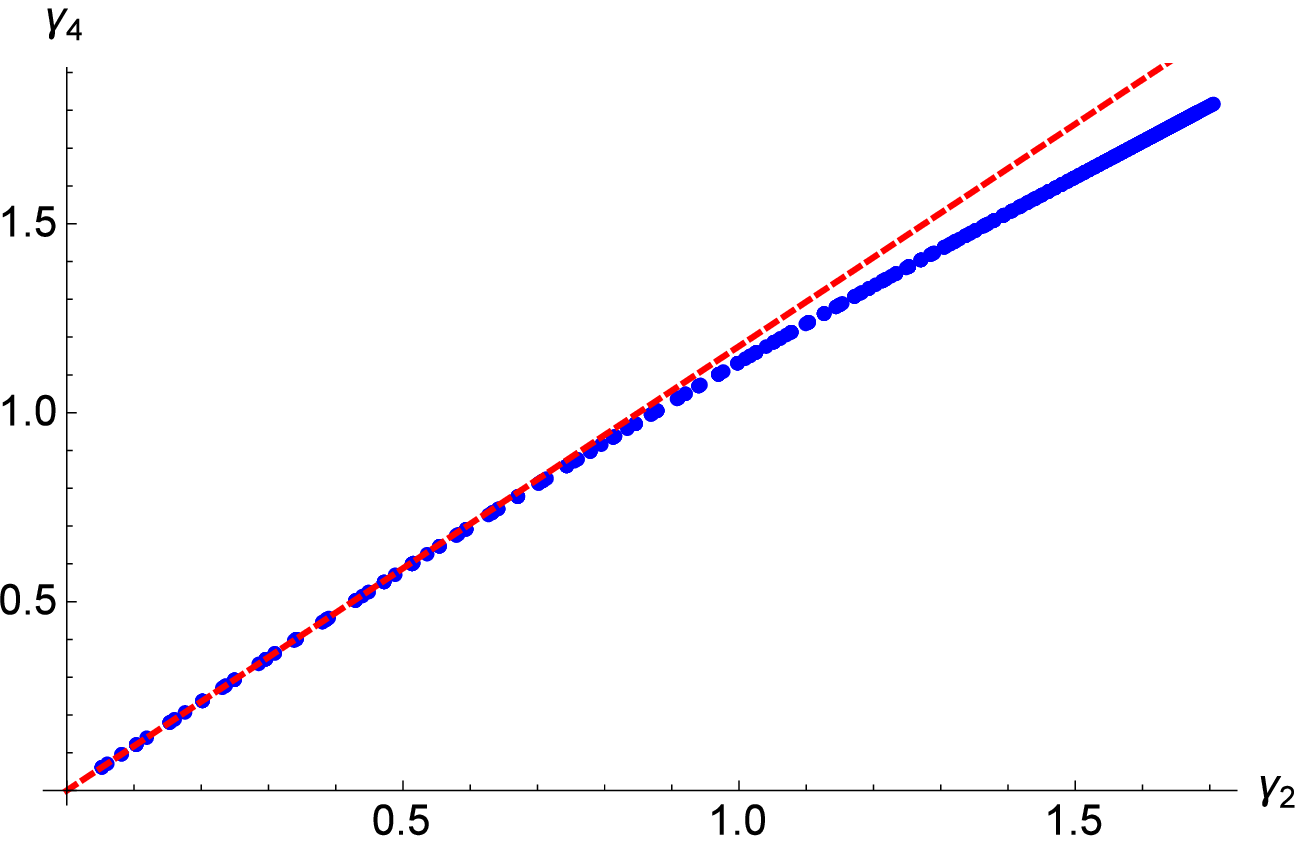}
\end{center}
\caption{
Similar plots as fig.~\ref{fig:mfdN2} and \ref{fig:mfdN3}
for the $SU(4)$ case.
}
\label{fig:mfdN4}
\end{figure}

Let us consider the higher $N$ cases.
In fig.~\ref{fig:mfdN3}
we give the similar plots as fig.~\ref{fig:mfdN2} for the $SU(3)$ case.
We now find both similarity and dissimilarity from the $SU(2)$ case.
The similarity is that
the image is still very narrow and therefore looks like one-dimensional.
The dissimilarity is that
the image is no longer straight line,
namely it is curved in the strong coupled regime.
To see this in more detail,
let us see fig.~\ref{fig:mfdN3} [Right-Top],
which plots the image in the $(\gamma_0 ,\gamma_2 )$-plane.
The straight line shows again 
the one-loop relation $\gamma_2 =(25/18)\gamma_0$.
The plot tells us that
the curve of the image deviates from the straight line
when 
$\gamma_0 \geq 0.95$.
There are two surprising things on this plot.
First,
the curve is still very narrow after the deviation.
Second,
the matching with the straight line holds 
even in the regime, 
where we cannot naively trust 
the one-loop approximation for the dimensions.
Fig.~\ref{fig:mfdN3} [Left-Bottom] and \ref{fig:mfdN3} [Right-Bottom]
imply that
similar results hold also 
on the $(\gamma_0 ,\gamma_4 )$ and $(\gamma_2 ,\gamma_4 )$-planes,
respectively.
Fig.~\ref{fig:mfdN4} shows
the similar plots as fig.~\ref{fig:mfdN2} and \ref{fig:mfdN3}
for the $SU(4)$ case.
We easily see that
the result for $N=4$ is qualitatively the same as the $N=3$ case
though the deviations from the straight lines are slightly larger.
We have checked that
similar results hold for other values of $N$.

The above results for $N\geq 3$ are different from those in \cite{Beem:2013hha},
where the images were still the straight lines even for $N=3$ and $N=4$
while the common feature is that the images are very narrow.
As a conclusion,
all the interpolating functions constructed so far 
give very narrow lines
for the image of the conformal manifold and
we expect that this property is probably true also for the exact results.
However,
we do not have
a definite conclusion on 
whether the narrow lines are straight or curved
though it would be natural to be curved in the strongly coupled regime.

\subsection{Level crossing}
\label{sec:levelcross}
In this subsection
we compare
the dimension of the leading twist operator
with the one of the subleading leading operator
and study level crossing phenomenon
between the leading and subleading twist operators.
We use the word ``level crossing" in the following senses.
As we increase the coupling,
the dimensions of the leading and subleading operators approach each other
and the following two things may occur:
\begin{enumerate}
\item Operator mixing between them do not occur by some additional protected symmetries and their dimensions cross over.
\item Operator mixing occurs and the dimensions of new eigenstates repel.
\end{enumerate}
We refer to both of the above as level crossing.

In the $\mathcal{N}=4$ SYM,
the operator ${\rm Tr}\phi^I D^M \phi^I$ in \eqref{eq:twist2}
has the dimension $(2+M)$ at the classical level
while this becomes very large
for large 't Hooft coupling in the planar limit.
This is because this operator is dual to a massive string state
and therefore its dimension behaves 
as $\sim 1/\sqrt{\alpha'}\sim \lambda^{1/4}$ \cite{Gubser:1998bc}.
This implies that
the operator \eqref{eq:twist2} is no longer the leading twist operator
in the large-$\lambda$ regime of the planar limit.
Indeed we have a family of double trace operators with the same spins,
whose dimensions are protected in the planar limit.
The double trace operators 
consist of protected chiral primary operators
and their dimensions are independent of $\lambda$ in the planar limit
because of large-$N$ factorization.
Thus we expect the level crossing in the first sense
between the leading and subleading twist operators
in the planar limit.

To interpret this,
let us recall
the Wigner-von Neumann non-crossing rule known in quantum mechanics,
which states that 
levels of states with the same symmetry cannot cross each other.
Since the dilatation operator in the $\mathcal{N}=4$ SYM on $\mathbb{R}^4$
corresponds to the Hamiltonian on $\mathbb{R}\times S^3$,
we expect that
the dimensions obey 
the Wigner-von Neumann non-crossing rule.
Recently it was discussed \cite{Korchemsky:2015cyx} that
in the problem of the $\mathcal{N}=4$ SYM ,
$1/N$ plays roles as 
``interaction energy" of two level system 
in quantum mechanics.
The actual crossing between the Konishi and double trace operators 
in the planar limit
is consistent with the Wigner-von Neumann non-crossing rule
if we have additional symmetry in the planar limit.
Most promising candidate for such symmetry
is the one associated with the integrability, 
which is supposed to appear in the planar limit.
Since we do not expect such additional symmetry beyond the planar limit,
we expect that
the level crossing in the first sense does not occur for finite $N$
but the one in the second sense occurs for large but finite $N$.

Here we approach the level crossing problem by using our interpolating function
for the spin-0 case\footnote{
As far as we know,
there are no results on one-loop correction 
of anomalous dimensions of the twist-four operators with non-zero spin.
}.
The operator \eqref{eq:twist2} for $M=0$
is nothing but the Konishi operator ${\rm Tr}\phi^I \phi^I$,
whose dimension is two at the classical level and
$2(4\pi\lambda )^{1/4} + \, (\text{corrections})$  for large-$\lambda$ in the planar limit.
Next let us also consider the following operators 
\begin{equation}
{\rm Tr}{\phi^I\phi^I} {\rm Tr}{\phi^J\phi^J},\qquad 
{\rm Tr}{\phi^I\phi^J}{\rm Tr}{\phi^I\phi^J}\qquad 
{\rm Tr}{\phi^I\phi^I\phi^J\phi^J},\qquad 
{\rm Tr}{\phi^I\phi^J\phi^I\phi^J} ,
\label{dilaton}
\end{equation}
which have naively the same symmetry as the Konishi operator.
Particular linear combinations of these operators 
are eigenvectors of the dilatation operator,
which are dimension $4$ in the weak coupling limit. 
Therefore
the lowest dimension among those
is the dimension of the subleading twist operators 
in the weak coupling regime.
The dimensions of the operators \eqref{dilaton} at one loop is given by \cite{Arutyunov:2002rs,Beisert:2003tq,Alday:2013bha}
\begin{\eq}
\Delta_{\rm sub} (\tau )
= 4 
+\frac{Nw(N)}{2\pi} g +\mathcal{O}(g^2 ) ,
\label{sub_pert}
\end{\eq}
where $w(N)$ is roots of the equation
\begin{\eq}
w^4 -25w^3 +\left( 188-\frac{160}{N^2}\right) w^2 -\left( 384 -\frac{1760}{N^2}\right) w -\frac{7680}{N^2} =0 .
\label{rooteqn}
\end{\eq}
The dimension of the subleading twist operator
is described by the smallest root $w_- (N)$ of this equation,
which is always negative\footnote{
For example,
$w(2)$ $=$ $(-3,16)$, $w(3)$ $=$ $(-1.60752$ , $5.33333$, $6.94715$, $14.327)$,
$w(4)$ $=$ $( -1.00282$, $4.36878$, $8.08825$, $13.5458)$,
$w(5)$ $=$ $( -0.68319$, $4$,  $8.57743$, $13.1058)$ and so on.
}.

Now we construct
interpolating functions
for the subleading twist operator
by imposing a match with the weak coupling expansion \eqref{sub_pert}
and the $SL(2,\mathbb{Z})$ duality.
Since we know the perturbative expansion up to only one-loop,
we have only one coefficient to be tuned 
in the $S$-duality invariant interpolating functions.
In this situation,
we cannot construct 
the FPR-like interpolating function \eqref{eq:FPR_wo_gravity}
since 
we needs $m\geq 2$.
Hence we use the Alday-Bissi type interpolating function \eqref{eq:AB}
for the dimension of the subleading twist operator.
For general $s$,
we can easily construct the Alday-Bissi type interpolating function as
\begin{\eq}
\Delta_{\rm sub}^{(s)}(\tau )
= 4 +\frac{Nw_-}{2\pi} \left(\frac{\zeta (2s)}{E_s (\tau )}\right)^{1/s} .
\end{\eq}

\begin{figure}[t]
\begin{center}
\includegraphics[width=7.4cm]{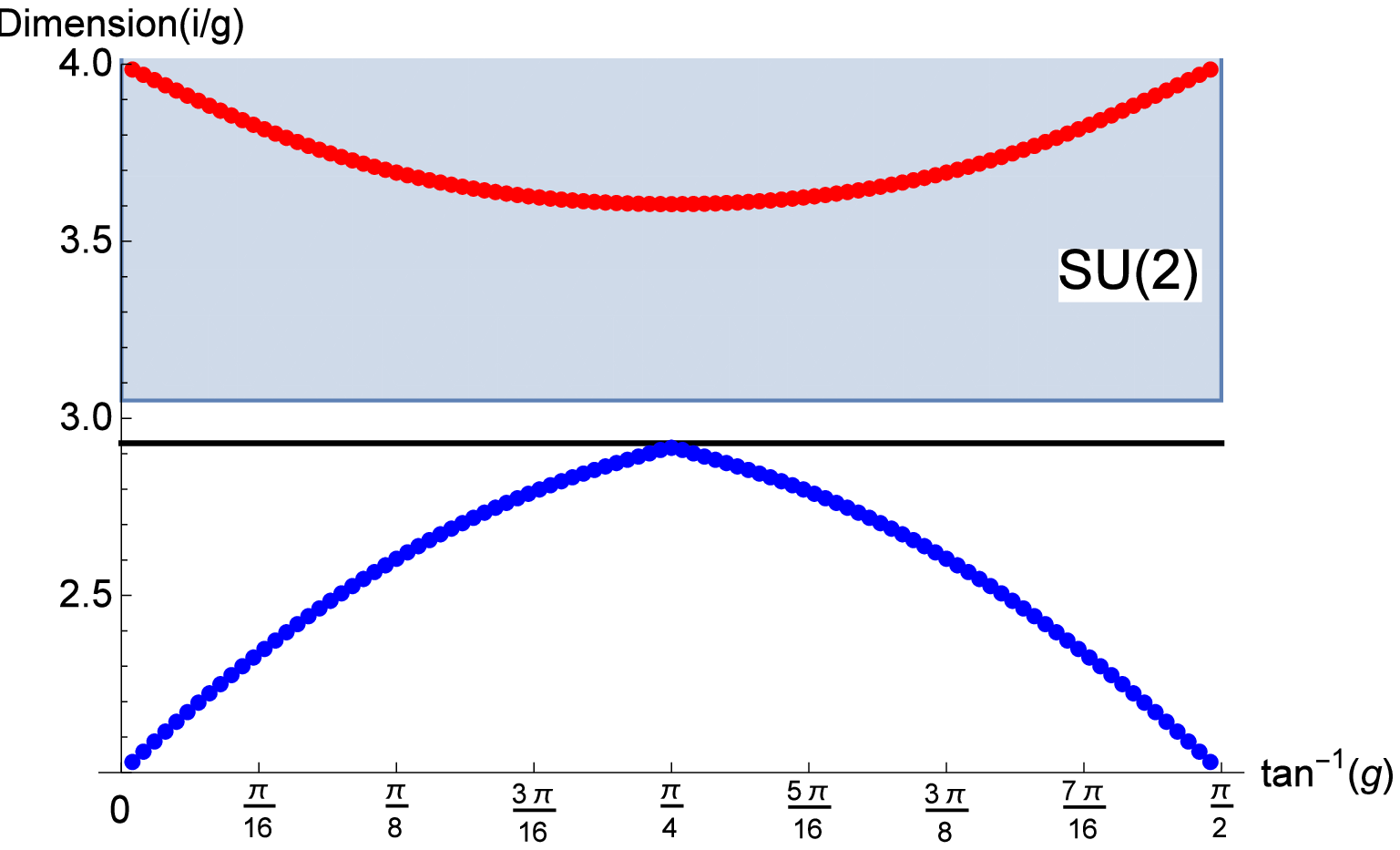}
\includegraphics[width=7.4cm]{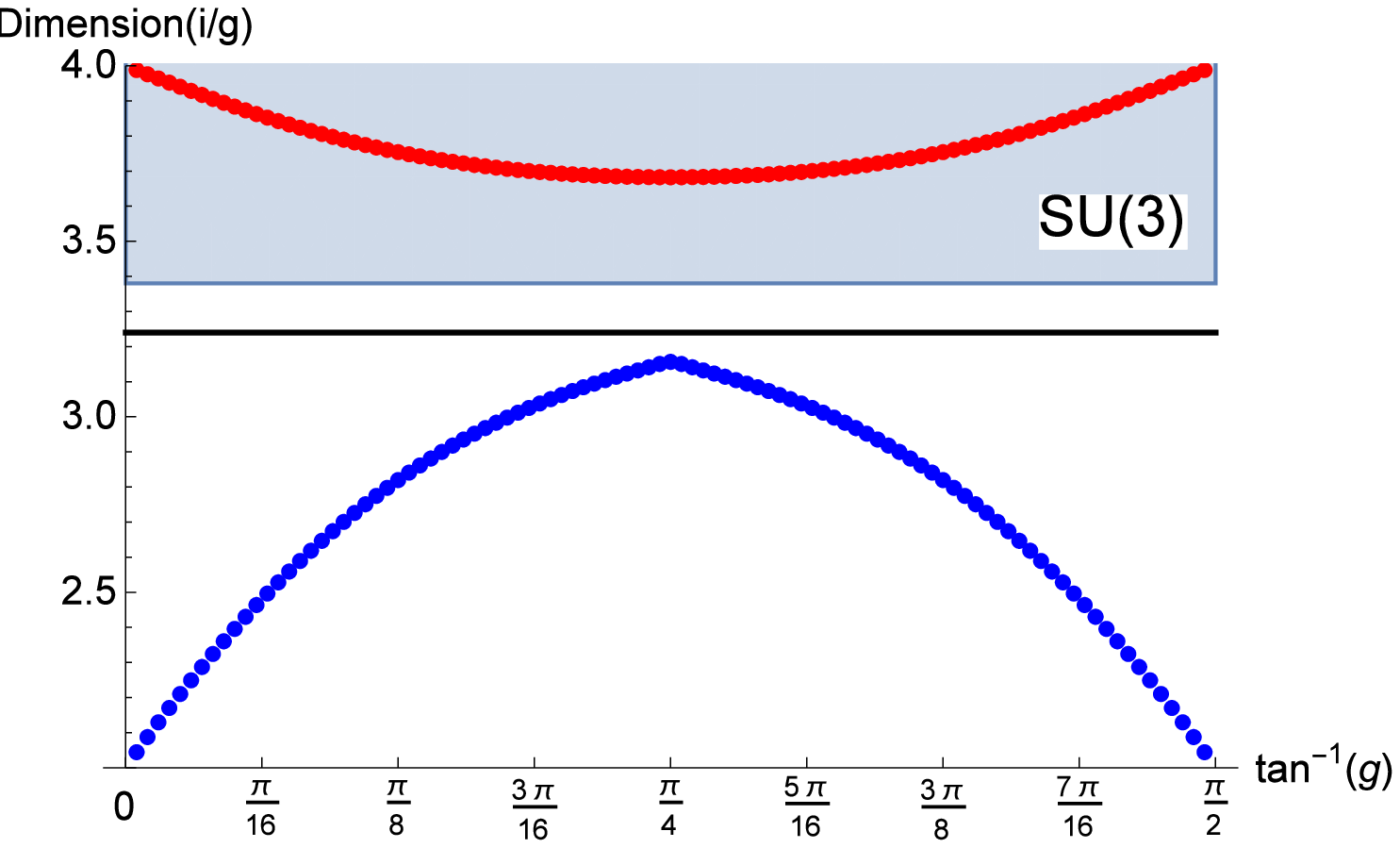}\\
\includegraphics[width=7.4cm]{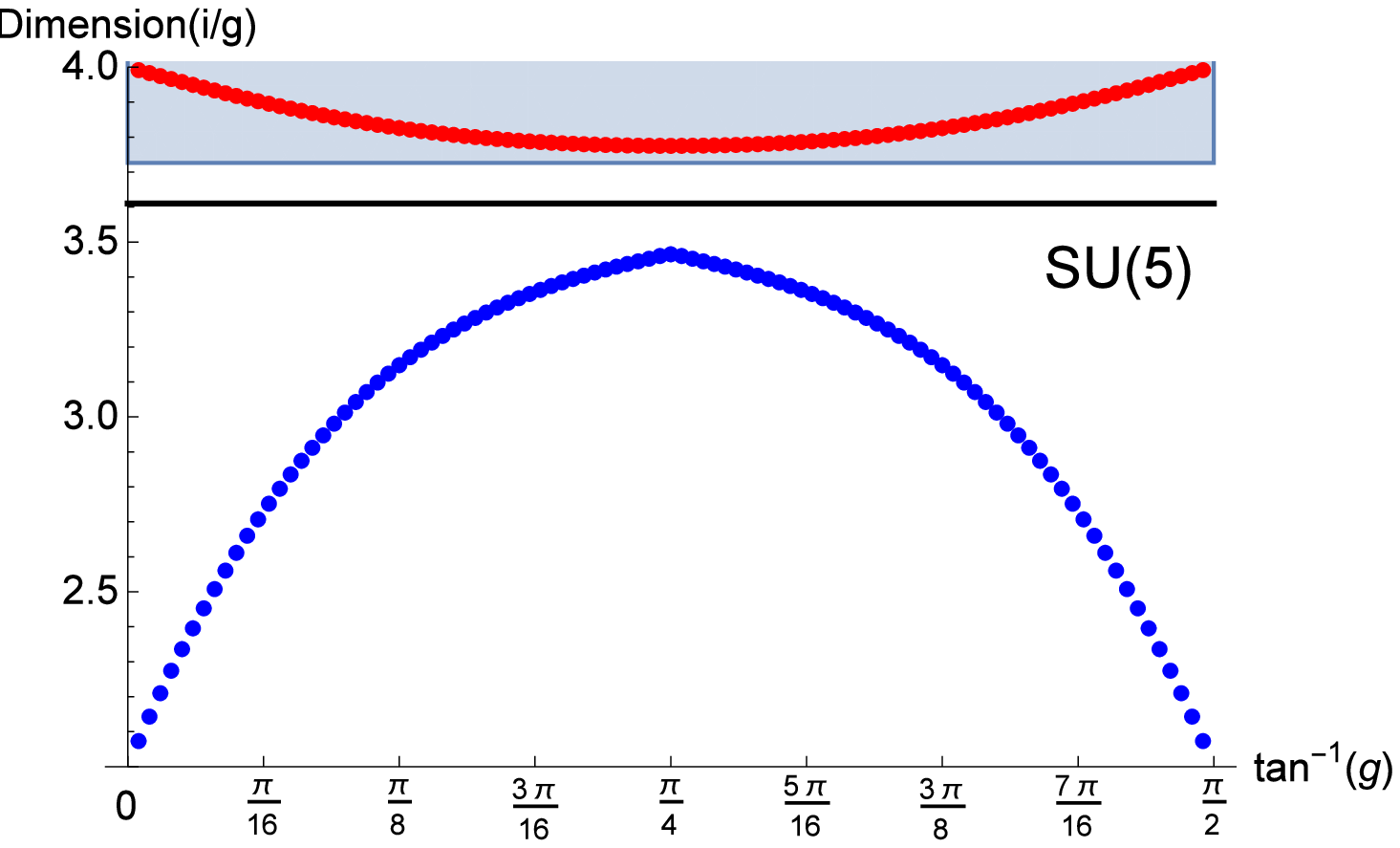}
\includegraphics[width=7.4cm]{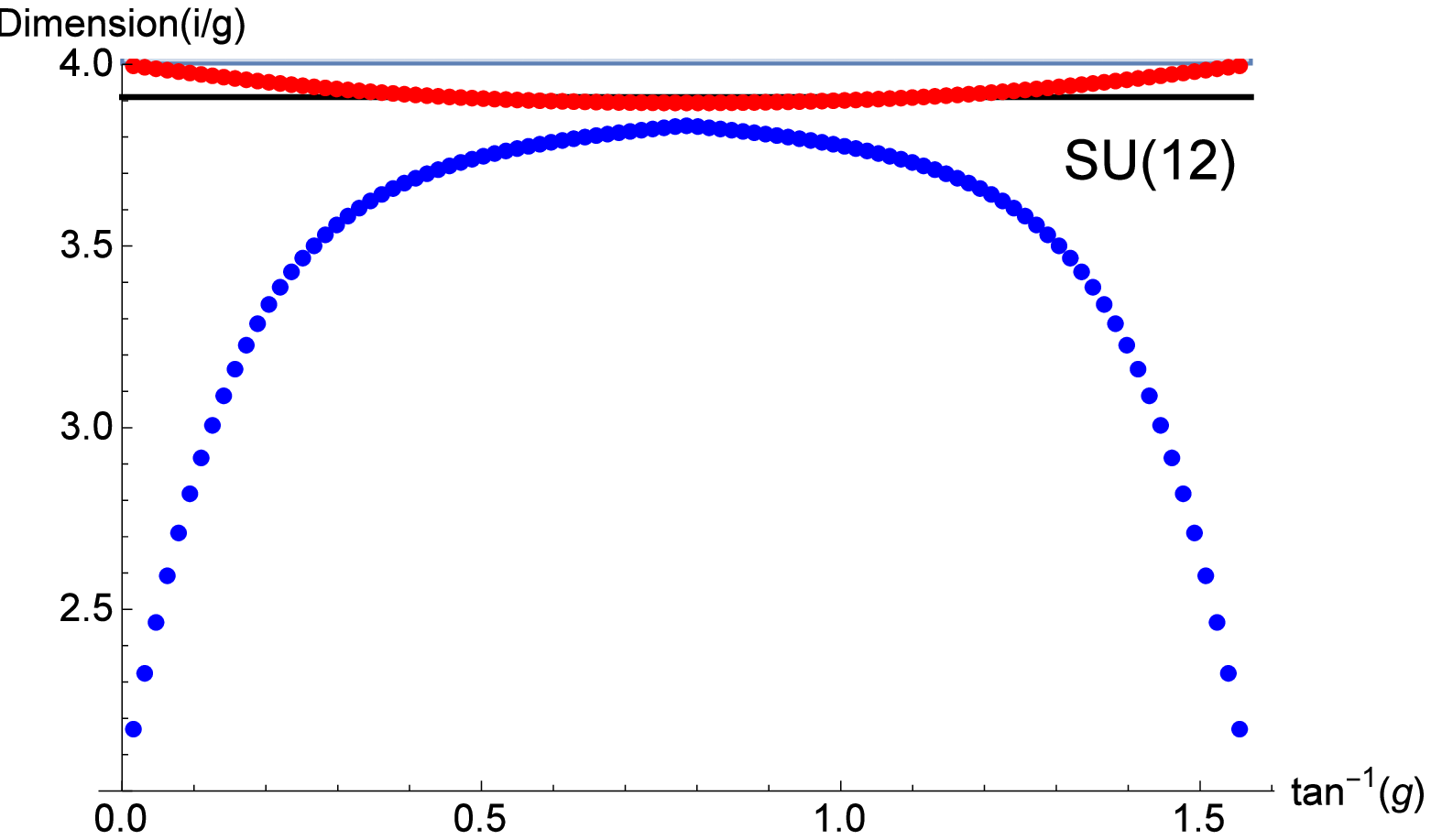}
\end{center}
\caption{
Comparison of the dimensions of the leading and subleading twist operators 
with spin-0 as functions of $g$ for $\theta =0$ and various $N$. The red line indicates the subleading interpolating function and the blue line represents the leading interpolating function.  
The shaded region and horizontal black solid line are 
the upper bounds and corner values for the leading twist operator
obtained by the $\mathcal{N}=4$ superconformal bootstrap, respectively.
}
\label{fig:levelcrossing}
\end{figure}

We do have 
infinitely many $s$-dependent $S$-duality invariant interpolating functions. Which value of $s$ is most appropriate?
For this purpose, let us consider
the small-$g$ expansion of the interpolating function:
\begin{\eq}
\Delta_{\rm sub}^{(s)}(\tau )
\simeq 4 +\frac{Nw_-}{2\pi}g
\left( 1 
+\frac{\sqrt{\pi}\Gamma (s-1/2)\zeta (2s-1) }{\Gamma (s) \zeta (2s)} g^{2s-1} \right)^{-1/s} ,
\end{\eq}
which is true up to non-perturbative corrections.
From this expression
we can easily see that after the one-loop correction, the next term is a $\mathcal{O}(g^{2s})$ correction and this fact is useful in constraining $s$.
First of all, we need $2s\in\mathbb{Z}$
to get integer powers of $g$ in the weak coupling expansion.
Next when $s$ is too large,
we have large jumps of powers in the weak coupling expansion
and therefore want a small value of $s$ as possible.
Since the interpolating function is well-defined for $s>1$,
we conclude that the most appropriate value of $s$ is $s=3/2$.

\begin{figure}[t]
\begin{center}
\includegraphics[width=7.4cm]{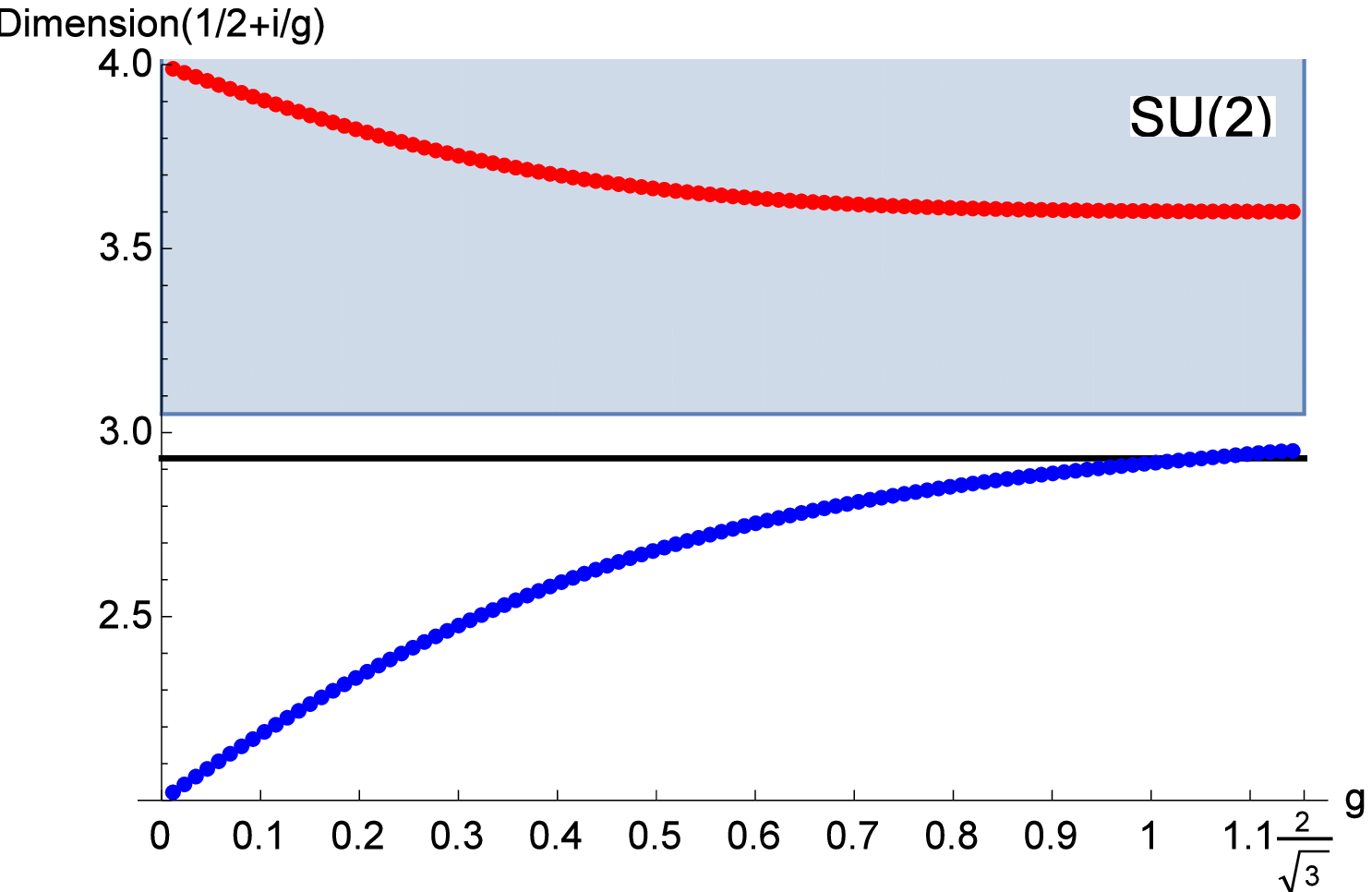}
\includegraphics[width=7.4cm]{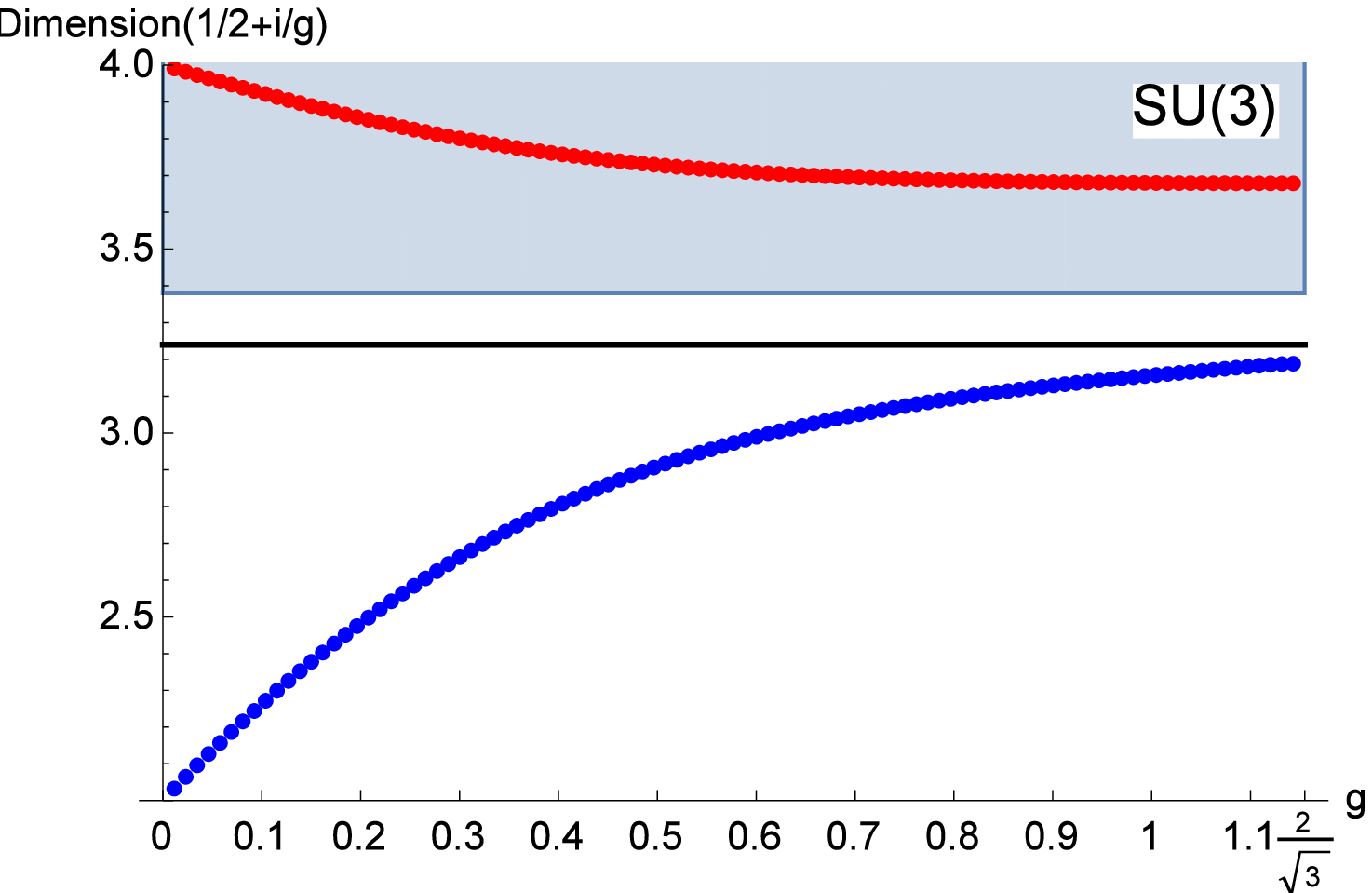}\\
\includegraphics[width=7.4cm]{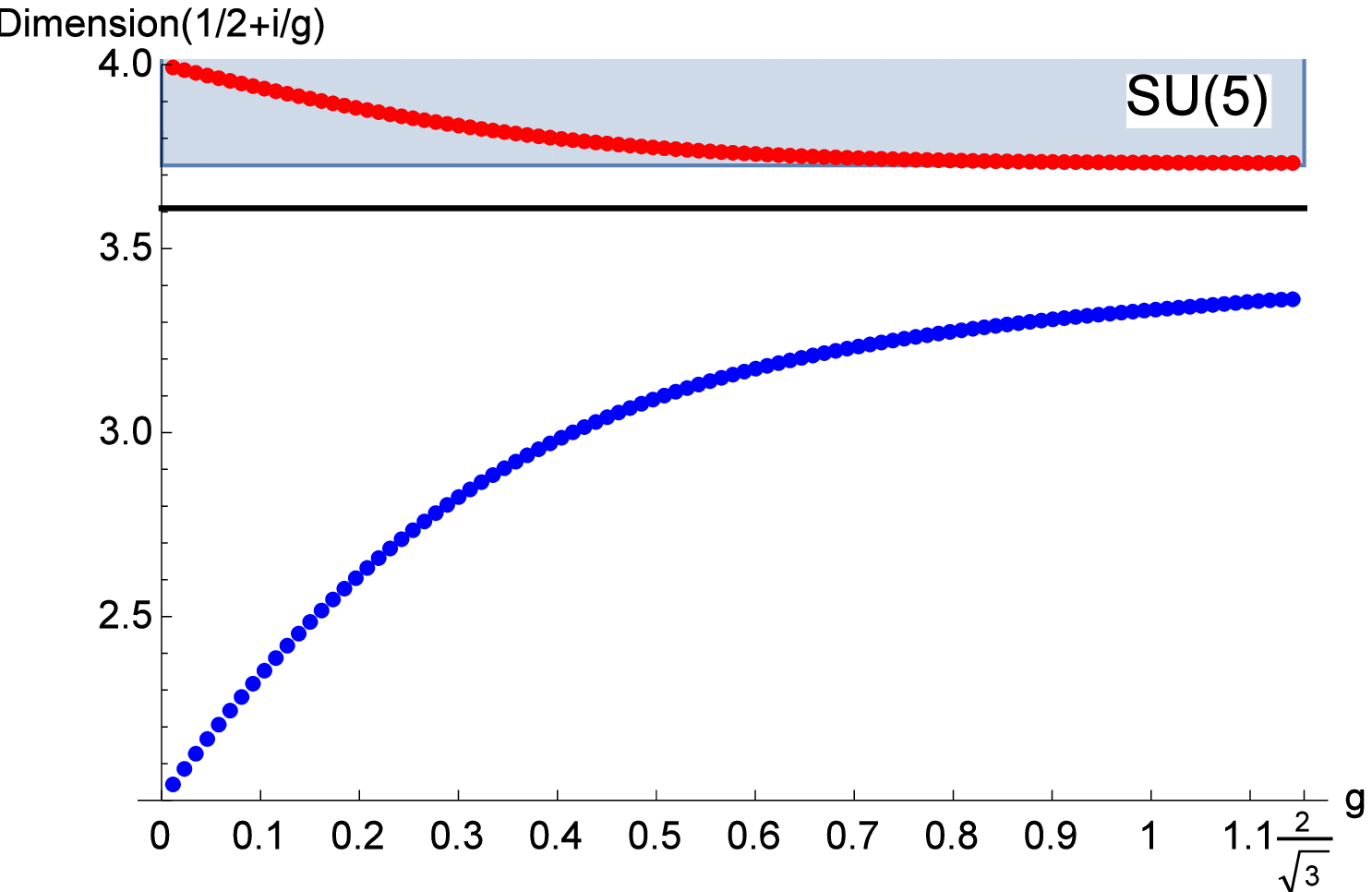}
\includegraphics[width=7.4cm]{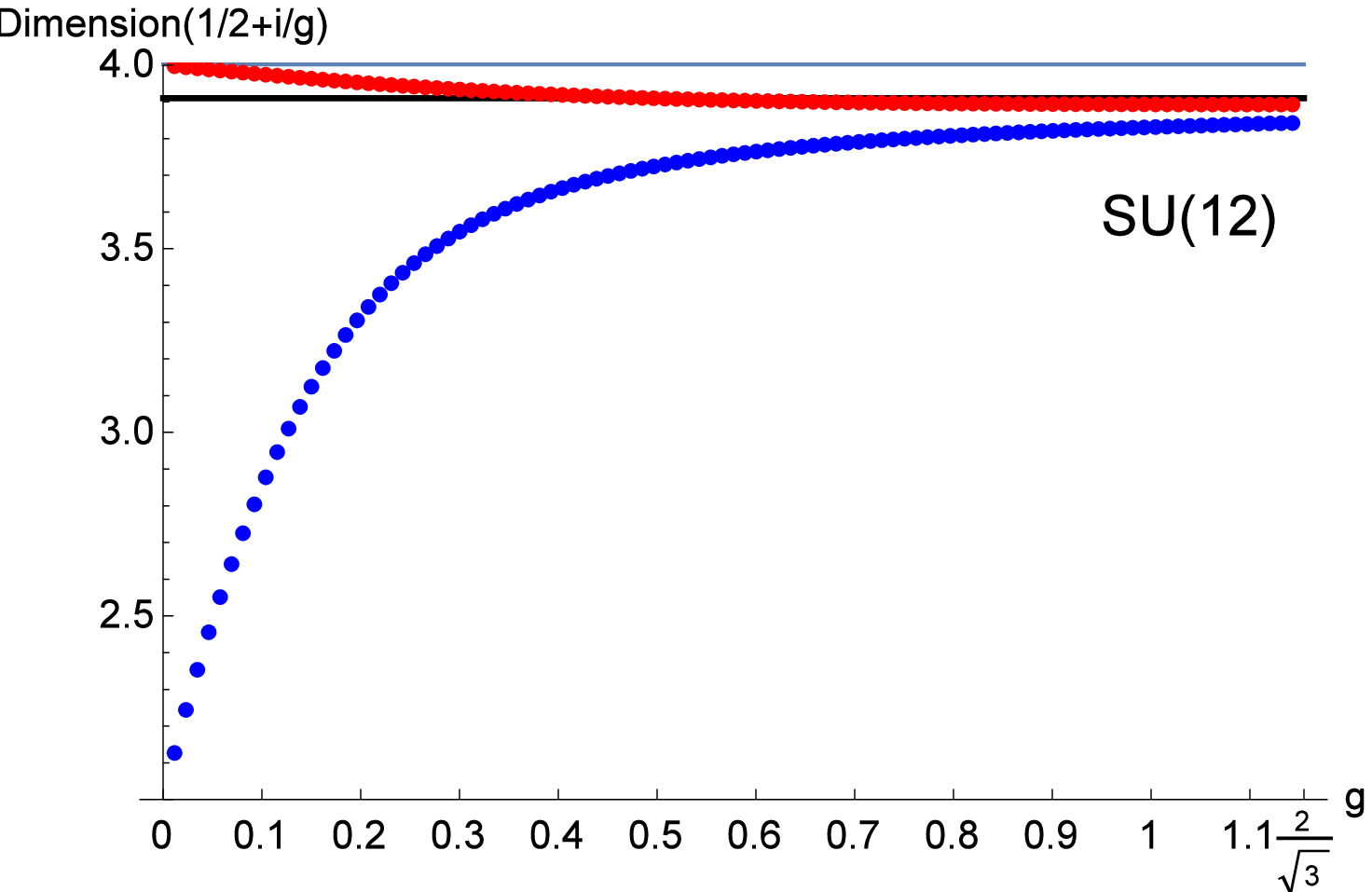}
\end{center}
\caption{
Similar plots for $\theta =\pi$ as fig.~\ref{fig:levelcrossing}.
}
\label{fig:levelcrossing_pi}
\end{figure}

In fig.~\ref{fig:levelcrossing}
we compare\footnote{
Note that we are not plotting the anomalous dimensions
but the dimensions themselves.
} 
the interpolating function $\Delta_{\rm sub}^{(3/2)}(\tau )$
for the subleading twist operator
with $F_4^{(30,1/4)}(\tau )+2$
for the leading twist operator constructed in the last section 
for $\theta =0$.
We easily see from the figure that
the interpolating functions do not cross each other
for all the values of $N$.
We have checked that this is true for other values of $N$,
which are not present here.
In fig.~\ref{fig:levelcrossing_pi}
we give similar plots for $\theta =\pi$ as in fig. \ref{fig:levelcrossing}
in order to test that
the above is true also for different values of $\theta$.
Again the interpolating functions do not cross each other
for all $N$.
For $SU(12)$,
we observe that
the dimension of the leading twist operator becomes very close to the subleading one around $\tau =\tau_S$ and $\tau =\tau_{TS}$.
Indeed we have found similar results for larger $N$.
Thus we conclude that
the level crossing in the first sense (actual crossing)
does not occur for general finite $N$
but 
the one in the second sense (small operator mixing) occurs 
for large but finite $N$.

Interestingly
the interpolating function for the subleading twist operator
has minimum at the duality invariant points.
It would be interesting to find any physical interpretations for that.
Finally we have not studied the problem in this subsection
for non-zero spin cases.
This is because there are no available results 
for the weak coupling expansions of the subleading twist operators
but the one-loop computations for non-zero spins
should  not be hard.
It is nice if one can perform the one-loop computations
and the same analysis as in this subsection.

\section{Results on Konishi operator in the planar limit}
\label{sec:results_planar}
In this section
we analyze the dimension of the Konishi operator in the planar limit.
While the Konishi operator is 
the leading twist operator in the weak coupling regime,
this has very large dimension in the classical string regime.
We approximate the dimension of the Konishi operator in the planar limit
in terms of the standard FPR described in sec.~\ref{sec:standardFPR}.

\subsection{Previous results}
\label{sec:previous_planar}
First we briefly review previous results
on the dimension of the Konishi operator in the planar limit:
\begin{\eq}
\lambda =gN=\frac{g_{\rm YM}^2 N}{4\pi} ={\rm fixed},\quad N\rightarrow\infty .
\end{\eq}
In this limit
there is a 7-loop computation in 
the weak coupling expansion \cite{Bajnok:2012bz}:
\begin{\eqa}
&&\Delta_{\rm Konishi} (\lambda ) \NN\\
&&= 2+\frac{3\lambda}{\pi}  -\frac{3\lambda^2}{\pi^2}  +\frac{21\lambda^3}{4\pi^3} 
 +\Biggl[ -39 +9\zeta (3) -\frac{45\zeta (5)}{2}  \Biggr] \frac{\lambda^4}{4\pi^4} \NN\\
&& +\Biggl[  \frac{945 \zeta (7)}{32}-\frac{135 \zeta (5)}{16}-\frac{81 \zeta (3)^2}{16}+\frac{27 \zeta
   (3)}{4}+\frac{237}{16} \Biggr] \frac{\lambda^5}{\pi ^5} \NN\\
&&+\Biggl[   -262656 \zeta (3)-20736 \zeta (3)^2+112320 \zeta (5)+155520 \zeta (3) \zeta (5)  \NN\\
&&+75600 \zeta (7)-489888 \zeta (9)-7680 
\Biggr] \frac{\lambda^6}{4096 \pi ^6} \NN\\
&& +48\Biggl[-8784 \zeta (3)^2+2592 \zeta (3)^3-4776 \zeta (5)-20700 \zeta (5)^2+24 \zeta (3) (357 \zeta (5)-1680 \zeta (7)+4540) \NN\\
&&-26145 \zeta (7)-17406 \zeta (9)+152460 \zeta (11)-44480 \Biggr] \left( \frac{\lambda}{4\pi} \right)^7 
+\mathcal{O}(\lambda^8 ) . 
\label{eq_Konishiw}
\end{\eqa}
There are also some holographic computation of the Konishi operator\footnote{
\cite{Gromov:2014bva} conjectured next order from a numerical computation as:
$\Delta_{\rm Konishi} (\lambda )
= 2 (4\pi\lambda )^{1/4} -2+\frac{2}{(4\pi\lambda )^{1/4}}
+\frac{-3\zeta (3) +\frac{1}{2}}{(4\pi\lambda )^{3/4}}
+\frac{\frac{15\zeta (5)}{2} +6\zeta (3) -\frac{1}{2}}{(4\pi\lambda )^{5/4}}
$.
}:
\begin{\eq}
\Delta_{\rm Konishi} (\lambda )
= 2 (4\pi\lambda )^{1/4} -2+\frac{2}{(4\pi\lambda )^{1/4}}
+\frac{-3\zeta (3) +\frac{1}{2}}{(4\pi\lambda )^{3/4}} .
\label{eq:planarS}
\end{\eq}
There are some numerical computations 
using the Thermodynamic Bethe Ansatz (TBA) \cite{Gromov:2009zb,Frolov:2010wt}
and we will compare this with our results.

\subsection{Comparison with Thermodynamic Bethe Ansatz}
\begin{figure}[t]
\begin{center}
\includegraphics[width=7.4cm]{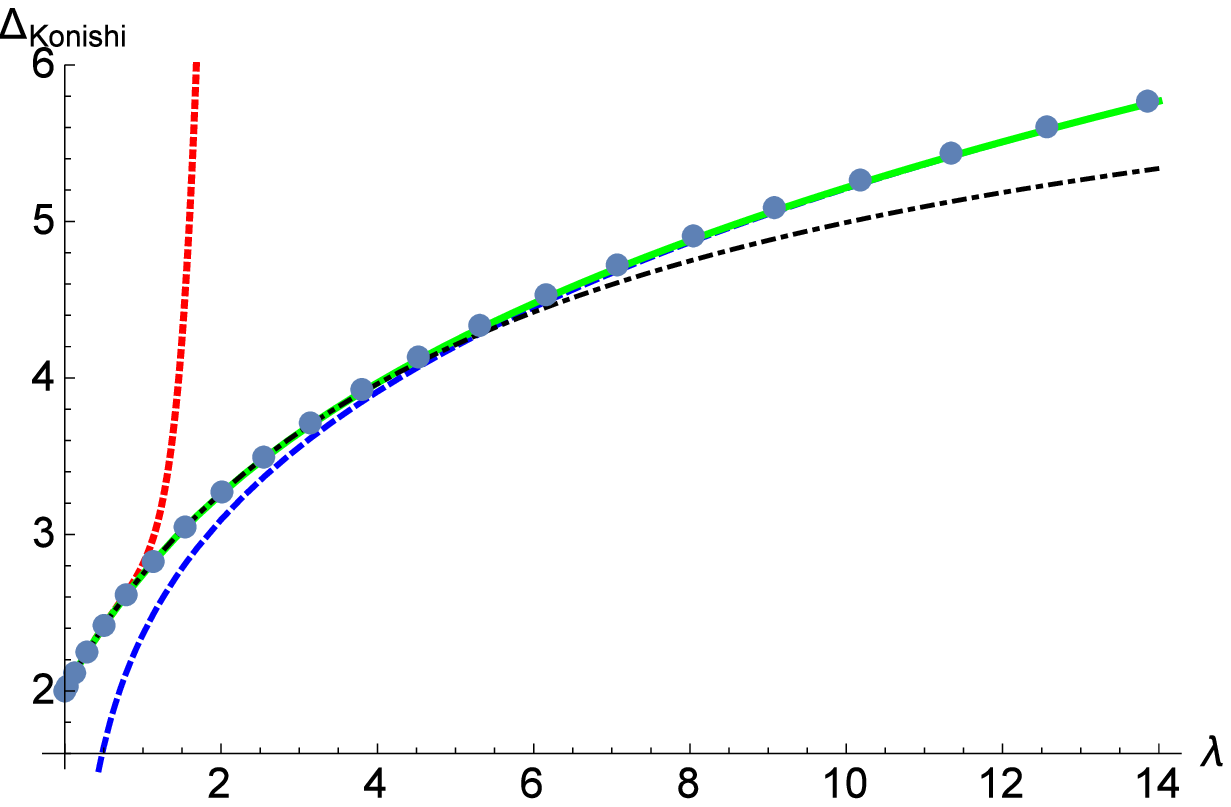}
\includegraphics[width=7.4cm]{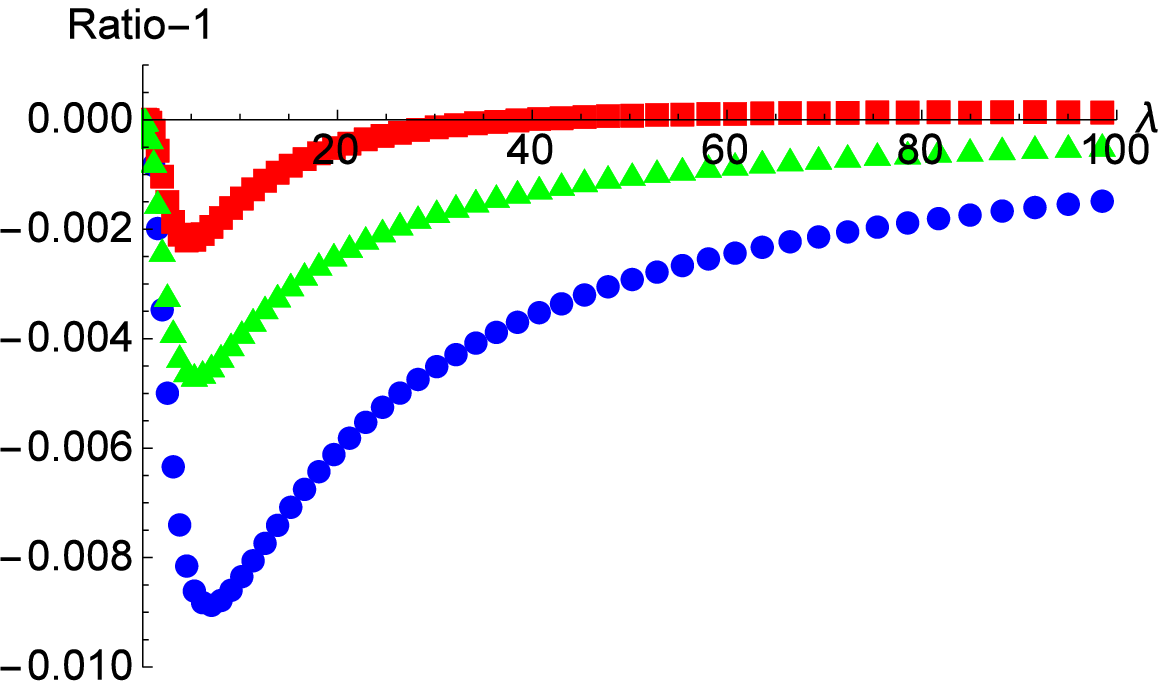}
\end{center}
\caption{
[Left]
Results on the dimension of Konishi operator in the planar limit.
The symbol denotes the numerical result obtained by TBA
the lines denote
the weak coupling expansion (red dotted), strong coupling expansion (blue dashed),
one point Pad\'e approximation $P_{(3|3)}(\lambda )$ (black dot-dashed) and 
FPR with $(m,n,\alpha )=(15,2,1/28)$ (green solid).\newline
[Right] 
The relative errors of the interpolating functions from the TBA result.
Blue circle, red square and green triangle denote
$(m,n,\alpha )=(15,2,1/36)$, $(m,n,\alpha )=(15,2,1/32)$ and
$(m,n,\alpha )=(15,2,1/28)$, respectively.
}
\label{fig:Konishi_planar}
\end{figure}
By using the weak and strong coupling expansions,
we can construct the interpolating functions.
Here we do not use the modular invariant interpolating function (3.11)
but use the standard FPR \eqref{eq:FPR},
which is consistent with the weak coupling expansion \eqref{eq_Konishiw}
and the holographic result \eqref{eq:planarS}
since the dimension of the Konishi operator is not expected
to be modular invariant.
For this purpose,
note that the weak coupling expansion of $\Delta_{\rm Konishi}(\lambda )$ 
is the power series expansion of $\lambda$
while the holographic result is the one of $\lambda^{-1/4}$.
However, if we work with $\Delta_{\rm Konishi}(\lambda ) +2$,
then the holographic result becomes
the power series expansion\footnote{
We expect this property also for higher orders
since $\alpha' \sim \lambda^{-1/2}$.
} of $\lambda^{-1/2}$.
Thus we consider the FPR-type interpolating functions
for $\Delta_{\rm Konishi}(\lambda ) +2$ 
rather than $\Delta_{\rm Konishi}(\lambda )$.
More precisely
we rewrite the power series expansions of $\Delta_{\rm Konishi}(\lambda ) +2$ 
in terms of $x=\sqrt{\lambda}$ instead of $\lambda$ and
consider the interpolating functions $F_{m,n}^{(\alpha )}(x)$
for the expansions\footnote{
	Note that
	because of this parametrization,
	the interpolating functions may have half-odd power of $\lambda$
	in their small-$\lambda$ expansions.
}.
Then the interpolating function approximates 
the dimension $\Delta_{\rm Konishi}(\lambda )$ by
\begin{\eq}
\Delta_{\rm Konishi}(\lambda )
\simeq F_{m,n}^{(\alpha )}(x=\sqrt{\lambda} ) -2 .
\end{\eq}

One of subtleties here is that
we can construct enormous number of interpolating functions
as in the leading twist operators and
appropriate choice of $(m,n,\alpha )$ is a priori unclear.
However, 
we have expectations on the appropriate choice at least for $(m,n)$.
First for $m$,
since the weak coupling expansion in the planar limit 
is expected to be convergent \cite{'tHooft:1982tz},
we expect that
interpolating functions with larger $m$ give better approximations.
For $n$, we have very few choices, namely $n=0,1,2$.
Although the large-$\lambda$ expansion would be asymptotic
and its optimized order is unclear,
it is natural to expect that the optimized order is larger than $n=2$.
Thus we should take $(m,n)$ to be large as possible.
It is still unclear what is an appropriate value of $\alpha$.
However, this is not so problematic
because the interpolating functions with $(m,n)=(15,2)$ 
are weakly dependent on $\alpha$ as we will see shortly.

In fig.~\ref{fig:Konishi_planar} [Left]
we compare our result with numerical result obtained 
by Thermodynamic Bethe Ansatz (TBA) \cite{Gromov:2009zb}.
We also draw 
the one-point Pad\'e approximant of the weak coupling expansion
defined as
\begin{\eq}
P_{(m|n)}(\lambda )
= \frac{\sum_{k=0}^m c_k \lambda^k}{1+\sum_{k=1}^n d_k \lambda^k} ,
\end{\eq}
where the coefficients are determined by the correct reproduction of the small-$g$ expansion up to $\mathcal{O}(g^{m+n})$.
The one-point Pad\'e approximant in fig.~\ref{fig:Konishi_planar} [Left]
is the so-called diagonal Pad\'e with $m=n=3$.
We easily see that
our interpolating function agrees with the TBA result 
in whole region of $\lambda$.
The one-point Pad\'e approximation is good up to around $\lambda =5$
but deviates from the TBA result in stronger coupling region.
In fig.~\ref{fig:Konishi_planar} [Right]
we plot the relative errors of the approximations by the interpolating functions 
from the TBA result
to study precision of their approximations.
We find that
all the interpolating functions
have errors less than $1\%$ in the whole region.

\subsection{Analytic property}
\begin{figure}[t]
\begin{center}
\includegraphics[width=8cm]{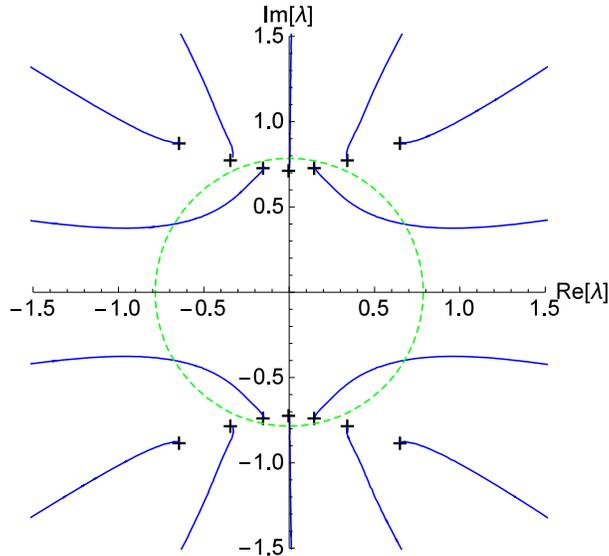}
\end{center}
\caption{Analytic property of the FPR with $(m,n,\alpha )=(15,2,1/32)$ as a complex function of $\lambda$.
The symbol ``$+$" denotes zeros of the rational function $(F_{m,n}^{(\alpha )})^{1/\alpha}$,
which give ends of branch cuts of $F_{m,n}^{(\alpha )}$
The green dashed line denotes $|\lambda |=\pi /4$,
which is expected radius of convergence in the weak coupling expansion.
}
\label{fig:Konishi_analytic}
\end{figure}

Here we study analytic properties of the interpolating functions
following the spirit of \cite{Honda:2015ewa}.
It has been proposed in \cite{Honda:2015ewa} that
when interpolating functions are to
give good approximations along an axis in parameter space, 
then their branch cuts have the following interpretations:
\begin{enumerate}
\item The branch cuts are particular to the FPR and the artifact of
  the approximation by the interpolating function.  
Namely, this type of branch cut is not helpful in 
  extracting any physical information.

\item The physical quantity, which we try to approximate by the FPR, 
  has an actual branch cut near the branch cut of the FPR.  
  Namely, the branch
  cut of the FPR approximates the ``correct" branch cut of the
  physical quantity. 

\item There is an anti-Stokes line near the branch cut
across which perturbation series of the physical
  quantity changes its dominant part. 
  \end{enumerate}
Here we expect that
some zeros and poles of the interpolating functions
approximate analytic properties of the dimension of the Konishi operator.

In fig.~\ref{fig:Konishi_analytic}
we summarize the  
analytic property\footnote{
We define locations of branch cuts as $(F_{m,n}^{(\alpha )})^{1/\alpha}\in (-\infty ,0)$.
The ends of branch cuts are given by infinity, zeros or poles of $(F_{m,n}^{(\alpha )})^{1/\alpha}$.
}
of the FPR with $F_{15,2}^{1/32}(x )$ 
as a complex function of $\lambda$.
The symbols denote
ends of branch cuts of the interpolating function,
which are characterized by zeros and poles\footnote{
The poles of $(F_{15,2}^{(1/36 )})^{36}$ are not present 
in fig.~\ref{fig:Konishi_analytic}
because they are located out of the scale.
} of 
the associated rational function $(F_{m,n}^{(\alpha )})^{1/\alpha}$.
The green dashed line denotes $|\lambda |=\pi /4$,
which is the expected radius of convergence in the weak coupling expansion
from previous works \cite{Beisert:2006ez,Volin:2008kd,Marboe:2014gma}.
From this figure,
we observe that
many ends of branch cuts are located 
around the circle $|\lambda |=\pi /4$.
This is indeed reasonable
because one expects that
the dimension of Konishi operator
has some singularities 
around radius of convergence of the weak coupling perturbative expansion
and
so does the interpolating function 
if the interpolating function approximates the proper analytic properties.

\section{Conclusion and discussions}
\label{sec:conclusion}
In this paper
we have mainly studied the dimensions of the unprotected leading twist operators
in the 4d $SU(N)$ $\mathcal{N}=4$ SYM.
We have constructed the class of interpolating functions \eqref{eq:main_interpolation}
to approximate the dimensions.
The interpolating functions
are consistent with the previous results 
on the perturbation theory \eqref{eq:weak}, holographic computation \eqref{eq:result_gravity} and full S-duality.
and
give the predictions 
for arbitrary value of $N$ and the complex gauge coupling $\tau$ in the fundamental region,
which particularly includes the duality-invariant points $\tau =i$ and $\tau =e^{i\pi /3}$.
We have used our interpolating functions 
to test the recent conjecture by the $\mathcal{N}=4$ superconformal bootstrap \cite{Beem:2013qxa}, 
which states that 
the upper bounds on the dimensions are saturated 
at either one of the duality-invariant points $\tau =i$ or $\tau =e^{i\pi /3}$. 
It has turned out that
our interpolating functions have the maximum at $\tau =e^{i\pi /3}$. 
In the regime where we expect reasonable approximations by the interpolating functions, the maximal values
are close to the conjectural values in \cite{Beem:2013qxa}.
Thus we conclude that
our interpolating function approach strongly supports 
the conjecture of \cite{Beem:2013qxa}
as well as \cite{Beem:2013hha,Alday:2013bha}.
In order to construct the interpolating functions
we have used the available four-loop or three-loop results of the weak coupling expansions.
Obviously,
if higher orders of the weak coupling perturbative series
become available,
then we can obtain more precise interpolating functions which are expected to better approach the corner value at $\tau =e^{i\pi /3}$.
It would be nice 
if one can obtain the higher order results and
repeat our analysis in this paper to construct better interpolating functions.

In terms of the interpolating functions,
we have drawn 
the image of the conformal manifold in the space of the dimensions $(\Delta_0 ,\Delta_2 ,\Delta_4 )$.
We have found that
the image is almost a line as in \cite{Beem:2013hha}
despite the conformal manifold being two-dimensional.
For the $SU(2)$ case,
the line is almost straight, 
whose slope is the ratio of the one-loop anomalous dimensions as in \cite{Beem:2013hha}.
For higher $N$,
we have seen that the line is curved contrast to \cite{Beem:2013hha}.

We have also constructed interpolating functions for the subleading twist operator
and studied the level crossing phenomenon
between the leading and subleading twist operators.
We have checked that
the interpolating functions for the both operators
do not have actual crossing with each other for finite $N$.
For large but finite $N$,
we have found that
the dimension of the leading twist operator becomes 
very close to the subleading one around $\tau =i$ and $\tau =e^{\pi i /3}$.
This implies the small mixing between the two operators. 
To construct the interpolating function for the sub-leading twist operator,
we have used only the one-loop result.
Computing two-loop order would give 
more insights to the level crossing problem.
In this paper 
we have considered only the spin-0 case as there are no one-loop computations for the subleading twist operators with non-zero spins.
It would be nice if one can perform the one-loop computations
and repeat the same analysis for non-zero spin cases.

We have also studied the dimension of Konishi operator in the planar limit. 
We have found that our interpolating functions match 
with the numerical result obtained by Thermodynamic Bethe Ansatz very well.
Furthermore
we have discussed the analytic property of the relatively best interpolating function in the spirit of \cite{Honda:2015ewa}. 
It has turned out that
analytic property of the interpolating function
reflects the expectations on radius of convergence
from the weak coupling perturbation theory.

The key to our interpolating functions is their modular invariance.
It would be illuminating
if we study other modular invariant observables by our interpolating functions.
More challenging direction is to construct 
interpolating functions for modular forms,
which is not modular invariant but have particular transformation properties 
under $SL(2,\mathbb{Z})$ transformations.
It would be also interesting to consider other theories,
which enjoy $SL(2,\mathbb{Z})$ duality.
Indeed 
many theories with the  $SL(2,\mathbb{Z})$ duality
were recently found 
by torus compactifications of 6d $(1,0)$ theories have the S-duality \cite{DelZotto:2015rca}.

\subsection*{Acknowledgements}
M.~H. thanks Christopher Beem, Leonardo Rastelli and Balt C.~van Rees
for kindly sending him numerical data in their previous work \cite{Beem:2013qxa}.
We are grateful to 
Dileep P.~Jatkar for helpful comments on the draft.
We would like to thank
Ofer Aharony, Lorenzo Di Pietro, Mikhail Isachenkov, Dileep P.~Jatkar,
Zohar Komargodski, Shota Komatsu, Shiraz Minwalla, 
Ashoke Sen and Tarun Sharma
for useful discussions.
The work of S.~T. was supported by a separate India Israel (ISF/UGC) grant, as well as the Infosys Endowment for the study of the Quantum Structure of Space Time.

\appendix
\section{On numerical computation of non-holomorphic Eisenstein series}
\label{app:eisen}
In this Appendix
we briefly explain
how to numerically compute the non-holomorphic Eisenstein series.
The non-holomorphic Eisenstein series $E_s (\tau )$
has the following expansion (see e.g. sec.5.3 of \cite{Klevang})
\begin{\eqa}
E_s (\tau )
&=& \zeta (2s) ({\rm Im}\tau )^s
 +\frac{\sqrt{\pi}\Gamma (s-1/2)}{\Gamma (s)} \zeta (2s-1) ({\rm Im}(\tau ))^{1-s} \NN\\
&&+\frac{4\pi^s}{\Gamma (s)} \sqrt{{\rm Im}(\tau )}
\sum_{k=1}^\infty \sigma_{1-2s}(k) k^{s-\frac{1}{2}} 
K_{s-\frac{1}{2}} \left( 2\pi k {\rm Im}(\tau ) \right)  \cos{\left( 2\pi k {\rm Re}(\tau )\right)} ,
\label{eq:eisenN}
\end{\eqa}
where $\sigma_s (k)$ is the divisor function
\begin{\eq}
\sigma_s (k) =\sum_{d|k} d^s .
\end{\eq}
In terms of $(g,\theta )$,
this is written as
\begin{\eqa}
E_s (\tau )
&=& \zeta (2s) g^{-s}
 +\frac{\sqrt{\pi}\Gamma (s-1/2)}{\Gamma (s)} \zeta (2s-1) g^{s-1} \NN\\
&&+\frac{4\pi^s}{\Gamma (s)} g^{-\frac{1}{2}}
\sum_{k=1}^\infty \sigma_{1-2s}(k) k^{s-\frac{1}{2}} 
K_{s-\frac{1}{2}} \left( \frac{2\pi k}{g} \right)  
\cos{\left( k\theta\right)} .
\end{\eqa}
This representation is suitable for numerical computation.
When the summation does not converge well,
we practically compute
the summation at another point connected 
by the $SL(2,\mathbb{Z})$ symmetry.

\section{Another FPR-like modular invariant interpolating function including Alday-Bissi's one}
\label{app:anotherFPR}
We can also construct
the following interpolating function,
which is inspired by the FPR
and a generalization of the Alday-Bissi's interpolating function
but a different form:
\begin{\eq}
\tilde{F}_m^{(s,\alpha )} (\tau ) 
= \Biggl[ \frac{\sum_{k=1}^p c_k E_{s+k} (\tau )}{1+\sum_{k=1}^q d_k E_{s+k} (\tau )} \Biggr]^\alpha .
\end{\eq}
We determine the coefficients $c_k$ and $d_k$ such that
expansion of $\tilde{F}_m^{(s,\alpha )}$ around $g=0$ agrees with
the one of $\gamma_M (g)$ up to $\mathcal{O}(g^{m+1} )$.
For $q=0$,
this is nothing but the Alday-Bissi's interpolating function and
we consider $q\neq 0$ case below.

Matching at $\mathcal{O}(g)$ leads us to
\begin{\eq}
\alpha (-p +q ) =1,\quad
\left( \frac{c_{s+p}\zeta (2s+2p)}{d_{s+q}\zeta (2s+2q)} \right)^\alpha = s_1 .
\end{\eq}
Imposing matching of other orders leads
\begin{\eq}
p+q = m .
\end{\eq}
Therefore we get
\begin{\eq}
p = \frac{1}{2}\left( m -\frac{1}{\alpha }\right) ,\quad
q = \frac{1}{2}\left( m +\frac{1}{\alpha }\right) .
\end{\eq}
We also require
\begin{\eq}
p,q \in \mathbb{Z}_{\geq 1} ,
\end{\eq}
which implies
\begin{\eq}
\alpha = \left\{ \begin{matrix}
\frac{1}{2\ell +1}  & {\rm for} & m:{\rm odd} \cr
\frac{1}{2\ell }  & {\rm for} & m:{\rm even} \end{matrix} \right. ,\quad
{\rm with}\ \ell \in\mathbb{Z} .
\end{\eq}
In the main text
we do not consider this type of interpolating functions.
But it would be interesting 
to compare the interpolating function
with the bootstrap.

\section{$s$-dependence of other interpolating functions}
\label{app:s-dep}
\begin{figure}[t]
\begin{center}
\includegraphics[width=7.4cm]{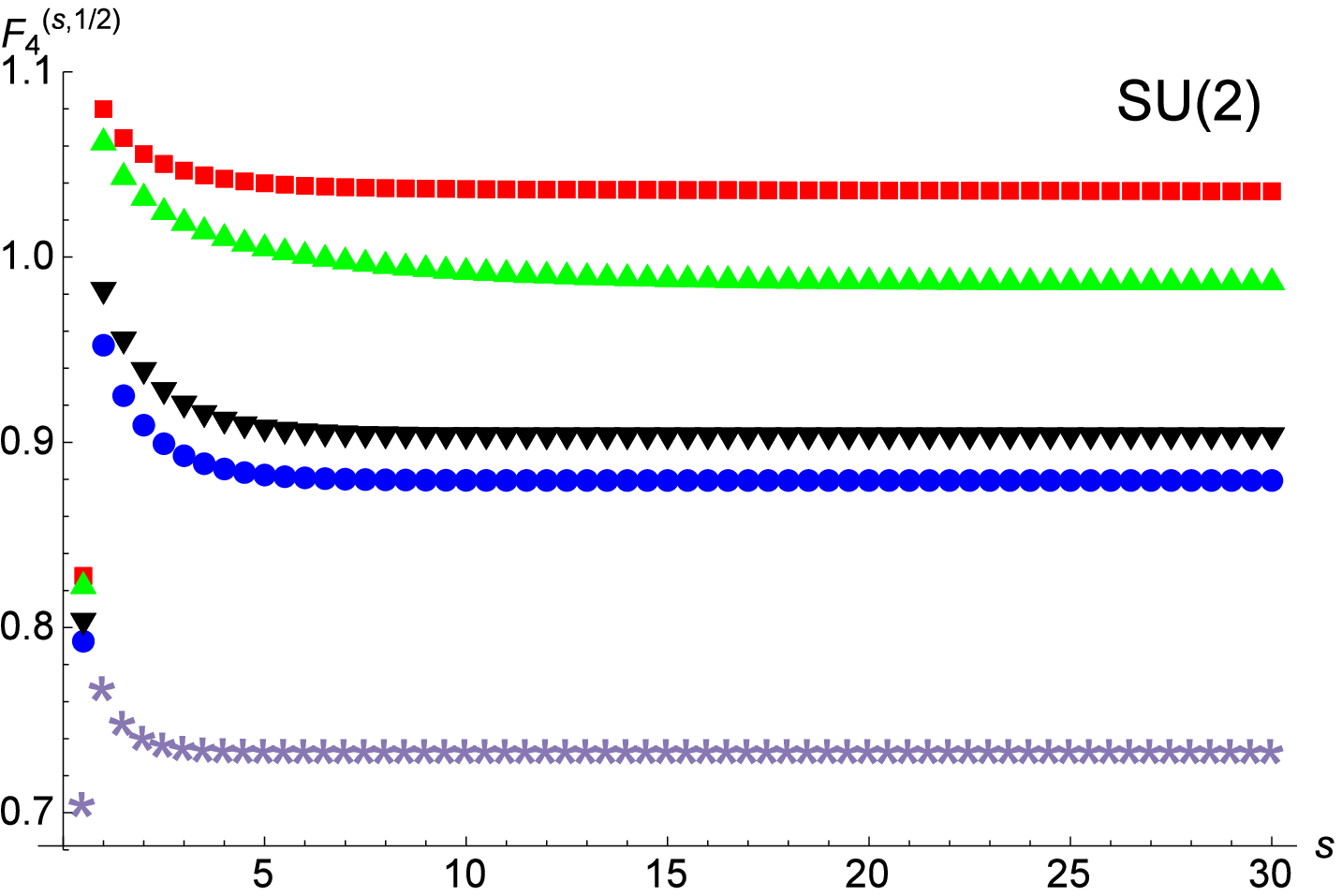}
\includegraphics[width=7.4cm]{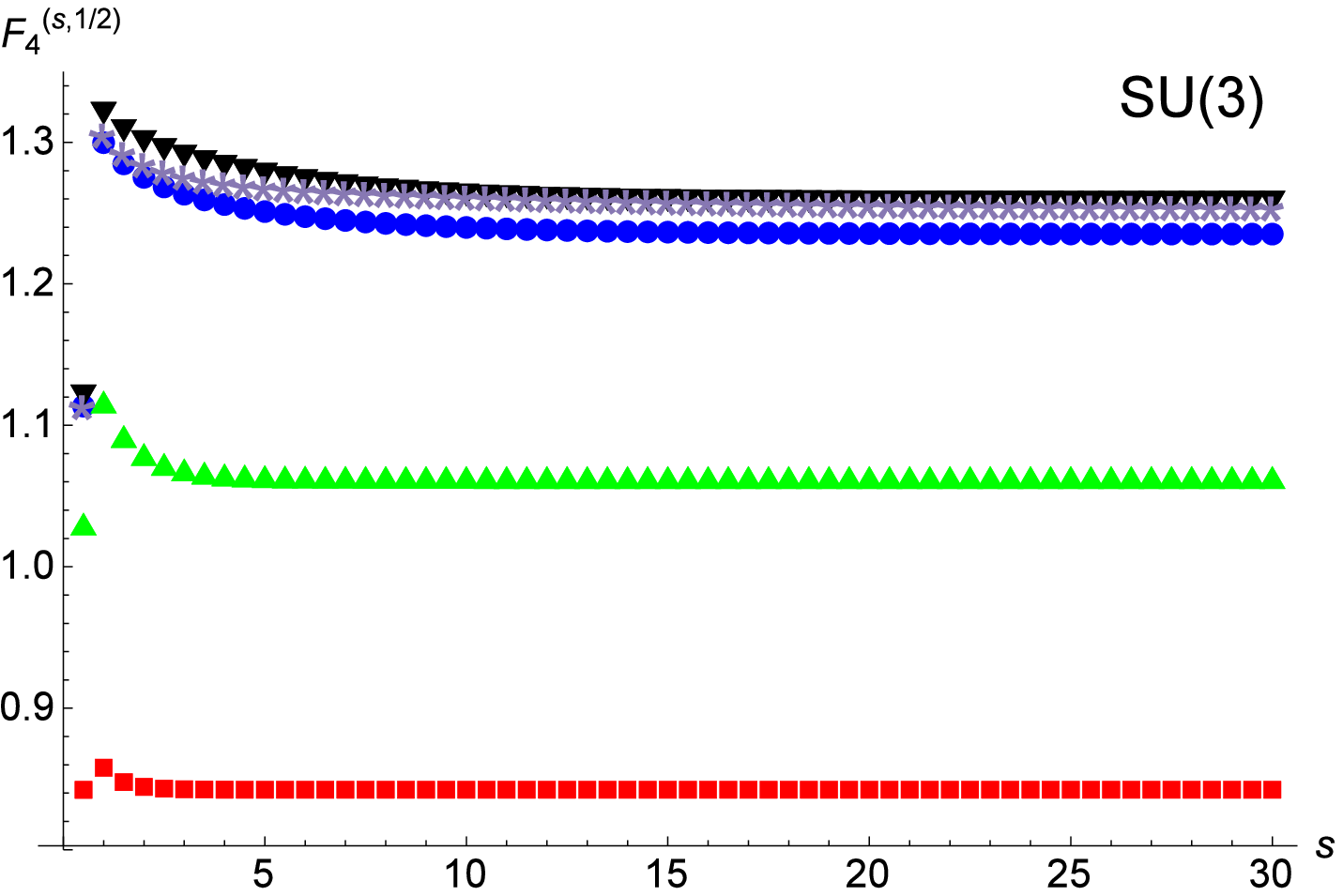}\\
\includegraphics[width=7.4cm]{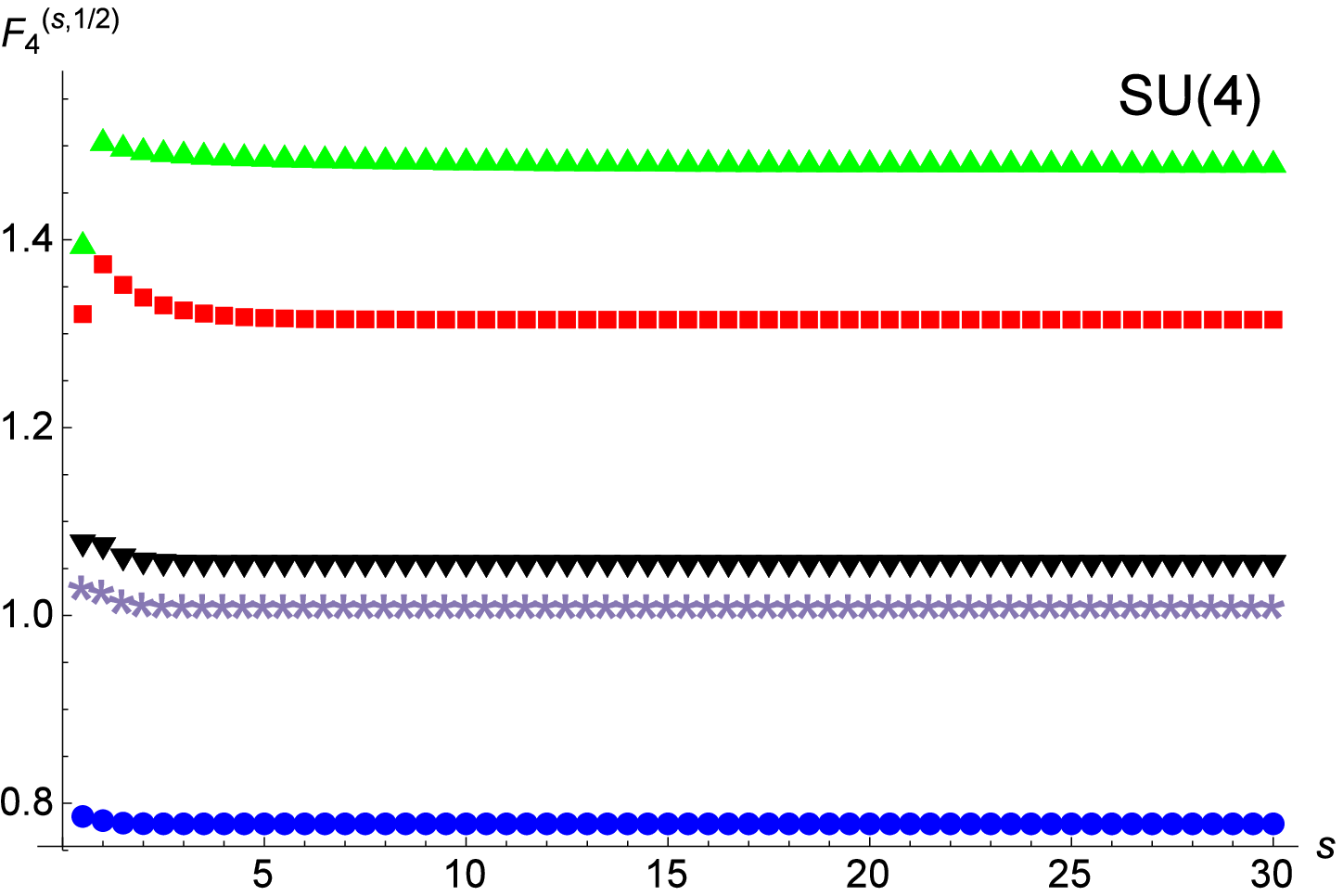}
\includegraphics[width=7.4cm]{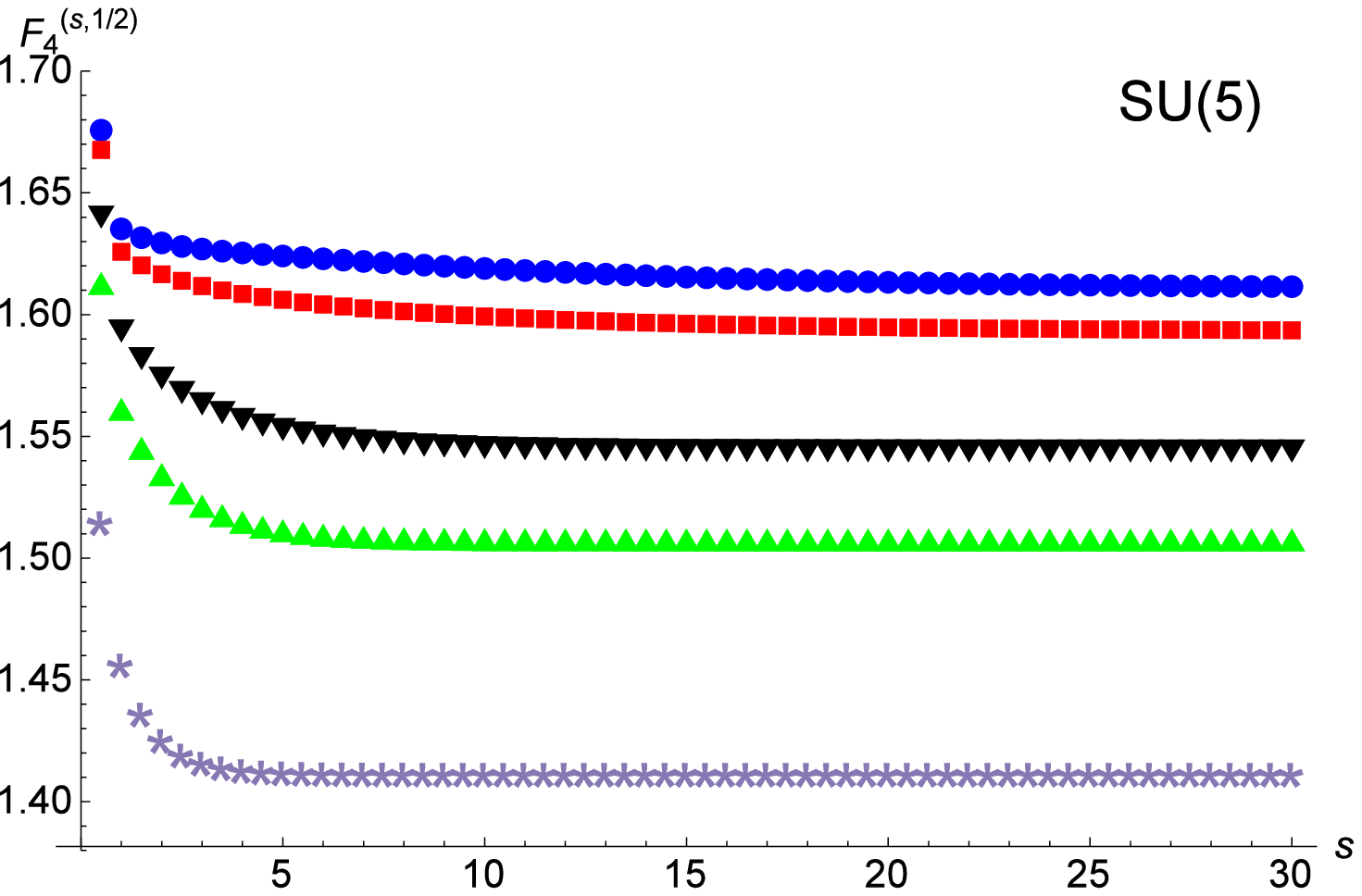}
\end{center}
\caption{$s$-dependence of 
the interpolating function $F_4^{(s, 1/2)}(\tau )$
for the spin-0 leading twist operator
at randomly chosen five points $\tau =(r_1 ,\cdots ,r_5 )$.
($r_1$: blue circle, $r_2$: red square, $r_3$: green triangle,
$r_4$: black inverse triangle, $r_5$: purple asterisk)
[Left-Top] $SU(2)$ case,
$(r_1 ,r_2 ,r_3 ,r_4 ,r_5 )$ $\simeq$
$(0.3377 + 0.4745i , 0.8231 +0.9956i ,0.5698 +0.5929i ,
0.8193 +0.7029i, 0.7449 + 0.34278i)$.
[Right-Top] $SU(3)$ case,
$(r_1 ,r_2 ,r_3 ,r_4 ,r_5 )$ $\simeq$
$(0.1185 +0.8770i, 0.1236 +0.3748i ,0.09718 +0.6059i,
0.6656 +0.8062i, 0.015689 + 0.9403i )$.
[Left-Bottom] $SU(4)$ case,
$(r_1 ,r_2 ,r_3 ,r_4 ,r_5 )$ $\simeq$
$(0.9907 +0.2714i, 0.3436 +0.3845i, 0.2747 +0.07620i,
0.03894 + 0.4271i, 0.9893 +0.4049i )$.
[Right-Bottom] $SU(5)$ case.
$(r_1 ,r_2 ,r_3 ,r_4 ,r_5 )$ $\simeq$
$(0.4757 +0.9538i, 0.7171 +0.8886i, 0.3572 +0.07676i,
0.5935 +0.5138i, 0.2600 +0.4709i )$.
}
\label{fig:s_dependence0_alpha12}
\end{figure}

\begin{figure}[t]
\begin{center}
\includegraphics[width=7.4cm]{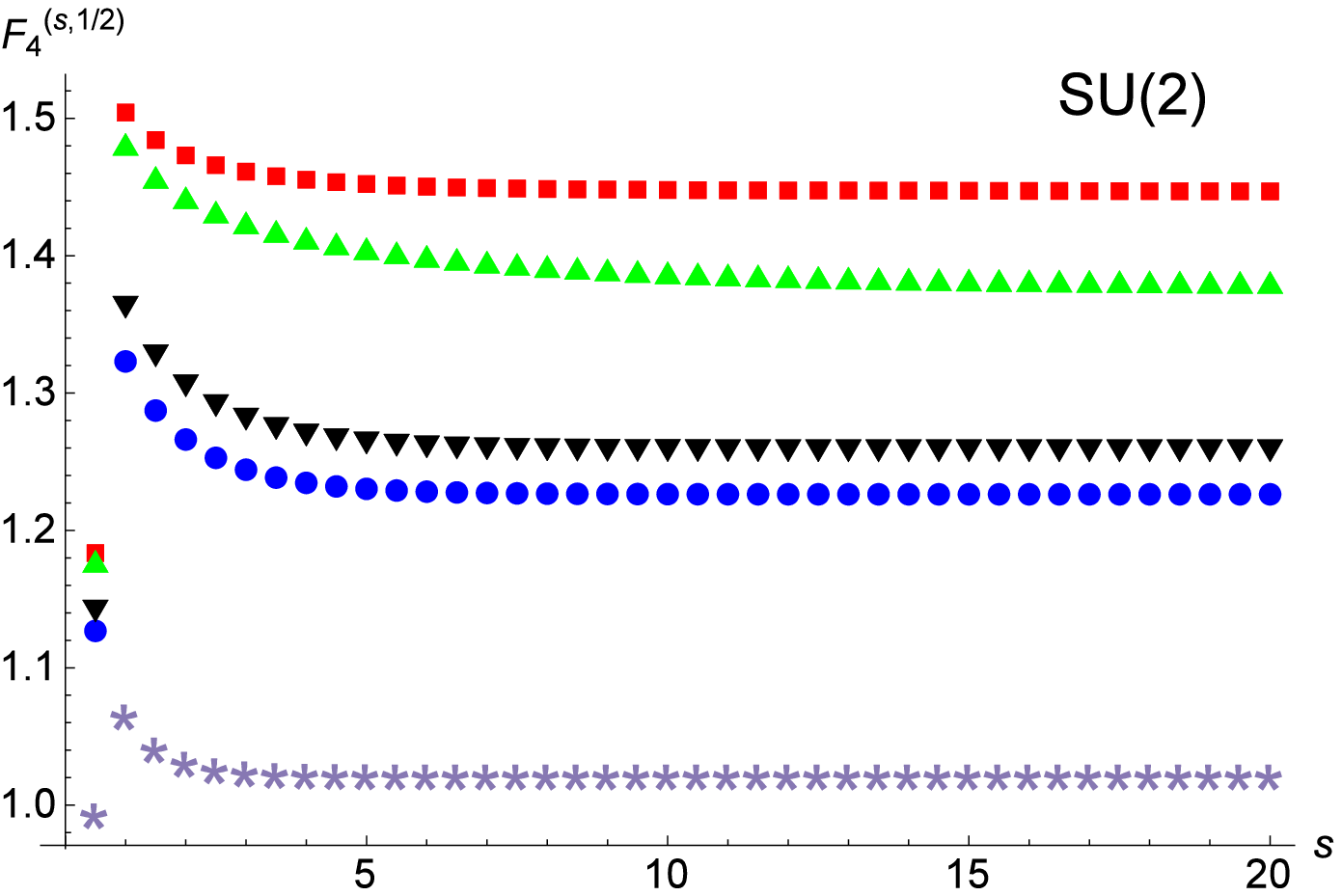}
\includegraphics[width=7.4cm]{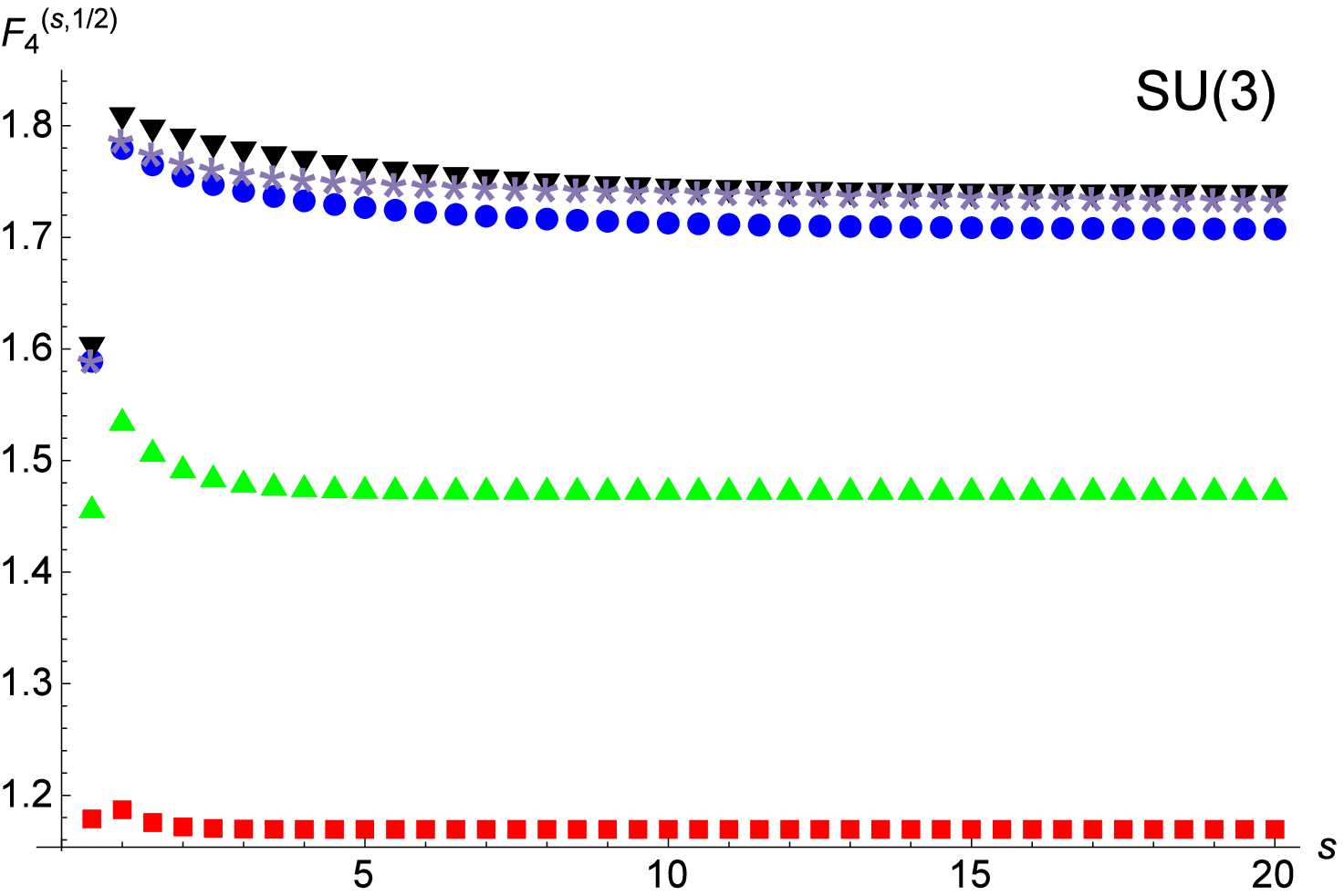}\\
\includegraphics[width=7.4cm]{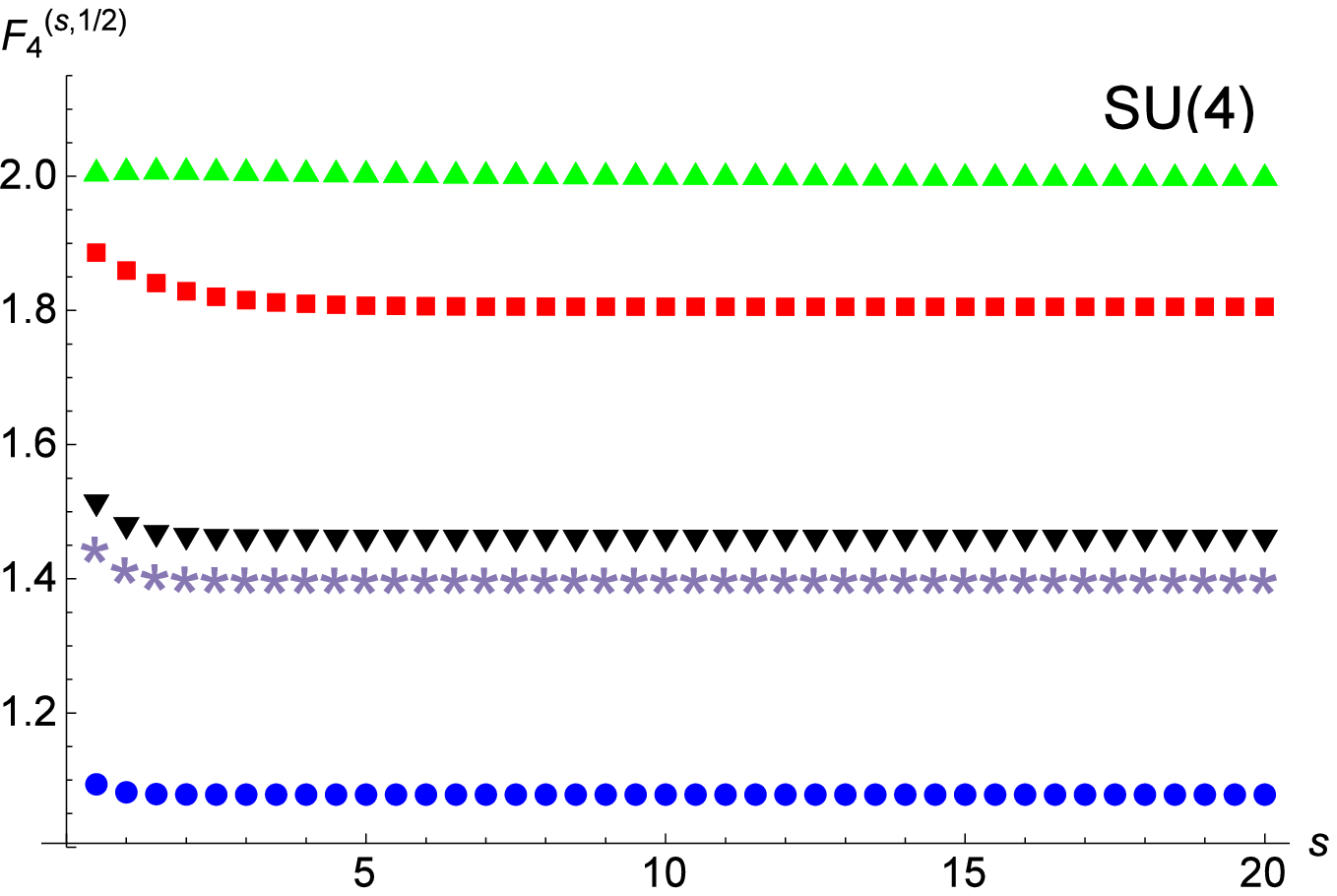}
\includegraphics[width=7.4cm]{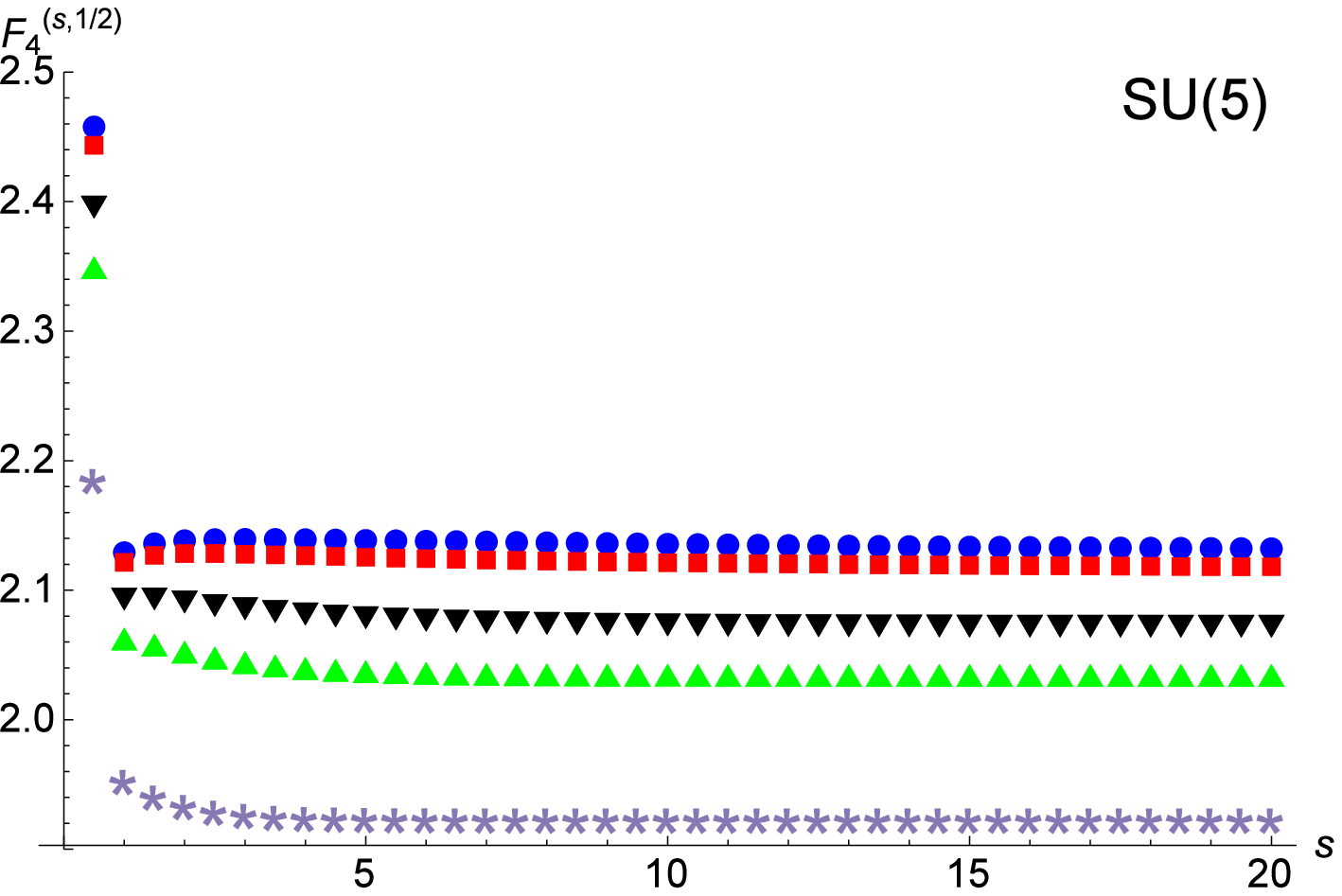}
\end{center}
\caption{
Similar plots as fig.~\ref{fig:s_dependence0_alpha12}
for the interpolating function $F_4^{(s, 1/2)}(\tau )$ of the spin-2 case.
[Left-Top]
$(r_1 ,r_2 ,r_3 ,r_4 ,r_5 )$ $\simeq$
$(0.3377 +0.4745 i, 0.8231 +0.9956i, 0.5698 +0.5929i,
0.8193 +0.7029i, 0.7449 +0.3428i )$.
[Right-Top]
$(r_1 ,r_2 ,r_3 ,r_4 ,r_5 )$ $\simeq$
$(0.1185 +0.8770i, 0.1236 + 0.3748i, 0.09718 +0.6059i,
0.6656 +0.8062i, 0.01569 +0.9403i  )$.
[Left-Bottom]
$(r_1 ,r_2 ,r_3 ,r_4 ,r_5 )$ $\simeq$
$(0.9907 +0.2714i, 0.3436 +0.3845i, 0.2747 +0.07620i,
0.03894 +0.4271i, 0.9893 +0.4049i  )$.
[Right-Bottom]
$(r_1 ,r_2 ,r_3 ,r_4 ,r_5 )$ $\simeq$
$(0.4757 +0.9538i, 0.7171 +0.8886i, 0.3572 +0.07676i,
0.5935 +0.5138i, 0.2600 +0.4709i )$.
}
\label{fig:s_dependence2_alpha12}
\end{figure}

\begin{figure}[t]
\begin{center}
\includegraphics[width=7.4cm]{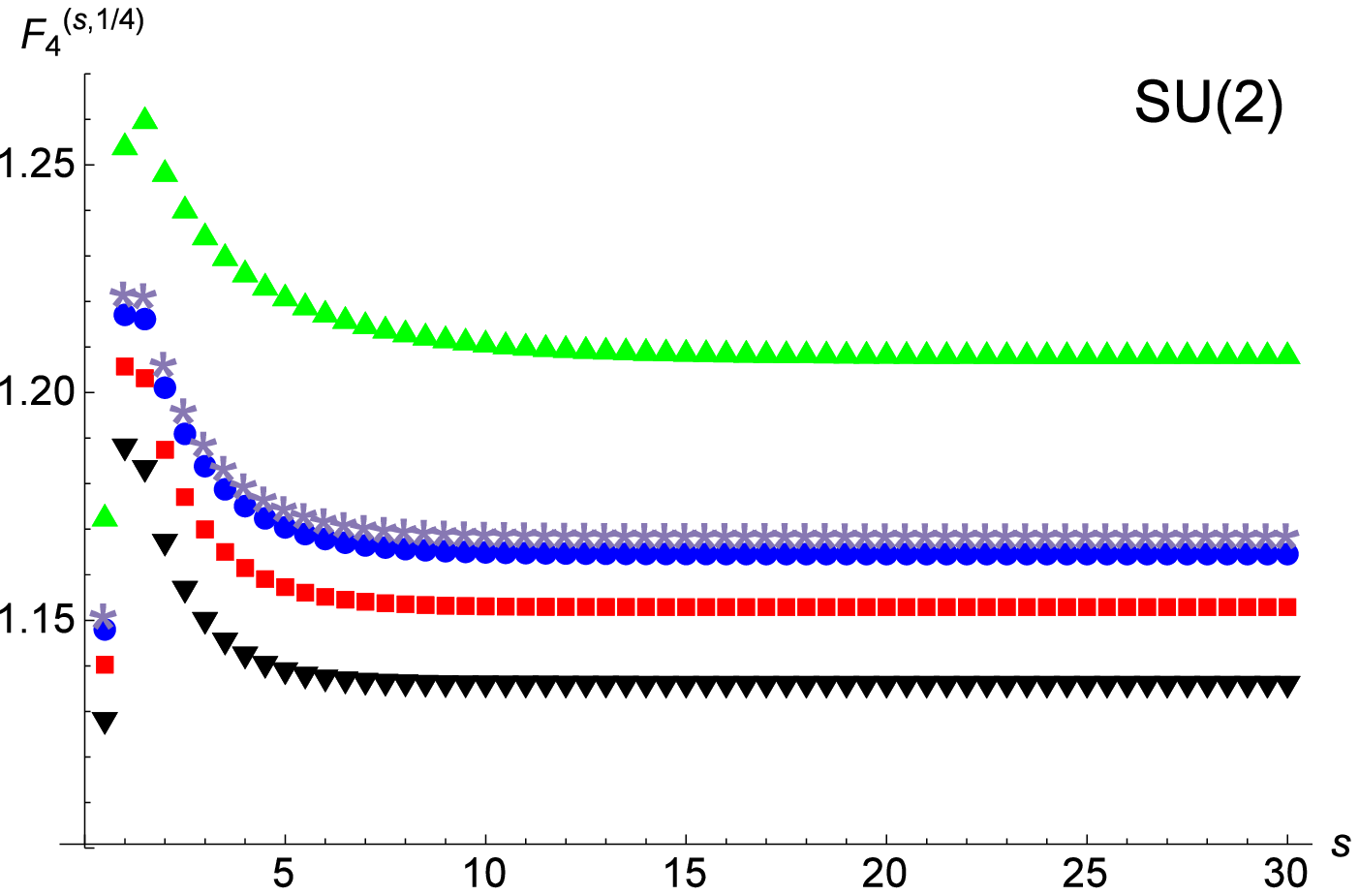}
\includegraphics[width=7.4cm]{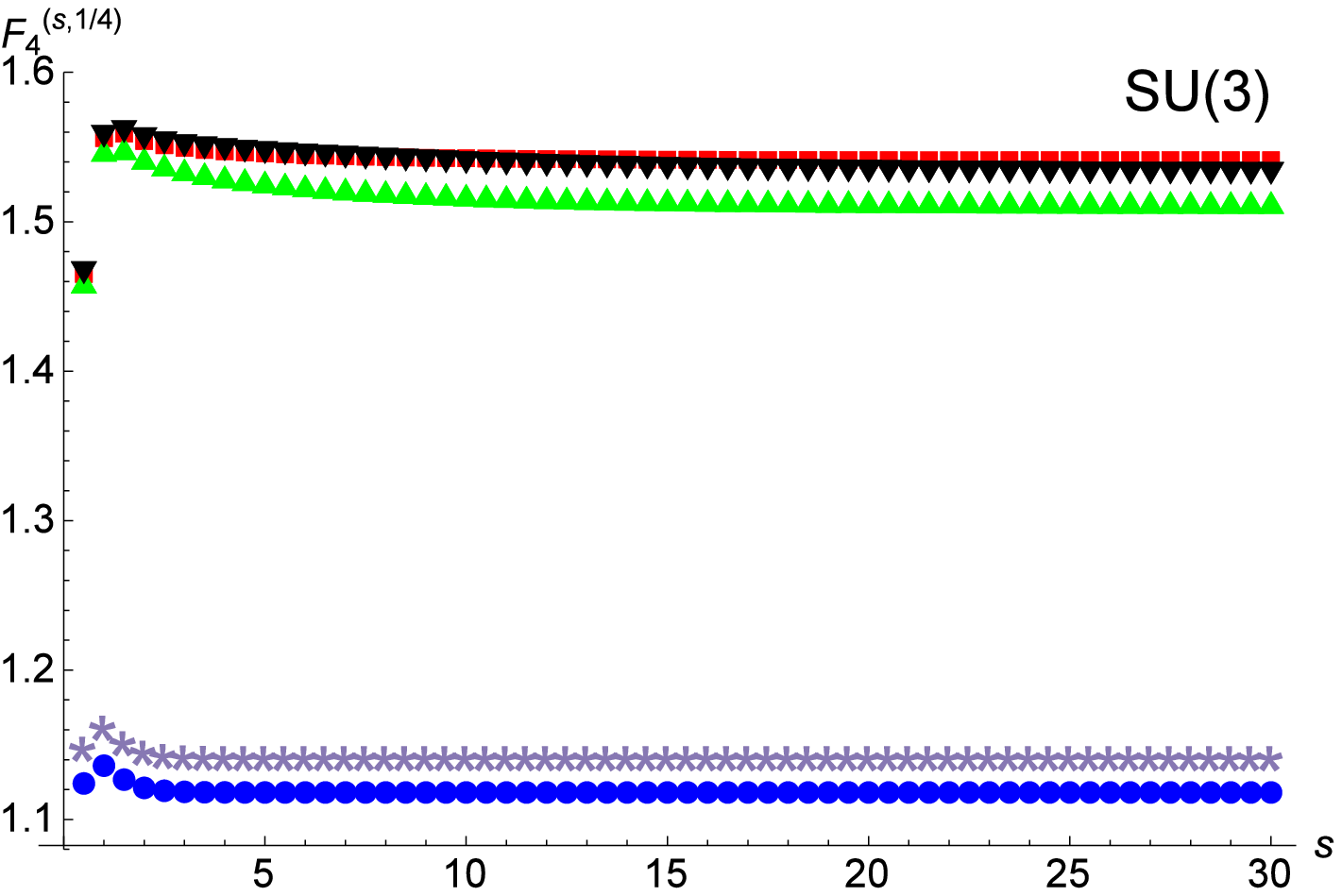}\\
\includegraphics[width=7.4cm]{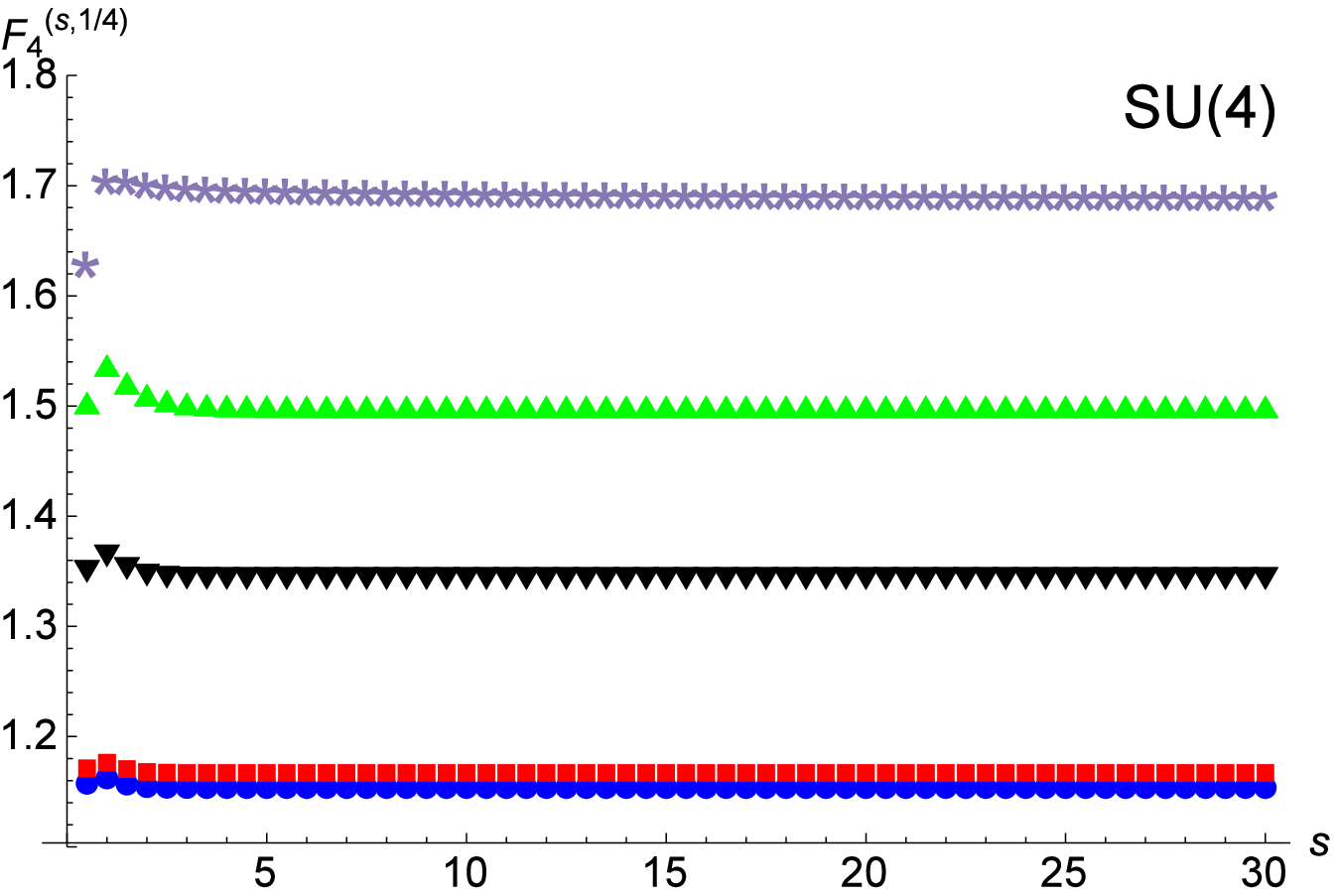}
\includegraphics[width=7.4cm]{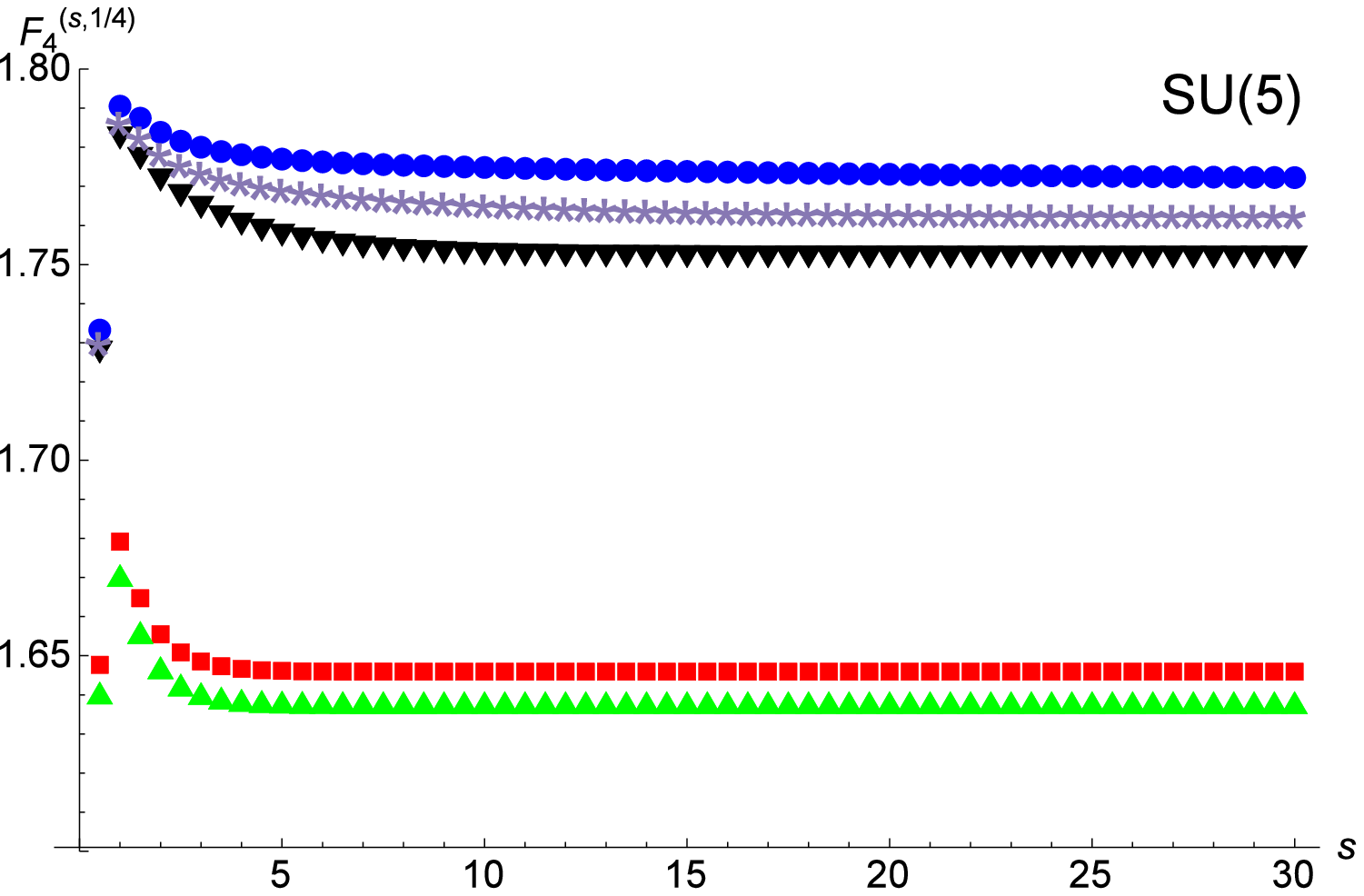}
\end{center}
\caption{
Similar plots as fig.~\ref{fig:s_dependence0_alpha12} and \ref{fig:s_dependence2_alpha12}
for the interpolating function $F_4^{(s, 1/4)}(\tau )$ of the spin-2 case.
[Left-Top] 
$(r_1 ,r_2 ,r_3 ,r_4 ,r_5 )$ $\simeq$ 
$(0.3422 +0.2076i, 0.3290 +0.08323i, 0.4127 +0.5518i,
0.2687 +0.6024i,  0.390 +0.3754i )$.
[Right-Top] 
$(r_1 ,r_2 ,r_3 ,r_4 ,r_5 )$ $\simeq$
$(0.7971 +0.2347i, 0.6482 +0.9407i, 0.2732 +0.8653i,
0.4621 +0.2902i, 0.07740 +0.4130i  )$.
[Left-Bottom] 
$(r_1 ,r_2 ,r_3 ,r_4 ,r_5 )$ $\simeq$
$(0.1233 +0.2514i, 0.1242 +0.2581i,  0.8283 +0.4928i,
0.7952 +0.1749i,  0.4156 +0.9410i )$.
[Right-Bottom] 
$(r_1 ,r_2 ,r_3 ,r_4 ,r_5 )$ $\simeq$
$(0.5265 +0.5726i, 0.02538 +0.5655i,  0.7295 +0.3379i,
0.3772 +0.6248i, 0.4563 + 0.4726i )$.
}
\label{fig:s_dependence2}
\end{figure}

\begin{figure}[t]
\begin{center}
\includegraphics[width=7.4cm]{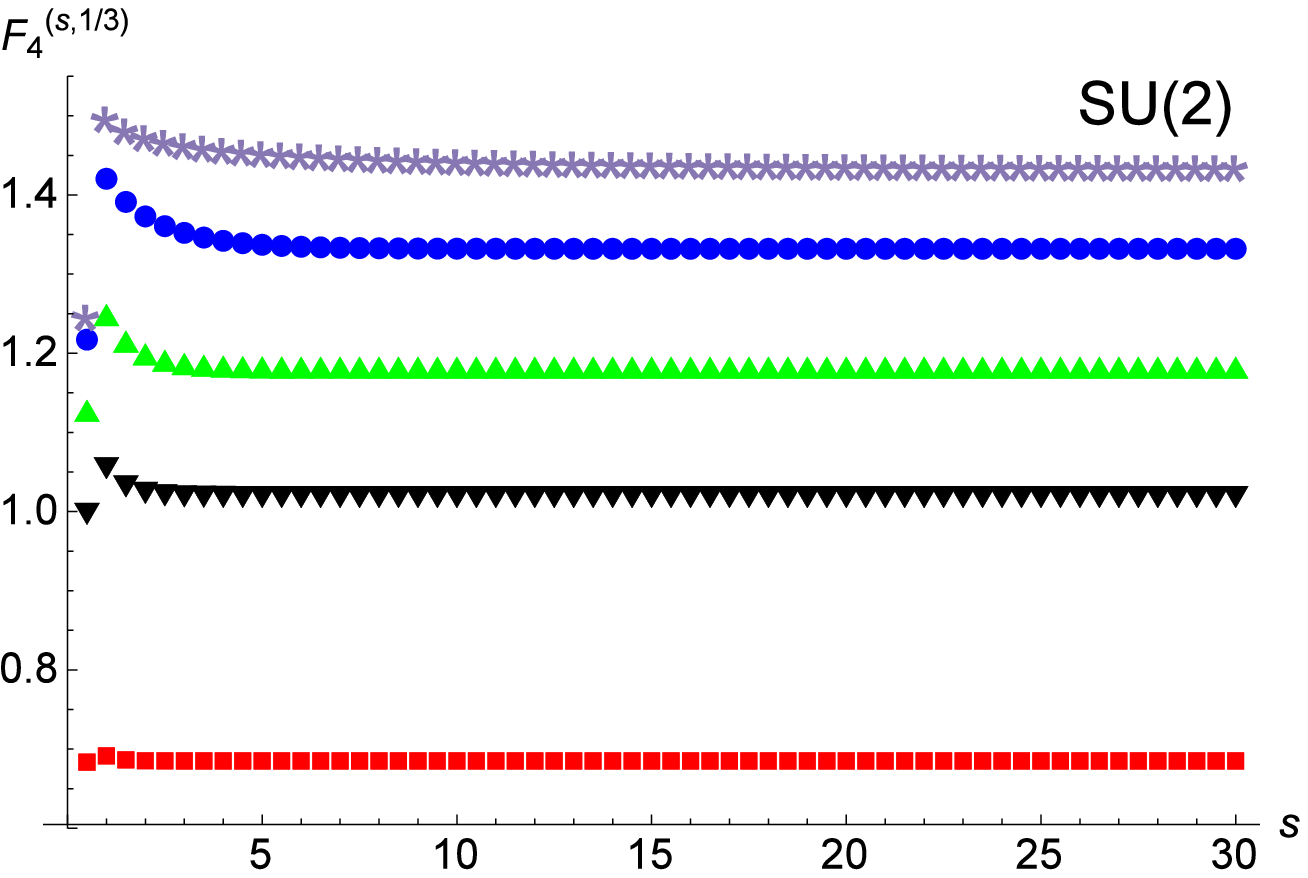}
\includegraphics[width=7.4cm]{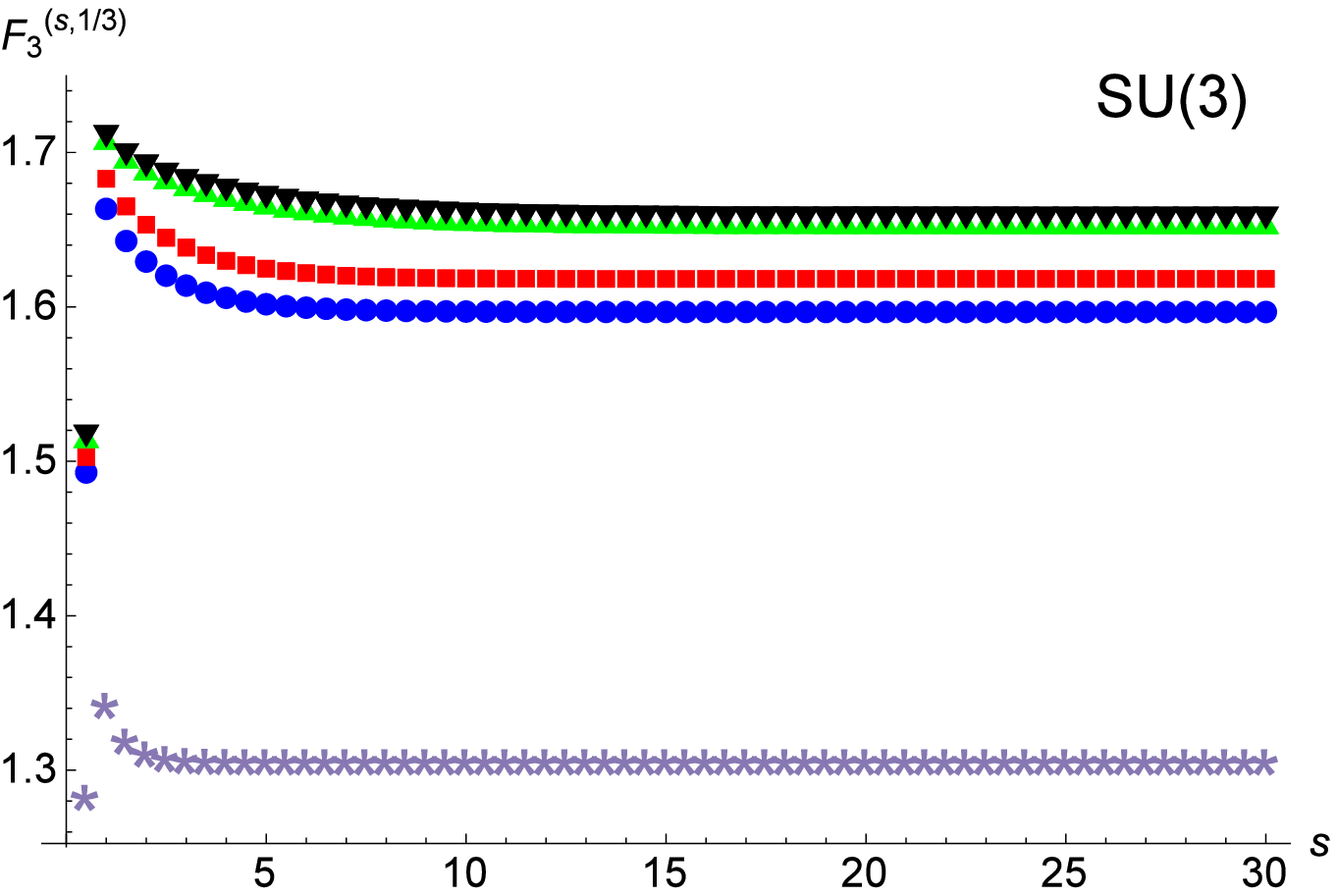}\\
\includegraphics[width=7.4cm]{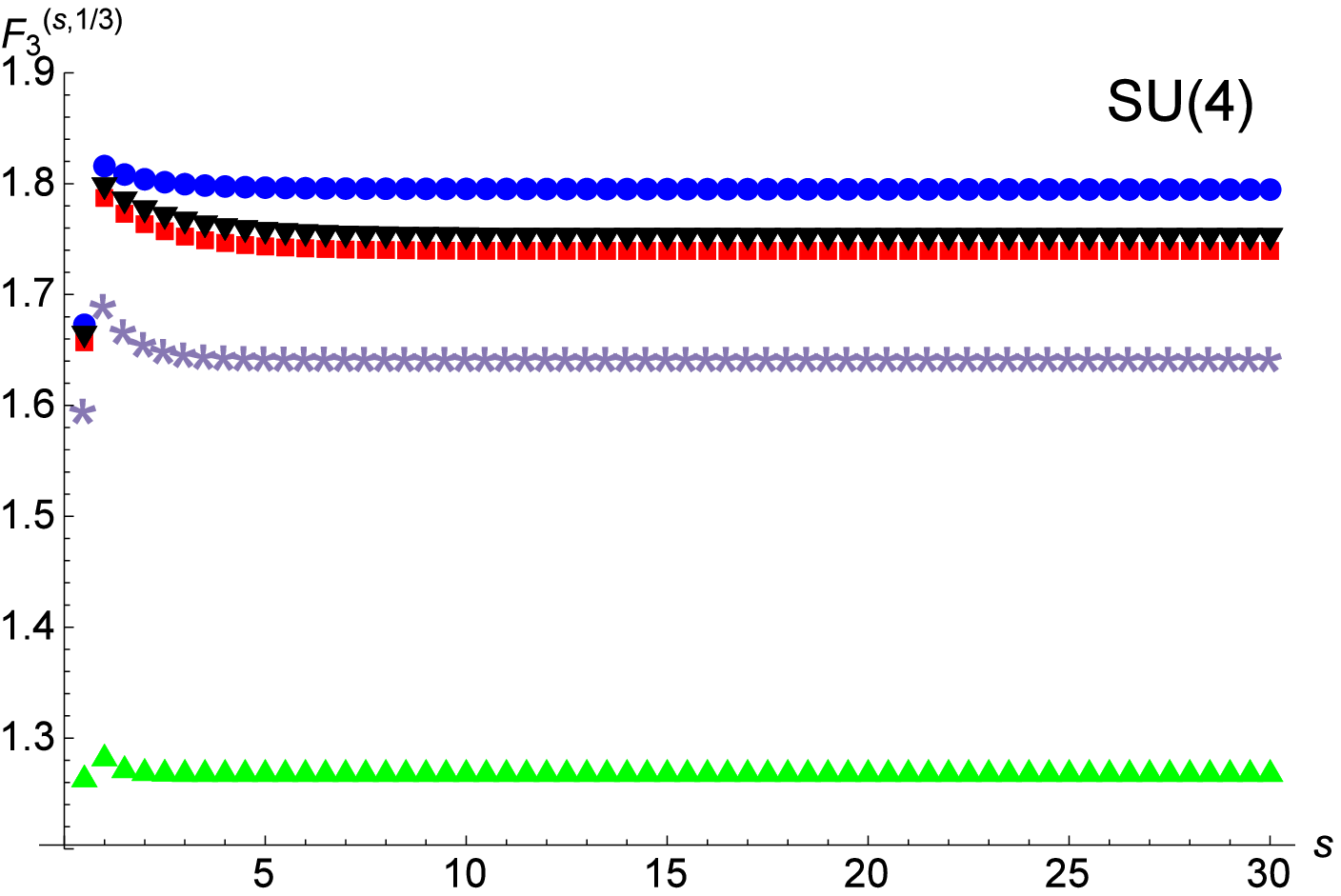}
\includegraphics[width=7.4cm]{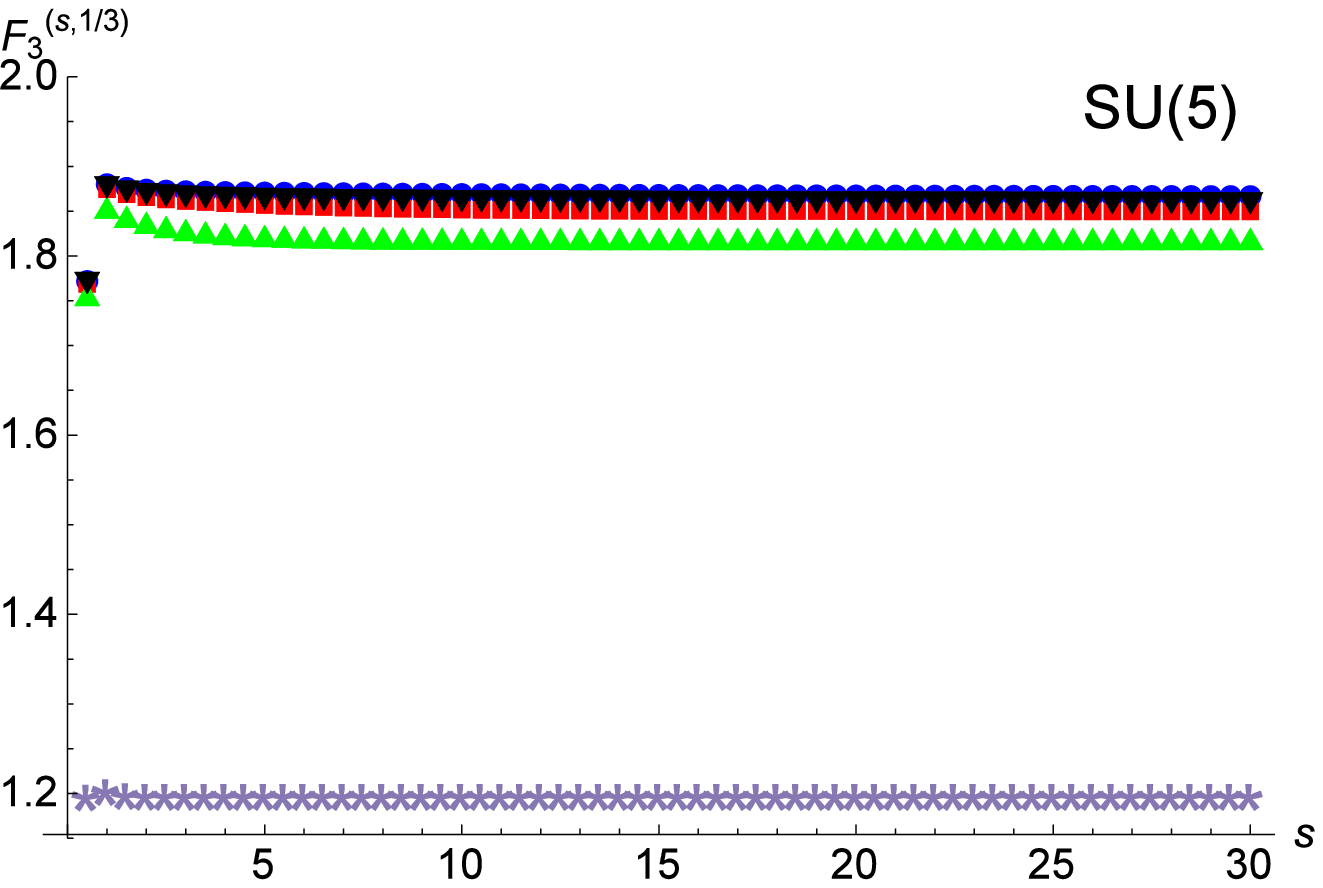}
\end{center}
\caption{
Similar plots as fig.~\ref{fig:s_dependence0_alpha12}, \ref{fig:s_dependence2_alpha12} and \ref{fig:s_dependence2}
for the interpolating function $F_3^{(s, 1/3)}(\tau )$ of the spin-4 case.
[Left-Top] 
$(r_1 ,r_2 ,r_3 ,r_4 ,r_5 )$ $\simeq$ 
$(0.1569 +0.7136i$, $0.001963 +0.2593i$, $0.7151 +0.3154i$,
$0.2043 +0.3075i$, $0.2278 +0.8969i  )$.
[Right-Top] 
$(r_1 ,r_2 ,r_3 ,r_4 ,r_5 )$ $\simeq$
$(0.4293 +0.1589i$, $0.7140 +0.6887i$, $0.6629 +0.7537i$,
$0.6183 +0.7089i$, $0.4724 +0.1022i  )$.
[Left-Bottom] 
$(r_1 ,r_2 ,r_3 ,r_4 ,r_5 )$ $\simeq$
$(0.06265 +0.9844i$, $0.6319 +0.5036i$, $0.9058 +0.2708i$,
$0.6299 +0.2444i$, $0.2900 +0.2536i )$.
[Right-Bottom] 
$(r_1 ,r_2 ,r_3 ,r_4 ,r_5 )$ $\simeq$
$(0.4256 +0.9369i$, $0.4141 +0.6878i$, $0.9963 +0.7780i$,
$0.7001 +0.9990i$, $0.3334 + 0.02433i )$.
}
\label{fig:s_dependence4}
\end{figure}

In this appendix
we present $s$-dependence of various interpolating functions, in fig.~\ref{fig:s_dependence0_alpha12},\ref{fig:s_dependence2_alpha12},
\ref{fig:s_dependence2} and \ref{fig:s_dependence4}.
We find that
all the results are similar to fig.~\ref{fig:s_dependence}.


\section{Saturation of interpolating function 
for weak coupling and at duality invariant points 
for large-$s$}
\label{app:s-dep_fp}
The coefficients in the interpolating function are determined only 
by the perturbative part of the Eisenstein series:
\begin{\eq}
\left. E_s (\tau ) \right|_{\rm perturbative}
= \zeta (2s) g^{-s}
 +\frac{\sqrt{\pi}\Gamma (s-1/2)}{\Gamma (s)} \zeta (2s-1) g^{s-1} .
\end{\eq}
In particular for large $s$, 
the coefficients are determined by the first term.
Since $\zeta (2s)$ is almost unity for sufficiently large $s$,
the interpolating functions for large $s$ is approximately given by
\begin{\eq}
F_m^{(s,\alpha )} (\tau ) 
\simeq \Biggl[ \frac{\sum_{k=1}^p c_k g^k +\mathcal{O}(g^{2s+1}) }
                             {\sum_{k=1}^q d_k g^k +\mathcal{O}(g^{2s+1})} \Biggr]^\alpha \, ,
\end{\eq}
and the coefficients $c_k$ and $d_k$ are almost independent of $s$ for very large $s$.
Because of this,
the interpolating function in weak coupling regime 
is almost independent of $s$ for large-$s$.

For values of the interpolating functions at the duality invariant points and for large $s$, we can further solidify our independent of $s$ claim.
For this purpose, we should know 
values of $E_s (i)$ and $E_s (e^{\frac{i\pi}{3}})$ for large $s$.
By using $\zeta_a (s)\simeq a^{-s}$ with $a\geq 0$ for large $s$,
we find
\begin{\eq}
E_s (i)\simeq 2 ,\quad E_s (e^{\frac{i\pi}{3}}) \simeq 3\left( \frac{\sqrt{3}}{2} \right)^s ,\quad
{\rm for}\ s\gg 1 .
\end{\eq}
Therefore the interpolating functions at the duality invariant points are given by
\begin{\eq}
F_m^{(s,\alpha )} (i ) 
\simeq \Biggl[ \frac{\sum_{k=1}^p c_k  }{\sum_{k=1}^q d_k  }\Biggr]^\alpha ,\quad
F_m^{(s,\alpha )} (e^{\frac{i\pi}{3}} ) 
\simeq \Biggl[ \frac{\sum_{k=1}^p c_k (\sqrt{3}/2)^k   }{\sum_{k=1}^q d_k  (\sqrt{3}/2)^k}
\Biggr]^\alpha ,\quad 
{\rm for}\ s\gg 1 .
\end{\eq}
Thus the interpolating function at the duality fixed points
is independent of $s$ for large-$s$.

\section{$S$-duality interpolating functions 
with $\lambda^{1/4}$ in the classical string limit}
\label{app:lambda14}
In the main text,
we have approximated the dimensions of the leading twist operators
by the interpolating functions,
which are consistent with the weak coupling expansions,
holographic results and full $S$-duality.
As the holographic results,
we have used the results \eqref{eq:result_gravity}
for the double trace operators,
which are the subleading twist operators in the weak coupling regime.
This is because the operator \eqref{eq:twist2} is
dual to the massive string state
and acquires the very large dimension in the classical string regime.

However,
one may wonder 
if one can construct another modular invariant interpolating functions,
which have the same weak coupling expansions
but different behaviours say as $\sim\lambda^{1/4}$ in the classical string regime.
This may not make sense physically
since the dimension of \eqref{eq:twist2} would not be modular invariant
but this may be useful in future for constructing interpolating functions
for other modular invariants with different behaviours 
in the classical string regime.
In this appendix,
we try to construct a class of modular interpolating functions,
with the same weak coupling expansion and 
$\lambda^{1/4}$ behaviour in the classical string limit.

To be specific, let us consider the Konishi operator.
The dimensions of the Konishi operator behaves as
\eqref{eq:weak} for weak coupling and as 
\eqref{eq:planarS} for large-$\lambda$ in the planar limit.
Now we would like to construct interpolating functions,
which are consistent with 
\eqref{eq:weak}, \eqref{eq:planarS} and the full $S$-duality.
As a conclusion,
we failed to construct a single interpolating function satisfying these properties
but we find that a linear combination of multiple interpolating functions which are slight modifications of \eqref{eq:main_interpolation},
satisfies the properties.

First we discuss that
the interpolating functions of the type \eqref{eq:main_interpolation},
which have been used for the leading twist operators in the main text,
cannot satisfy the above properties.
Indeed we have explicitly checked this for various cases and
we can also show this for large-$s$ in the following way.
Recall that 
the coefficients $c_k$ and $d_k$ in \eqref{eq:main_interpolation}
are determined in terms of
only perturbative part of the Eisenstein series $E_s (\tau )$.
Although the perturbative part of $E_s (\tau )$
has $\mathcal{O}(g^{-s})$ and $\mathcal{O}(g^{s-1})$ parts,
only the $\mathcal{O}(g^{-s})$ part is relevant 
to determine the coefficients for large-$s$.
Thus $c_k$ and $d_k$ are effectively determined by
\begin{\eq}
 \Biggl[ \frac{\sum_{k=1}^p \zeta (2s+2k) c_k g^{-(s+k)} }
 {\sum_{k=1}^q \zeta (2s+2k) d_k g^{-(s+k)} } \Biggr]^\alpha .
\end{\eq}
While $c_k$ and $d_k$ are nontrivial function of $N$ in general,
we know that
their planar limits behave as $\mathcal{O}(N^{q-k} )$ \footnote{We would expect it to be $\mathcal{O}(N^{-(s+k)})$ but since we normalize $d_{s+q}=1$, we multiply each coefficient by $N^{s+q}$.}
since the anomalous dimension is $\mathcal{O}(1)$ in the planar limit.
Hence in the planar limit, the interpolating function becomes
\begin{\eq}
 \Biggl[ \frac{\sum_{k=1}^p \bar{c}_k \lambda^{-(s+k)} }
 {\sum_{k=1}^q \bar{d}_k \lambda^{-(s+k)} } \Biggr]^\alpha , 
\end{\eq}
where 
\begin{\eq}
\bar{c}_k = \lim_{N\rightarrow\infty} \zeta (2s+2k)N^{k-q} c_k ,\quad
\bar{d}_k = \lim_{N\rightarrow\infty} \zeta (2s+2k)N^{k-q} d_k.
\end{\eq}
Since this function becomes $\mathcal{O}(1)$ for large-$\lambda$,
the interpolating function \eqref{eq:main_interpolation} for large-$s$
cannot have the $\lambda^{1/4}$-law in the classical string limit.
Thus we shall consider different types of interpolating functions.

Alternatively
let us consider the following type of interpolating functions
\begin{\eq}
I_m^{(s,t,\alpha )} (\tau ) 
= \Biggl[ \frac{ c_1 E_{s+t+1}(\tau ) +\sum_{k=2}^p c_k E_{s+k} (\tau )}
{\sum_{k=1}^q d_k E_{s+k} (\tau )} \Biggr]^\alpha ,
\label{eq:inter_t}
\end{\eq}
where $c_1 =\mathcal{O}( N^{q-t-1})$ 
and
the only difference from \eqref{eq:main_interpolation}
is the presence of the new parameter $t$ in the first term of the numerator.
Note that $t$ should be integer
to get weak coupling expansion with only integer powers of $g$.
By a similar argument as above,
we find that 
the planar limit of this interpolating function for large-$s$ is given by
\begin{\eq}
\left. I_m^{(s,t,\alpha )} (\tau ) \right|_{\rm planar}
= \Biggl[ \frac{ \bar{c}_1 \lambda^{-(s+t+1)} +\sum_{k=2}^p \bar{c}_k \lambda^{-(s+k)} }
 {\sum_{k=1}^q \bar{d}_k \lambda^{-(s+k)} } \Biggr]^\alpha 
= \Biggl[ \frac{ \bar{c}_1 \lambda^{-t} +\sum_{k=2}^p \bar{c}_k \lambda^{-k+1} }
 {\sum_{k=1}^q \bar{d}_k \lambda^{-k+1} } \Biggr]^\alpha , 
\end{\eq}
where 
$\bar{c}_1= \lim_{N\rightarrow\infty} \zeta (2s+2)N^{t+1-q} c_1$,  $\bar{c}_k=\lim_{N\rightarrow\infty} \zeta (2s+2k)N^{k-q} c_k$ and
$\bar{d}_k = \lim_{N\rightarrow\infty} \zeta (2s+2k)N^{k-q} d_k$.
When $t$ is negative,
the leading order of this function in the large-$\lambda$ expansion
is $\mathcal{O}(\lambda^{-\alpha t})$.
Thus the interpolating function of the class \eqref{eq:inter_t}
can have the $\lambda^{1/4}$-law in the classical string limit
by appropriately choosing $\alpha$ and $t$.
Indeed, 
if the `to be matched' term is $\mathcal{O}(\lambda^{c})$,
then we have to solve for $ -\alpha t = c $. 

However
the interpolating function $I_m^{(s,t,\alpha )} (\tau )$
with $-\alpha t=1/4$ cannot correctly reproduce
the subleading order of \eqref{eq:planarS}, namely $\mathcal{O}(1)$.
The reason is that
if we consider $I_4^{(s,t,\alpha )} (\tau )$ with large-$s$,
then the $\lambda^{1/4}$-law uniquely\footnote{
We could consider the same form of the interpolating function as \eqref{eq:inter_t}
but imposing three of the coefficients for match with the holographic result and
four of them with the weak coupling expansion.
Then we can also take $(\alpha ,t)=(1/8,-2)$ 
but this case also does not have $\mathcal{O}(1)$ in the large-$\lambda$ expansion.
} requires $(\alpha ,t)=(1/4,-1)$ and
therefore the subleading order is $\mathcal{O}(\lambda^{-3/4})$
rather than $\mathcal{O}(1)$.
Thus we cannot single interpolating function
consistent with \eqref{eq:weak}, \eqref{eq:planarS} and the full $S$-duality
by the type \eqref{eq:inter_t}.
Alternatively we find that
the following linear combination of $I_m^{(s,t,\alpha )}$
satisfies the desired properties\footnote{
The value of $\alpha$ in the second term is constrained only by $m$
since the case with $t=0$ does not give new constraint to $\alpha$.
}:
\begin{\eq}
 w_1 I_4^{(s,-1,1/4)}(\tau ) +w_2 I_4^{(s,0,\alpha )}(\tau )
+ w_3 I_4^{(s,-1,-1/4)}(\tau ) ,\quad
{\rm  with}\ w_1 +w_2 +w_3 =1 .
\end{\eq}
where all the coefficients including $w_1 , \, w_2$ and $ w_3$ 
are fixed with $I_4^{(s,-1,1/4)}(\tau )$, $I_4^{(s,0,\alpha )}(\tau )$ and $I_4^{(s,-1,-1/4)}(\tau )$ respectively 
matching to the $\mathcal{O}(\lambda^{1/4})$, $\mathcal{O}(1)$ and  $\mathcal{O}(\lambda^{-1/4})$ coefficients. 
We could also work with variations of the above scheme 
with more terms from the large-$\lambda$ expansion in the planar limit and weak coupling expansion
though we do not explicitly write their constructions. 

\section{Explicit forms of interpolating functions}
\label{app:explicit}
In this appendix
we present explicit forms of the interpolating functions used in the main text. 
It should also be noted that 
we have to solve linear equations to solve for the unknown coefficients and 
thus it does not involve numerical approximation. 
Although we often write their coefficients with 6 digits of precision 
to avoid too long expressions,
we practically use analytic expressions or infinite digits in Mathematica files. 

\subsection{Leading twist operators}
\subsubsection{Spin-0}

\begin{\eqa}
F_4^{(30,1/2)}(\tau )
&=& \left( 0.911891N^2 E_{32}  
  -0.00719662N \left(-30.5858 N^2 -283.977\right) E_{31} \right)^{1/2} \NN\\
&& \Bigl(  E_{34} +N\left(0.0550286N^2  +1.39138\right) E_{31} 
   -0.000418683E_{32}\left(-246.029N^2-3407.72\right) \NN\\
&&   -\frac{0.00263066  \left(-333.757  N^2-851.93\right)}{N} E_{33} \Bigr)^{-1/2} 
,\NN\\
F_4^{(30,1/4)}(\tau )
&=& 0.95493N  \left( E_{31} \right)^{1/4}
\Bigl( 0.303964N^2E_{33}  -0.0161258N \left(-4.34157N^2 -186.647\right) E_{32} \NN\\
&& +\left(0.0519715 N^4 +1.66309N^2 \right)E_{31} +1.27324NE_{34} +E_{35} \Bigr)^{-1/4} .
\end{\eqa}

\subsubsection{Spin-2}

\begin{\eqa}
F_4^{(30,1/2)}(\tau )
&=& N^{1/2} \left( E_{31} \left( 0.225557N^2 +3.60609\right)
+1.75905N E_{32} \right)^{1/2}  \NN\\
&&\Bigl( E_{34} +0.0563893N^3 E_{31} 
 +0.0276488N^2 E_{32} +1.12708N E_{31}  \NN\\
&& +0.78253N E_{33}  +\frac{2.05002 }{N}E_{33} +1.34134E_{32}  \Bigr)^{-1/2} ,\NN\\
F_4^{(30,1/4)}(\tau )
&=& 1.32629N \left( E_{31}  \right)^{1/4} \NN\\
&& \Bigl( 0.315612 N^2 E_{33}
-6.221361 \times 10^{-6}  N \left(-8614.58 N^2 -399396\right)E_{32} \NN\\
&&  +N^2 \left(0.193391N^2 +1.54713 \right) E_{31}
+1.30861NE_{34} +E_{35}   \Bigr)^{-1/4} .
\end{\eqa}

\subsubsection{Spin-4}
\begin{\eqa}
F_3^{(30,1/3)}(\tau )
&=& 1.55972N(E_{31})^{1/3}
\Bigl(  N\left( 0.474295N^2  +1.36597\right) E_{31}  \NN\\
&& +0.0749153N^2 E_{32}   +0.987822N E_{33} +E_{34}
\Bigr)^{-1/3} .
\end{\eqa}

\subsection{Konishi operator in the planar limit}
\begin{\eqa}
F_{15,2}^{(1/28)}
&=& \frac{4}{( -0.798124 x^2-0.940303 x+1)^{1/28}}
 \Bigl( -0.147129 x^{16}-1.33546 x^{15}-5.10135 x^{14} \NN\\
 && -7.68337 x^{13}-8.77477 x^{12}-19.9647 x^{11}-4.97804 x^{10}-30.8794 x^9+7.2648 x^8-30.131 x^7 \NN\\
&&  +16.5478 x^6-18.2566 x^5+14.0806 x^4-6.28546 x^3+5.88638
   x^2-0.940303 x +1 \Bigr)^{1/28}
,\NN\\
F_{15,2}^{(1/32)}
&=& \frac{4}{(0.280107 x+1)^{1/32}} 
\Bigl( 1+ 0.0405549 x^{17}+0.510874 x^{16}+2.74729 x^{15}+9.80801 x^{14} 
 \NN\\
&& +8.27997 x^{13} +29.56 x^{12}+15.0662 x^{11}+53.7874 x^{10}+17.9808 x^9+64.1926 x^8 
+14.2419 x^7\NN\\
&& +50.8445 x^6+7.2371 x^5+25.8369 x^4+2.13986
   x^3+7.63944 x^2+0.280107 x   \Bigr)^{1/32}
 ,\NN\\
F_{15,2}^{(1/36)}
&=&
4\times \Bigl( 0.113713 x^{18}+1.1548 x^{17}+5.19491 x^{16}+39.7286 x^{14}+83.2077 x^{12} \NN\\
&&+117.012 x^{10} +113.748 x^8+75.8133 x^6+33.17 x^4+8.59437 x^2+1 \Bigr)^{1/36}.
\end{\eqa}

\newpage

\providecommand{\href}[2]{#2}\begingroup\raggedright\endgroup

\end{document}